\documentclass{aa}  

\usepackage[utf8]{inputenc}
\usepackage{comment}
\usepackage{enumitem}
\usepackage{graphicx}	
\usepackage{caption}
\usepackage{subcaption}
\usepackage{xcolor}
\usepackage{gensymb, amsmath,amssymb}
\usepackage{makecell}
\usepackage[hidelinks,colorlinks=true,linkcolor=blue,citecolor=blue,urlcolor=black]{hyperref}

\usepackage{txfonts}
\usepackage{lscape}

\begin{document}

    \title{CHEX-MATE: A LOFAR pilot X-ray -- radio study on five radio halo clusters}
    
    
    \titlerunning{CHEX-MATE: A LOFAR pilot X-ray -- radio study}
    \authorrunning{M. Balboni et al.}
    
    \author{M. Balboni \inst{\ref{iasf}, \ref{insubria}},
    F. Gastaldello \inst{\ref{iasf}}, A. Bonafede \inst{\ref{unibo},\ref{ira}}, A. Botteon\inst{\ref{ira}},  I. Bartalucci\inst{\ref{iasf}}, H. Bourdin\inst{\ref{infn_roma2},\ref{torvergata}}, G. Brunetti\inst{\ref{ira}}, R. Cassano\inst{\ref{ira}}, S. De Grandi\inst{\ref{brera}}, F. De Luca\inst{\ref{infn_roma2},\ref{torvergata}}, S. Ettori\inst{\ref{oas},\ref{infn_bo}}, S. Ghizzardi\inst{\ref{iasf}}, M. Gitti\inst{\ref{unibo},\ref{ira}}, A. Iqbal\inst{\ref{saclay}}, M. Johnston-Hollitt\inst{\ref{curtin}}, L. Lovisari\inst{\ref{iasf}}, P. Mazzotta\inst{\ref{infn_roma2},\ref{torvergata}}, S. Molendi\inst{\ref{iasf}}, E. Pointecouteau\inst{\ref{irap}}, G.W. Pratt\inst{\ref{saclay}}, G. Riva\inst{\ref{iasf}, \ref{unimi}}, M. Rossetti\inst{\ref{iasf}}, H. Rottgering\inst{\ref{leiden}}, M. Sereno\inst{\ref{oas},\ref{infn_bo}},
    R.J. van Weeren\inst{\ref{leiden}},
    T. Venturi\inst{\ref{ira}}, 
    I. Veronesi\inst{\ref{iasf}}}
    \institute{
        INAF - IASF Milano, via A. Corti 12, 20133 Milano, Italy \label{iasf}
        \and
        DiSAT, Universit\`a degli Studi dell’Insubria, via Valleggio 11, I-22100 Como, Italy \label{insubria}
        \and
        DIFA - Universit\`a di Bologna, via Gobetti 93/2, I-40129 Bologna, Italy\label{unibo}
        \and
        INAF - IRA, Via Gobetti 101, I-40129 Bologna, Italy\label{ira}
        \and
        IRAP, Université de Toulouse, CNRS, CNES, UT3-UPS, (Toulouse), France\label{irap}
        \and
        INAF, Osservatorio di Astrofisica e Scienza dello Spazio, via Piero Gobetti 93/3, 40129 Bologna, Italy\label{oas}
        \and
        INFN, Sezione di Bologna, viale Berti Pichat 6/2, 40127 Bologna, Italy\label{infn_bo}
        \and
        Universit\'e Paris-Saclay, Universit\'e Paris Cit\'e, CEA, CNRS, AIM, 91191, Gif-sur-Yvette, France \label{saclay}
        \and
        Leiden Observatory, Leiden University, PO Box 9513, 2300 RA Leiden, The Netherlands\label{leiden}
        \and
        INFN, Sezione di Roma 2, Università degli studi di Roma Tor Vergata, Via della Ricerca Scientifica, 1, Roma, Italy\label{infn_roma2}
        \and
        Università degli studi di Roma ‘Tor Vergata’, Via della ricerca scientifica, 1, 00133 Roma, Italy\label{torvergata}
        \and
        Dipartimento di Fisica, Universitá degli Studi di Milano, Via G. Celoria 16, 20133 Milano, Italy \label{unimi}
        \and
        INAF - Osservatorio Astronomico di Brera, via E. Bianchi 46, I-23807 Merate (LC), Italy\label{brera}
        \and
         Curtin Institute for Data Science, Curtin University, GPO Box U1987, Perth, WA 6845, Australia\label{curtin}
        }

   \abstract{
    The connection between the thermal and non-thermal properties in galaxy clusters hosting radio halos seems fairly well established. However, a comprehensive analysis of such a connection has been made only for integrated quantities (e.g. $L_X - P_{radio}$ relation). In recent years new-generation radio telescopes have enabled the unprecedented possibility to study the non-thermal properties of galaxy clusters on a spatially resolved basis. Here, we perform a pilot study to investigate the mentioned properties on five targets, by combining X-ray data from the CHEX-MATE project with the second data release from the LOFAR Two meter Sky survey. We find a strong correlation ($r_s \sim 0.7$) with a slope less than unity between the radio and X-ray surface brightness. We also report differences in the spatially resolved properties of the radio emission in clusters which show different levels of dynamical disturbance. In particular, less perturbed clusters (according to X-ray parameters) show peaked radio profiles in the centre, with a flattening in the outer regions, while the three dynamically disturbed clusters have steeper profiles in the outer regions. We fit a model to the radio emission in the context of turbulent re-acceleration with a constant ratio between thermal and non-thermal particles energy density and a magnetic field profile linked to the thermal gas density as $B(r) \propto n_{th}^{0.5}$. We found that this simple model cannot reproduce the behaviour of the observed radio emission.
    }
   
\maketitle
\section{Introduction}\label{sec:intro}

Galaxy clusters are excellent laboratories for studying non-thermal phenomena, such as particle acceleration processes on large scales. In the last decades, a growing number of studies evidenced the presence of radio synchrotron emission in many of these objects \citep[e.g.][for a review]{vanWeeren19}. In particular, the most puzzling cases of radio emission are the extended radio sources (e.g. radio relics, mini-halos, giant halos).
Giant radio halos are among the most extended cases of such emission. They are diffuse sources characterised by a steep spectral index ($\alpha < -1$, where $S_{\nu} \propto \nu^{\alpha}$) most frequently found in massive merging clusters.\\
Historically, two main scenarios have been proposed to explain radio halo origin: turbulent re-acceleration and hadronic models. 
In the former, relativistic electrons are re-accelerated via a Fermi-II like process by the turbulence injected in the Intra-Cluster Medium (ICM) by cluster-cluster merger events \citep{Brunetti2001,Petrosian2001}.
In the latter case, instead, synchrotron emitting electrons are produced as \emph{secondary} particles, i.e. as a decay product of heavy nuclei/particle collisions. These secondary models exploit cosmic ray protons (CRp) long lifetime over which they can diffuse on $\sim$Mpc scales and then, through collisions with ICM heavy nuclei, produce Cosmic Ray electrons (CRe) throughout all the cluster environment \citep[e.g.][]{Dennison1980, Blasi1999, PfrommerEsslin2004}. Such CRp can be produced mainly by AGN activity but also by accretion shocks and galactic outflows \citep{Brunetti14}.\\
The non-detection of gamma rays from $\pi^0$ decay from clusters \citep{Ackermann2016, Brunetti2017, Adam2021}, the connection of radio halos with mergers \citep{Cassano2010, Cassano2023, Cuciti21-2} and the spectral properties of halos \citep[e.g.][]{Blasi2001, Brunetti2008, Macario13,Bruno2021} suggest that the role of hadronic collisions is subdominant. 
Therefore, turbulence acceleration should be the driving mechanism for halo production and the required energy to obtain this (re-)acceleration is expected to be injected via merger events.\\
Indeed clusters form by accretion of sub-clusters and groups through mergers, which release up to $\sim 10^{64}$ ergs into the ICM in a cluster crossing time \citep[$\sim$ Gyr, ][]{Tormen2004}. Such energy is mainly dissipated as gas heating through shocks, enhancing the ICM thermal bremsstrahlung emission in the X-rays. However, a small fraction of this energy can be channelled into turbulence and re-accelerates relativistic electrons
triggering synchrotron radio emission \citep{Brunetti14}. As the clusters' mass sets the initial energy budget, it is expected that more massive systems host the most powerful radio halos which emit at higher energies, up to GHz frequencies.
Less energetic events, due to the merger of less massive systems, will give rise to less luminous halos, emitting only at low radio frequencies: the Ultra Steep Spectrum Radio Halos, a crucial class of objects to understand the radio halo formation mechanism and for which low frequency observations are providing a growing evidence of their existence. \citep[e.g.][]{Wilber2018,Duchesne2021, DiGennaro2021}.\\
\\
However, the details of the re-acceleration process are yet to be understood.
Based on which turbulence mechanism re-accelerates relativistic particles (transit time damping;  Alfven resonant scattering, etc.) the interplay between the various microphysical players (amplification of magnetic fields, heating of the plasma, etc) might lead to very different and complex scenarios \citep[e.g.][]{Brunetti-Lazarian07, Brunetti-Lazarian11, Miniati15, Brunetti-Lazarian16}.\\
%
Spatially resolved studies of radio and X-ray derived quantities are becoming possible for a large sample of objects in these years. They have the great potential to put constraints on the microphysics of the ICM and, in turn,  on the particle acceleration mechanisms \citep[e.g.][]{Govoni2001, Kamlesh2021-A2744}.\\
Recent works exploited the correlation among radio and X-ray surface brightness in the centre and in the periphery of clusters to derive constraints on the non-trivial distribution of the synchrotron emitting electrons \citep[e.g.][]{Biava21}. Similarly, from radio spectral index profiles it is now possible to constrain some of the crucial parameters of the model, such as the typical re-acceleration time \citep[$t_{acc}$, e.g.][]{Kamlesh2023-A2256}.  Specifically, LOFAR statistical studies on large samples follow the predictions of the simplest homogeneous models, with parameters that do not vary throughout the cluster volume \citep[e.g.][]{Cassano2023}. However when radio spectral index profiles are available that allow to constrain model parameters such as $t_{acc}$ this might not be the case \citep[e.g.][]{Kamlesh2023-A2256,Bonafede2022}. In this work, we found further evidence of that, with the model that poorly reproduces the observed quantities (Sec.~\ref{sec:modeling}).
\\
So far, spatially resolved studies have been performed using different radio and X-ray facilities, with different sensitivities and at different frequencies. All these effects can deeply impact the outcomes and make it difficult to compare results from different works. Therefore, a systematic radio and X-ray spatially resolved analysis is required to confirm or disprove the found relations and pose new tests for radio halos theoretical models.\\
With the present pilot work, which allows us to verify our analyses for future wider studies, we want to start the first systematic study of radio halo resolved brightness properties and their possible correlations with thermal ICM emission. In particular, we aim to study a representative sample of objects, for which we have uniform and homogeneous coverage in both radio and X-ray bands. This is possible by matching the Cluster HEritage project with XMM-Newton - Mass Assembly and Thermodynamics at the Endpoint of structure formation \citep[CHEX-MATE,][]{Arnaud21} and the LOFAR Two-meter Sky Survey Data Release 2 \citep[LoTSS DR2,][]{Shimwell2022} datasets, which provide the necessary data in the two requested bands and for a large sample of PSZ2 objects \citep{Botteon2022}, both in mass and in redshift.\\
We analyze the thermal -- non-thermal connection in a first subsample of five radio halo clusters, selected from a wider sample of 18 CHEX-MATE hosting radio halos among the 40 objects observed by LoTSS DR2 (the ongoing survey has already observed the totality of the CHEX-MATE sample in the Northern sky, 77 out of 82 objects).\\
%
The paper is organized as follows: in Section~\ref{sec:obs-data} we present observations and data reduction; in Section~\ref{Sec:sample} we introduce the studied objects; in Section~\ref{sec:data_analysis} we describe the data analysis; in Section~\ref{sec:discussion} we comment on our results. In Section \ref{sec:summary} we draw our conclusions.\\
As for the other papers of the CHEX-MATE collaboration, in this work, we assume a flat, $\Lambda$CDM Universe cosmology with $H_0 = 70 \rm ~ km/s/Mpc$ and $\Omega_{m,0}=0.3$ \citep{Arnaud21}.
\section{Observations and data reduction}\label{sec:obs-data}
\begin{table*}[ht!]
    \centering
    \begin{tabular}{c|c|c|c|c|c|c|c|c}
     PSZ2 Name & Abell Name & $z$ & \thead{${M_{500}}$ \\ $(10^{14} {M_{\odot}})$ } & \thead{$Y_{SZ}$ \\ $(10^{-3} ~ {\rm arcmin^{-2})}$} & \thead{ $P_{150 \rm MHz}$ \\ $\rm ( 10^{24} ~ W~Hz^{-1})$} & $c$ & $w ~ (\times 10^{-1})$ & M \\
    \hline
    PSZ2G046.88+56.48 & A2069 & 0.115 & 5.10 & $13.28 \pm  2.22$ & $8.36 \pm 0.98$ & $0.14_{-0.04}^{+0.04}$ & $0.33_{-0.18}^{+0.17}$ & 1.41 \\
    PSZ2G056.77+36.32 & A2244 & 0.095 & 4.34 & $9.14 \pm 1.68$ & $1.95 \pm 0.63$ & $0.49_{-0.06}^{+0.06}$ & $0.029_{-0.001}^{+0.004}$ & -0.81 \\
    PSZ2G066.41+27.03 & / & 0.575 & 7.69 & $1.21 \pm 0.16$ & $172.0 \pm 19.5$ & $0.15_{-0.02}^{+0.01}$ & $0.21_{-0.06}^{+0.04}$ & 0.74 \\    
    PSZ2G077.90-26.63 & A2409 & 0.147 & 4.99 & $4.97 \pm 0.86$ & $6.57 \pm 0.69$ &  $0.40_{-0.07}^{+0.06}$ & $0.046_{-0.010}^{+0.002}$ & -1.03 \\
    PSZ2G107.10+65.32 & A1758&  0.280 & 7.80 & $6.20 \pm 0.73$  & $36.9 \pm 1.1$&  $0.19_{-0.04}^{+0.03}$ & $0.45_{-0.10}^{+0.08}$ & 0.66 \\
    \end{tabular}
    \caption{Sample of the selected clusters present in LoTSS DR2 and CHEX-MATE observations, with their redshift, mass and $Y_{SZ}$ parameter within $\rm R_{500}$ \citep[from the Planck cluster catalogue PSZ2][]{Planck2016}, radio power at 150 MHz as found by \cite{Botteon2023} (since for PSZ2G077.90-26.63 and PSZ2G107.10+65.32 the radio power was not reported or reported for different regions we provide our estimate using the procedure described in Sec.~\ref{sec:radial_prof}), the light concentration ratio ($c$), centroid shift ($w$) and their combination (M) from \cite{Campitiello2022}.}
    \label{tab:1}
\end{table*}
\subsection{CHEX-MATE}

The CHEX-MATE project \citep{Arnaud21} is a three mega-second XMM-Newton Multi-Year Heritage Programme to obtain X-ray observations of a minimally-biased, signal-to-noise limited sample of 118 galaxy clusters detected by \textit{Planck} through the Sunyaev-Zel'dovich effect \citep{Planck2016}. The program aims to study the ultimate products of structure formation in time and mass, using a census of the most recent objects to have formed (Tier-1: $0.05 < {z} < 0.2$ ; ${M_{500}}$
\footnote{$ M_{500} = 500 \frac{4}{3} \rho_c R_{500}^3$ with $\rho_c$ the critical density and $R_{500}$ the radius within which the average cluster density is 500 $\rho_c$ }
$> 2 \times 10^{14}~M_{\odot}$), together with a sample of the highest-mass objects in the Universe (Tier-2: $z<0.6$; ${M_{500}}>7.25 \times 10^{14}~M_{\odot}$). The project acquired X-ray exposures of uniform depth that ensure a detailed mapping of the thermodynamic properties in the cluster volume where the non-thermal plasma is present. Therefore, the CHEX-MATE cluster sample is the best choice for a systematic and statistical analysis of cluster thermodynamic properties, allowing studies in a wide range of mass and redshift.
\subsection{X-ray data reduction}
The X-ray data used in this work have been reduced using the CHEX-MATE pipeline described in \cite{Bartalucci23} and here we report only the main steps.\\
The clusters were observed with the European Photon Imaging Camera (EPIC, \citealt{turner2001} and \citealt{struder2001}). Datasets were reprocessed using the Extended-Science Analysis System (E-SAS\footnote{\href{http://cosmos.esa.int/web/xmm-newton}{cosmos.esa.int/web/xmm-newton}}, \citealt{snowden08}) embedded in SAS version 16.1. Flare events have been removed by using the tools \emph{mos-filter} and \emph{pn-filter} analyzing the light curves in the [2.5-8.5] keV energy range. Time intervals with count rates exceeding $ 3 \sigma$ times the mean count rate have been excised. Point sources were filtered from the analysis following the scheme
detailed in Section 2.2.3 of \cite{ghirardini19} and further described in \cite{Bartalucci23}.\\
%
To produce the scientific images used for the analysis, we proceeded as follows. We extracted the photon-count images in the [0.7-1.2] keV band \citep[i.e. the band that maximizes the signal-to-noise of the cluster thermal emission][]{ettori10} for each camera and produced the exposure maps with the tool \emph{eexpmap}. 
The background affecting X-ray observations is due to a sky component, composed of the local Galactic emission and the Cosmic X-ray background \citep{kuntzsnowden2000}, and an instrumental component, due to the interaction of high-energy particles with the detector. We followed the strategy described in \cite{ghirardini19} to remove the latter one by producing background images accounting for the particle background and the residual soft protons, whereas we used a constant component in the profile to characterize the sky component as described in \cite{Bartalucci23}\\
The images, exposure, and background maps of the 3 cameras are merged to maximise the statistics following the procedure described in \cite{Bartalucci23}.
%
\subsection{LoTSS}

The LoTSS \citep{LoTSS2017} is a deep, $120 - 168$ MHz radio survey, that produces high resolution ($\sim 6"$) and high sensitivity ($\sim 100~\mu\rm{Jy~ beam^{-1}}$) images of the Northern sky. It had its first data release (LoTSS-DR1) in 2019 \citep[][released area  $\sim 400$ $\rm{deg^2}$ ]{LoTSS2019} and the second release in 2022 \citep{Shimwell2022} providing images and radio catalogs for $\sim 5,700~ \rm{deg^2}$ of the Northern sky. One of the main goals of this survey is to find new diffuse radio sources inside galaxy clusters, such as giant radio halos, to determine their origin and to test theoretical and numerical models. 
Thanks to its high sensitivity to diffuse sources at low frequencies, the LoTSS allows us to perform detailed studies of radio halos at low frequencies, where these objects show brighter emissions.\\
For this work, we will use the public data of the PSZ2 galaxy clusters covered in the LoTSS-DR2, available online \footnote{\url{https://lofar-surveys.org/planck\_dr2.html}} and that have been described in \cite{Botteon2022}.
\subsection{Radio data reduction}\label{sec:radio_data}
In the following, we report an overview of the calibration and imaging procedures applied to the radio data. The complete description of the reduction process is presented in \cite{Botteon2022}.\\
LoTSS pointings are typically obtained with an integration time of 8hr and in the 120-168 MHz frequency range. The collected data are processed with fully automated pipelines developed by the LOFAR Surveys Key Science Project team that aims to correct for direction-independent and direction-dependent effects. The pipelines are \textsc{prefactor} \citep{vanWeeren2016,Williams2016,deGasperin2019}, \texttt{ddf-pipeline} \citep{Tasse2021} and \texttt{facetselfcal} \citep{vanWeeren2021}. Before imaging, in order to correct for the inaccuracies of the LOFAR beam model, the images are scaled to align the flux density scale with the \cite{Roger1973} scale.\\
The imaging was done with WSClean v2.8 \citep{Offringa2014} adopting the \cite{Briggs1995} weighting scheme with \texttt{robust=-0.5} and applying Gaussian uv tapers in arcsec equivalent approximately to 25, 50, and 100 kpc at the cluster redshift. Finally, to better study the diffuse emission, discrete source components have been subtracted from visibility data and then new source subtracted images have been obtained with the same taper values.\\
Since we are particularly interested in studying the diffuse halo emission, in this work we use images with the further excision of compact source emission and with the lowest resolution. Additionally, prior to exploiting the final images, we perform a visual inspection checking for contaminating sources, both compact and extended. In particular, we search for residual source emission as a consequence of poor subtraction, such as tailed AGN, and for very low (high) surface brightness regions clearly not associated with the radio halo emission (e.g. diffuse emission clearly detached from the central halo, revived fossil plasma or cluster's sub-components). Hence, we manually mask these regions in both X-ray and radio images (see Fig.~\ref{fig:5-targets} for the excluded regions).
\section{Sample overview}\label{Sec:sample}
Among the 18 CHEX-MATE radio halo clusters, we selected five targets for a feasibility study: Abell 2069 (PSZ2G046.88+56.48), Abell 2244 (PSZ2G056.77+36.32), Abell 2409 (PSZ2G077.90-26.63), PSZ1G066.41+27.03. and Abell 1758N (PSZ2G107.10+65.32). This sample (i) allows a reasonable number of objects for a pilot study to tune the analysis for future wider studies, (ii) spans a similar range of mass and redshift of the CHEX-MATE--LoTSS DR2 sample and (iii) at the moment of the start of the project for these objects the X-ray data analysis was completed down to the spectral analysis.
These objects (see Tab.~\ref{tab:1}) span a broad range in redshift, mass, and merging status, as probed by the morphological parameter M \citep{Campitiello2022}. The latter quantity combines four different morphological indicators: light concentration ($c$), centroid shift ($w$) and the power ratios {$P_{20}$} and {$P_{30}$}. M is computed by summing the deviations of each parameter from the mean of its distribution
(computed over the whole CHEX-MATE sample) in units of standard deviation.
In this way, you can obtain a unique indicator of the cluster morphological status, where relaxed (perturbed) clusters have low (high) M values \citep[see][for further details on this parameter]{Campitiello2022}. By performing a Kolmogorov-Smirnov (KS) test we found a high probability ($\gtrsim 95\%$) for our five targets to be representative of the 18 clusters redshift and M distributions. The KS test on the mass
\footnote{All the derived masses reported in this work have been extracted from the MMF3 \textit{Planck} catalogue \citep{Planck2016}} 
distribution returns a lower value (70\%) than in the $z$ and M case, but still high enough to consider the sub-sample representative of the full sample.
Therefore, these objects represent an optimal explorative sample to test and tune our analyses for future wider studies.\\
General information about the five studied targets is found in Table~\ref{tab:1}, while radio and X-ray images are presented in Fig.~\ref{fig:5-targets}.
\begin{figure*}
    \captionsetup{labelformat=empty}
    \centering
    \includegraphics[width=8cm, height=6.3cm]{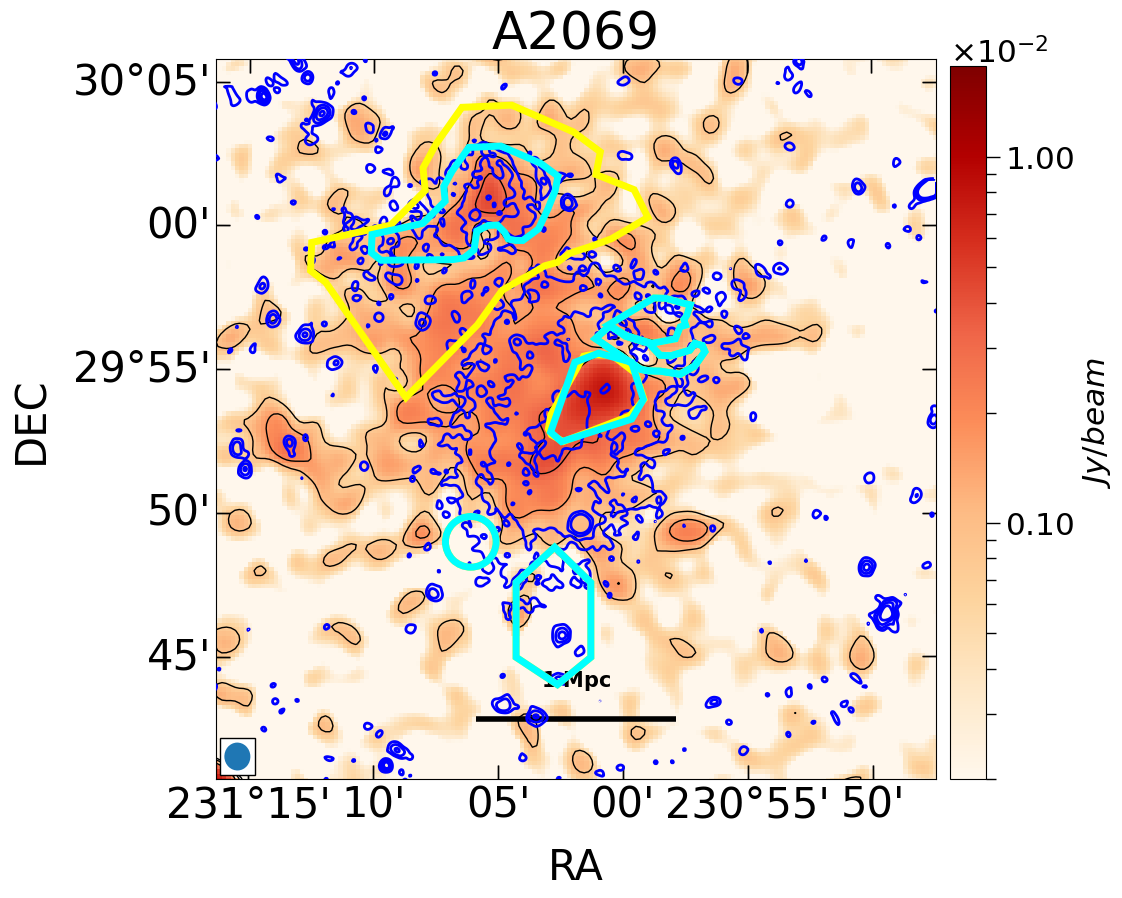}
    \includegraphics[width=8cm, height=6.3cm]{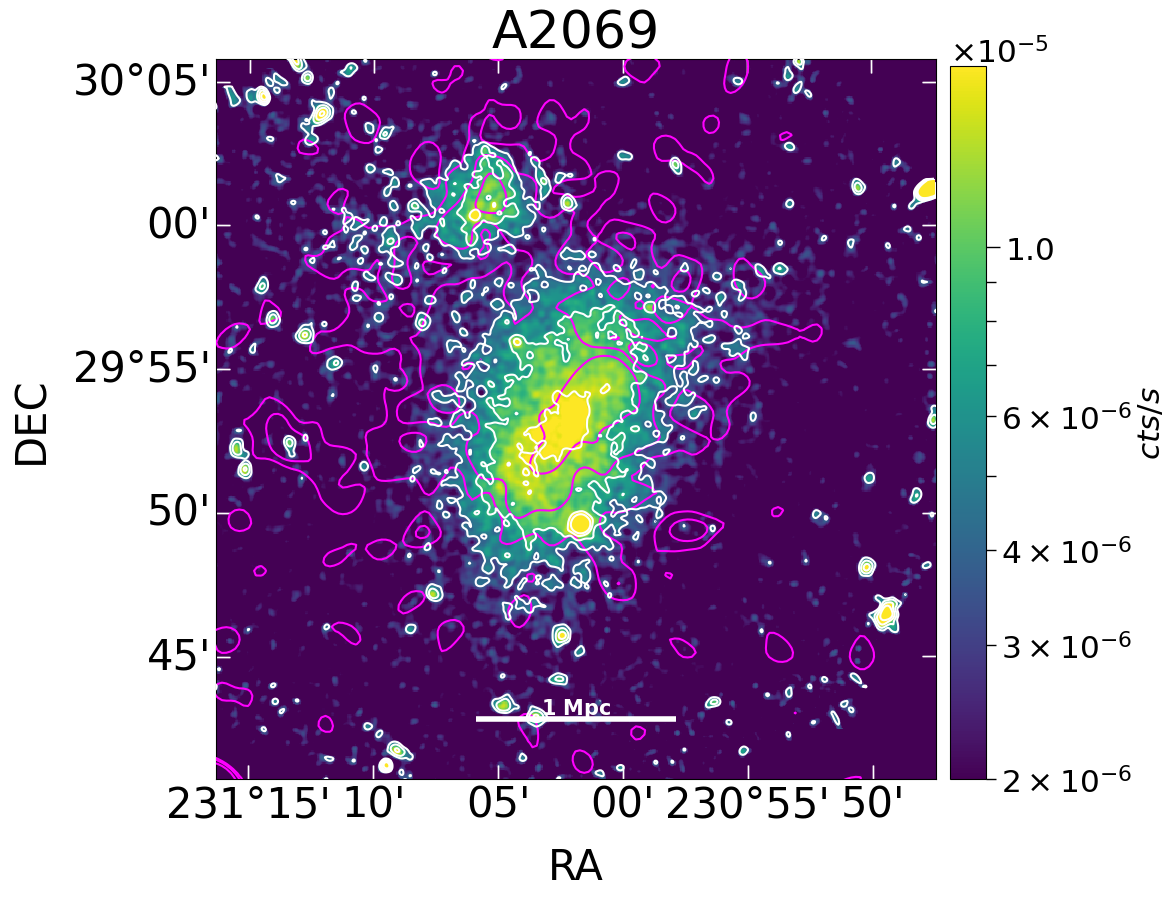}
    \includegraphics[width=8cm, height=6.3cm]{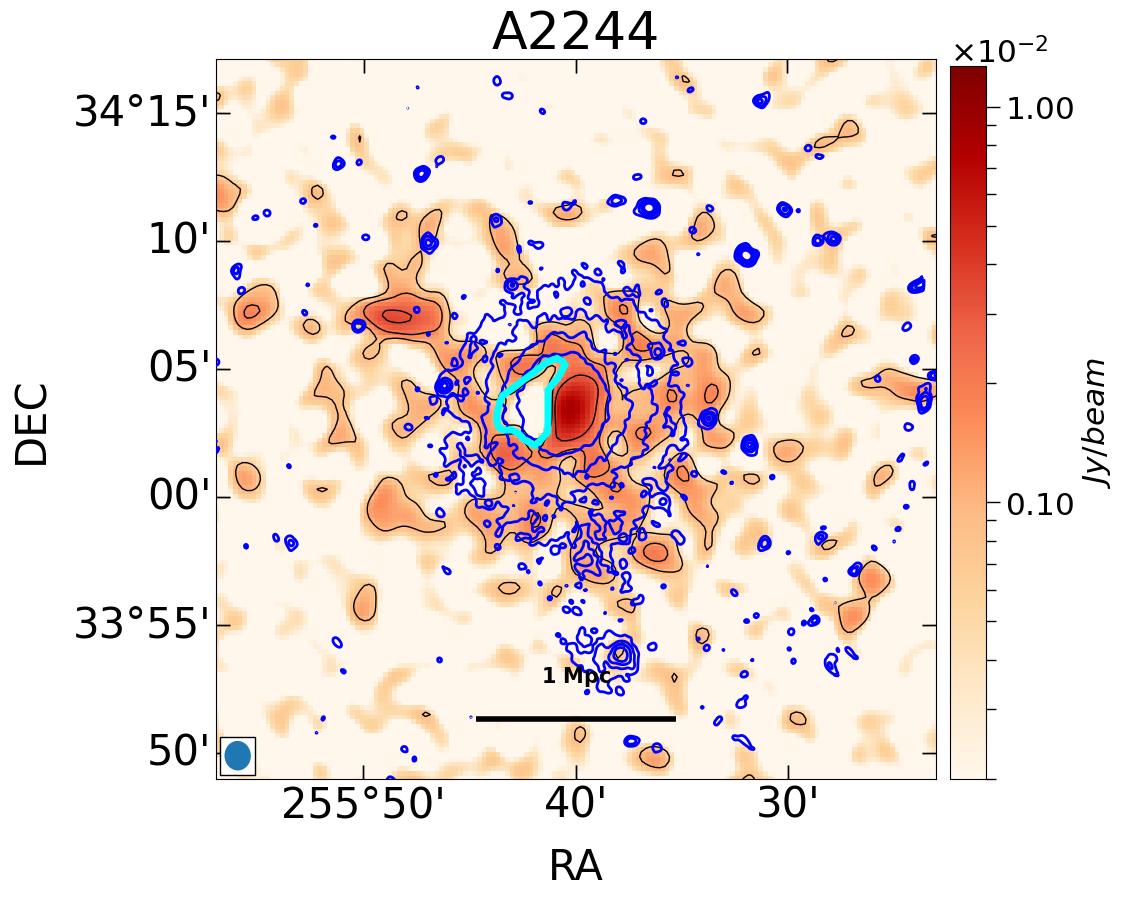}
    \includegraphics[width=8cm, height=6.3cm]{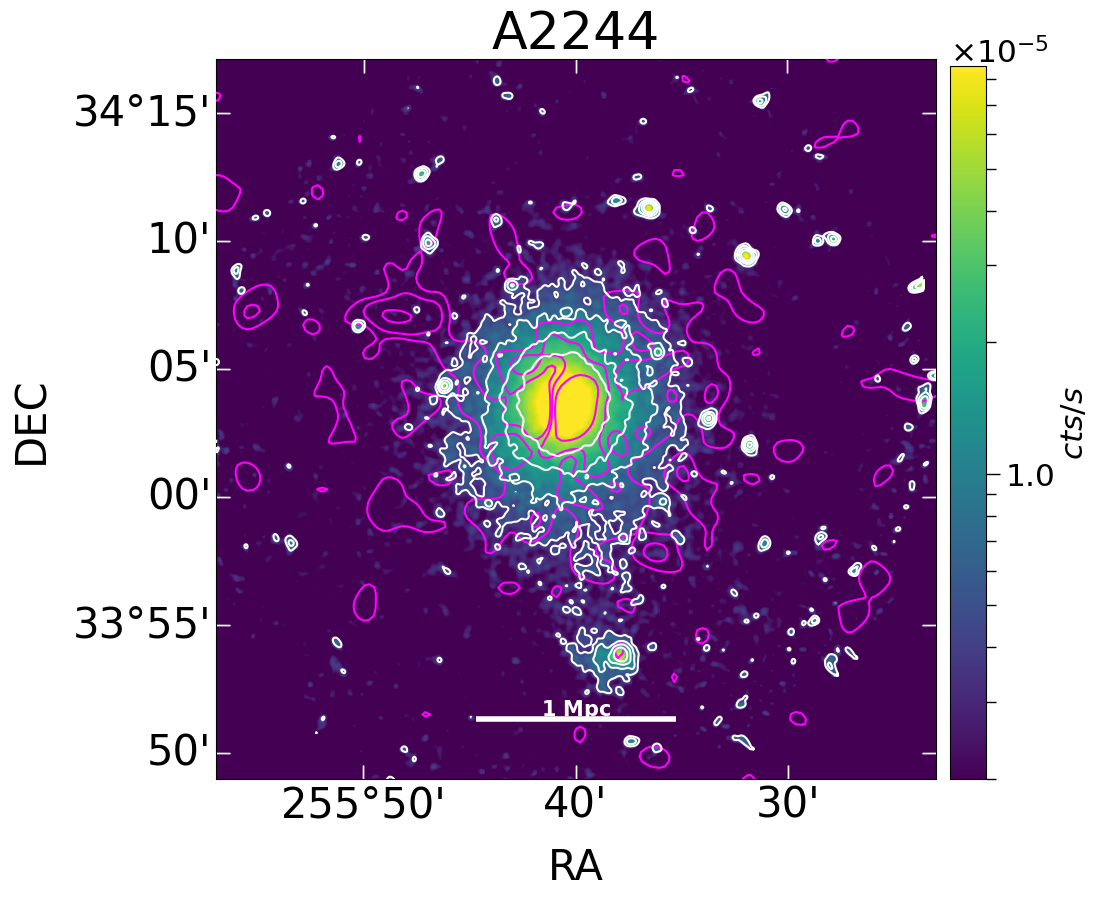}
    \includegraphics[width=8cm, height=6.3cm]{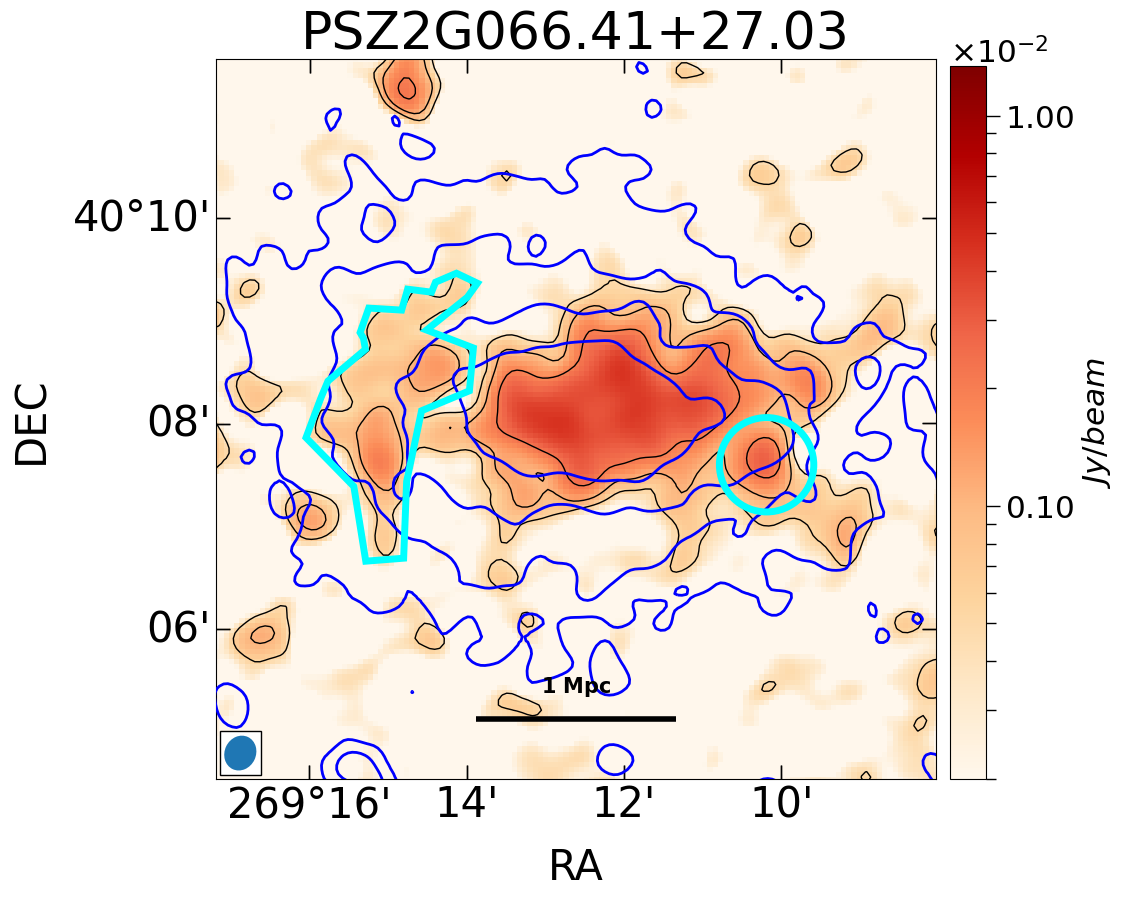}
    \includegraphics[width=8cm, height=6.3cm]{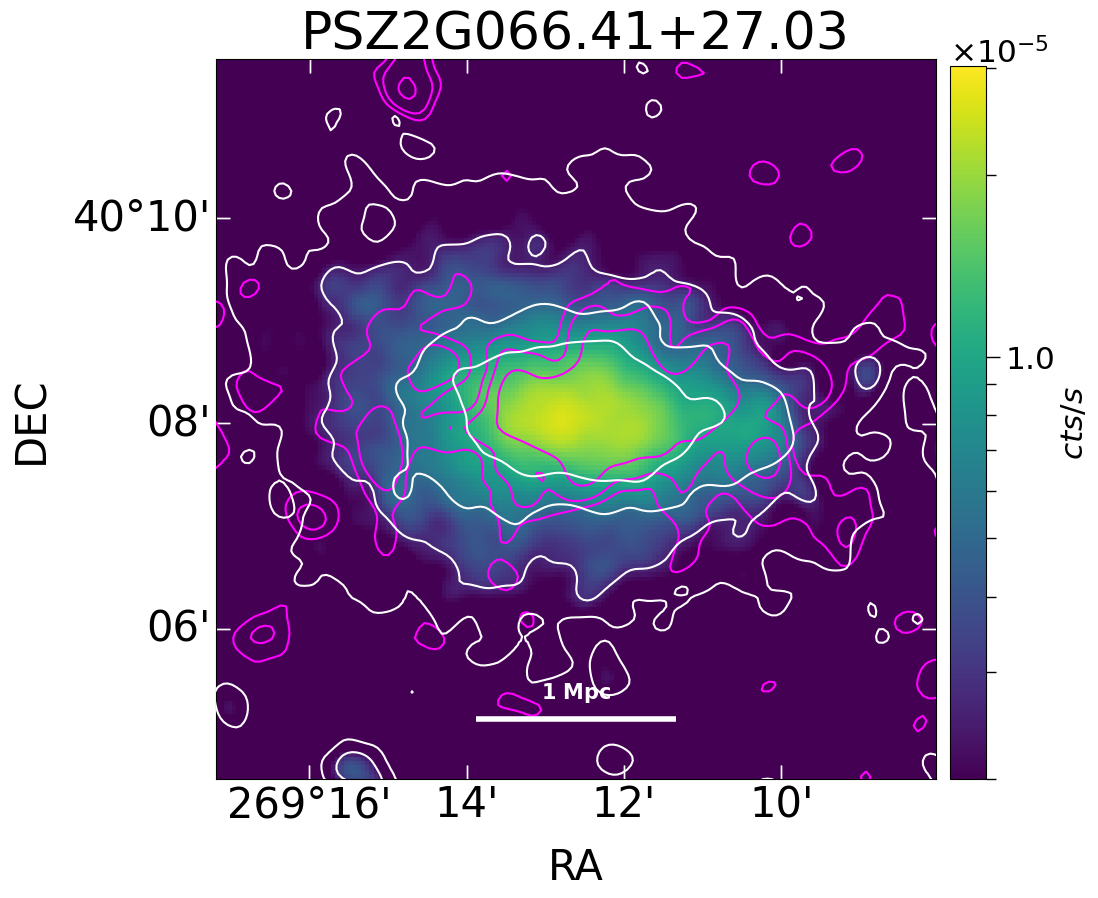}
    \includegraphics[width=8cm, height=6.3cm]{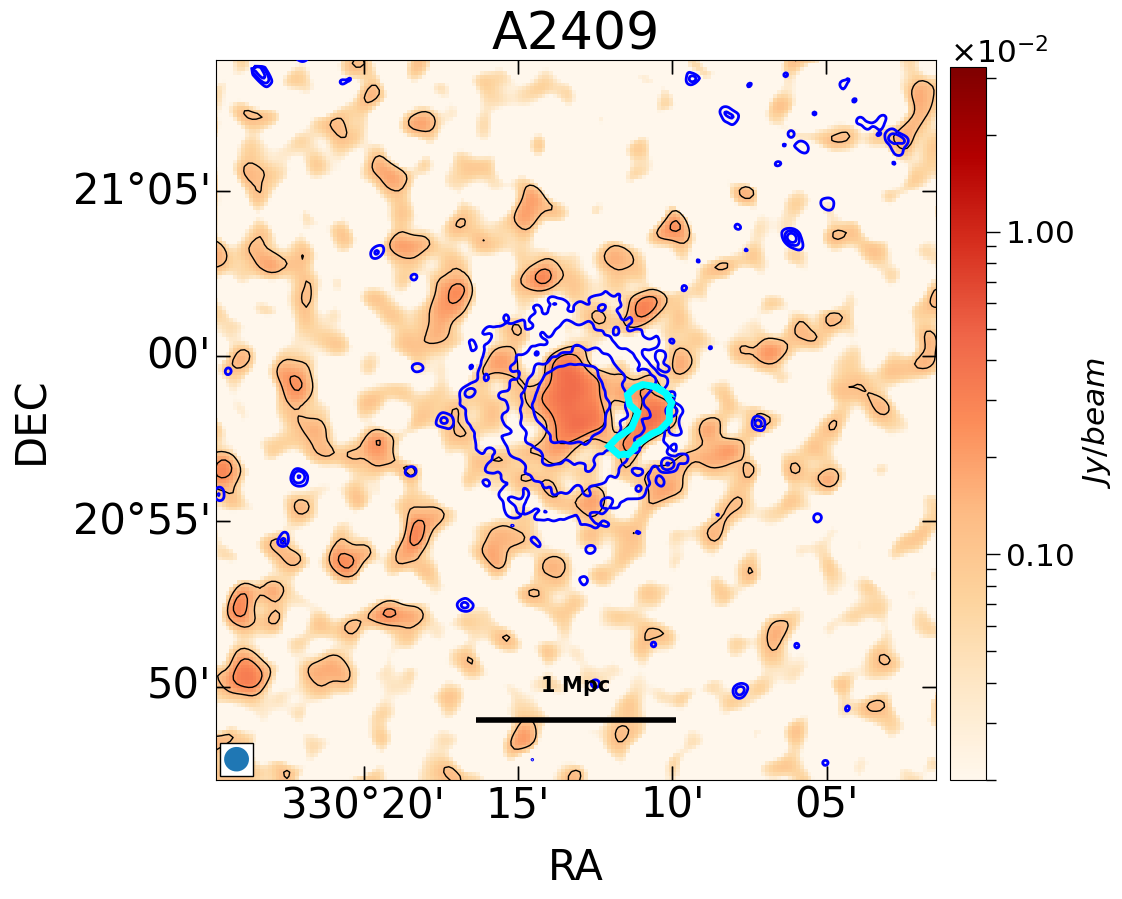}
    \includegraphics[width=8cm, height=6.3cm ]{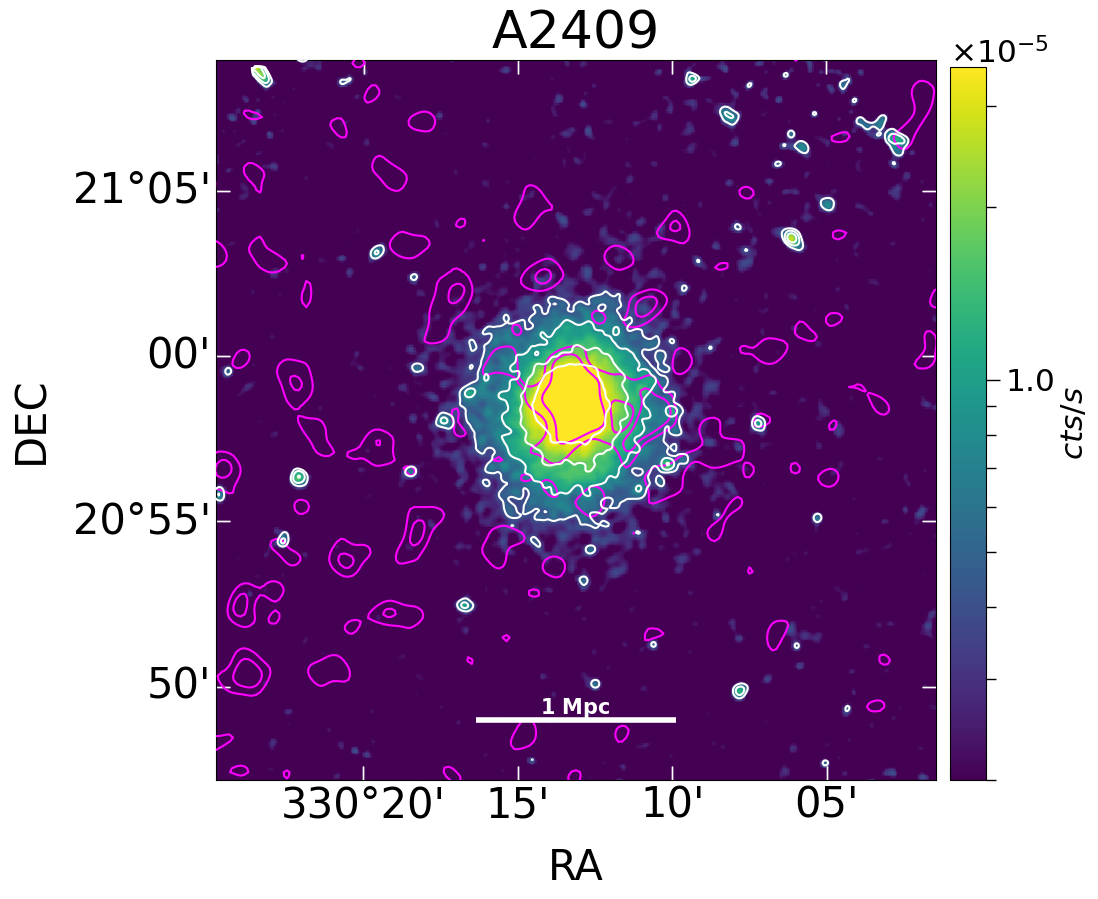}
\end{figure*}
\begin{figure*}
    \centering
    \includegraphics[width=8cm, height=8cm]{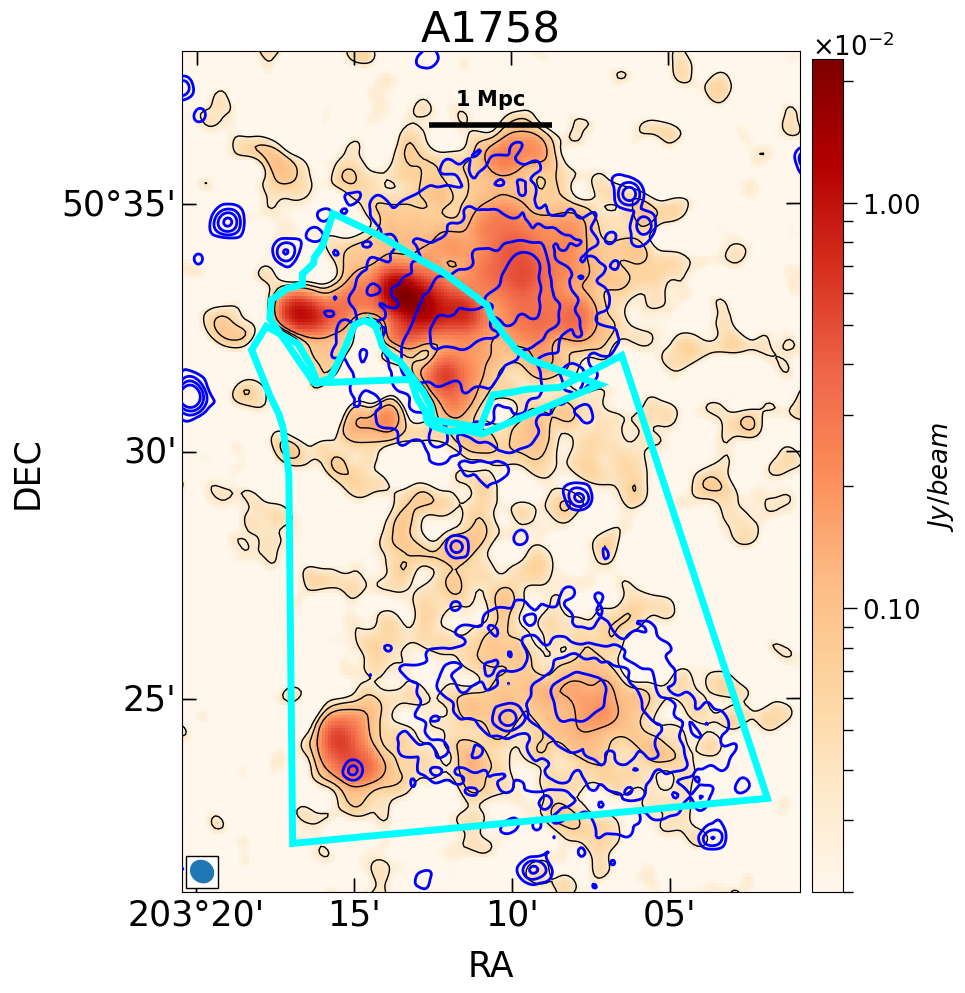}
    \includegraphics[width=8cm, height=8cm]{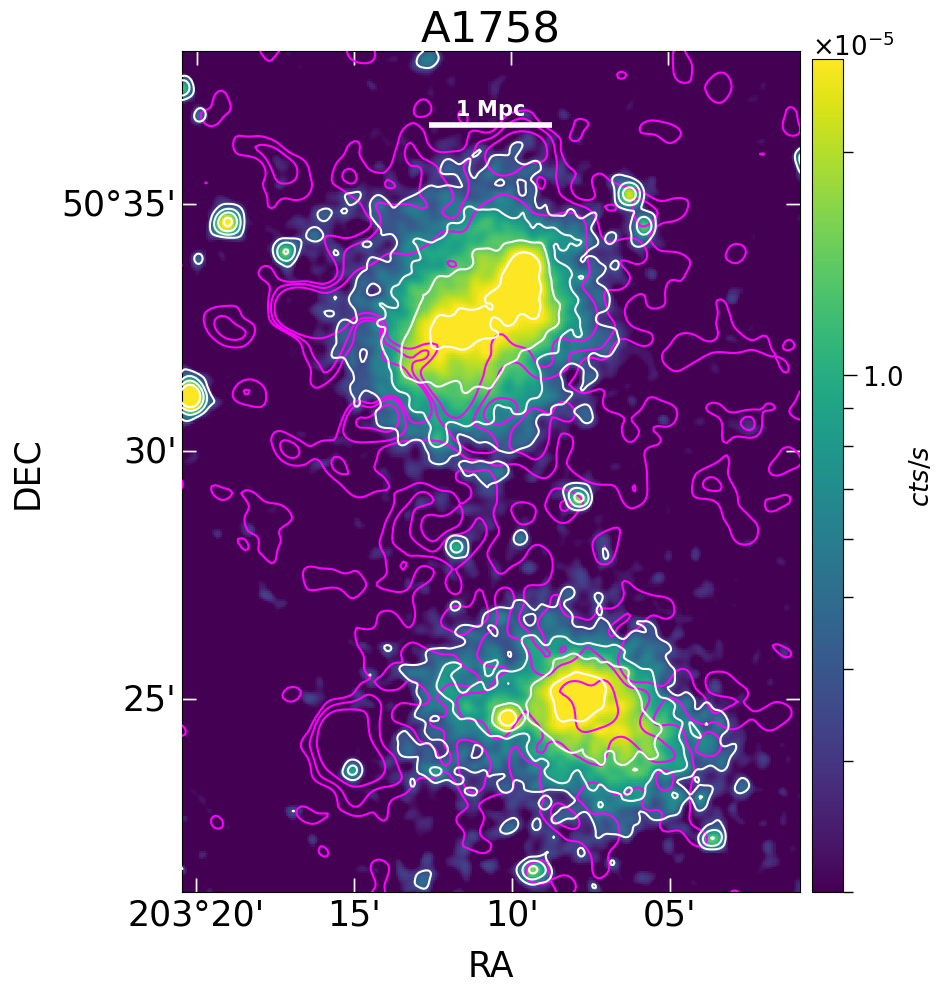}
    \caption{Radio (left) and X-ray (right) images of A2069, A2244, PSZ2G066.41+27.03, A2409 and A1758. Radio and X-ray contours are present in both images (radio contours in black and magenta; X-ray contours in blue and white). The radio contours are plotted at $2,4,8 \times {\rm \sigma_{RMS}}$, while the X-ray ones start at a level of $0.2 \rm ~ cts$ and are spaced with a factor of $2$. In cyan are highlighted the excluded regions (see Sec.~\ref{sec:ptp}). The radio image resolutions are, respectively: 58, 69, 20, 45 and 30 arcsec.
    For A2069 we also show a larger yellow region used to mask the entire subcomponent emission, as presented in Sec.~\ref{sec:disc_ptp}.}
    \label{fig:5-targets}
\end{figure*}
\subsubsection*{Abell 2069}
Abell 2069 (A2069) is a merging galaxy cluster with a $M_{500} = 5.10 \times 10^{14} {M_{\odot}}$ located at redshift z=0.115 which is part of the "A2069-supercluster" \citep{Einasto97}. It consists of two merging components (A2069-A and A2069-B), which are well detected by X-ray images. In the main component A2069-A, two bright elliptical galaxies are present, separated by a projected distance of about 55 kpc \citep{Gioia82}. In X-rays it shows an elongated thermal emission which seems to be connected with the X-ray emission of A2069-B. 
Faint and extended non-thermal emission has been detected
by \cite{Farnsworth13} who classified it as a radio halo. This result was also confirmed by \cite{Drabent15}. \cite{Drabent15} detected diffuse radio emission coming from A2069-B, where Chandra observations showed also the presence of a cold front \citep{Owers09}.
The radio diffuse emission from both A2069-A and A2069-B is detected also in the 144 MHz LOFAR images, as well as an emitting region among the two components. In this work, we show how this connecting region may be separated by the A2069-A halo emission through the radio--X-ray analysis (see also Sec.\ref{sec:disc_ptp}). The detailed analysis of the LOFAR observations of this target will be presented by Drabent et al. (in preparation).
\subsubsection*{Abell 2244}
Abell 2244 (A2244) is galaxy cluster located at z=0.095 and its SZ derived mass is ${M_{500} = 4.34 \times 10^{14}~M_{\odot}}$.
X-ray morphological indicators classify A2244 as a relaxed object, indeed the concentration parameter (c) and centroid shifts (w) measured by \cite{Zhang2023} would assign this object to the quadrant of relaxed objects as defined in \cite{Cassano2010}.
A similar classification was provided by \cite{Botteon2022} and \cite{Campitiello2022}. 
Despite the regular and centrally peaked X-ray morphology, \cite{Donahue2005} reported the lack of a central temperature gradient and a central entropy intermediate between the ones of cool cores and non-cool cores, all signs of a dynamical disturbance.
The likely candidate for such a disturbance is a group of galaxies that interacted with the main cluster leaving a consistent trail of gas in the South. Given the mass of the object as derived from the SZ signal of $M_{500} = 4.3 \times 10^{14} {M_{\odot}}$ and the mass of the group to be in the range $ M_{500}= 2-5 \times 10^{13} {M_{\odot}}$ as typical for a 1 keV system (the central temperature measured in the group from our X-ray analysis) we have a relatively off-axis merger with a mass ratio of 10-20.\\
High sensitivity LOFAR 144 MHz observations of A2244 revealed Mpc-scale diffuse emission, not associated with compact sources, likely caused by the turbulence generated by the group passage. \\
%
\subsubsection*{PSZ2G066.41+27.03}
PSZ2G066.41+27.03, is the highest redshift cluster of our sample (and the 2nd highest of all CHEX-MATE), z=0.575, and it is also quite massive, $M_{500} = 7.69 \times 10^{14} {M_{\odot}}$. No studies have been performed on this target so far.\\
PSZ2G066.41+27.03 shows powerful X-ray diffuse emission with a complex and elongated structure.
The morphological parameters of this object would classify it as clearly disturbed, both estimating them within 500 kpc and within $R_{500}$ \citep{Botteon2022,Campitiello2022,Zhang2023}.\\
The radio halo emission detected by LOFAR in PSZ2G066.41+27.03 was claimed by \cite{Botteon2022} for the first time. At these frequencies, the target shows a bright and extended diffuse radio emission, with few compact sources in the field.\\
\subsubsection*{Abell 2409}
Abell 2409 (A2409) is a low mass galaxy cluster with $M_{500} = 4.99 \times 10^{14} {M_{\odot}}$ and located at z=0.147. As for PSZ2G066.41+27.03, no dedicated studies have been performed on this target. 
Its morphological parameters classification is not so straightforward, with $\rm{c} = 0.4 \pm 0.07$ and $\rm{w} = (4.6 \pm 1) \times 10^{-3}$ that made \cite{Campitiello2022} classify this object as "mixed" (in agreement with the classification made by \citealt{Botteon2022} which puts A2409 close to the disturbed quadrant of \citealt{Cassano2010}). From the CHEX-MATE X-ray images, the overall shape is fairly roundish, however, some inhomogeneities at the very centre of the emission are observed.
The most evident one is a small elongated emission in the North-South direction and a probable surface brightness discontinuity in the South-West region. 
\\
LoTSS observations show relatively limited diffuse emission coming from A2409, which \cite{Botteon2022} classified as radio halo emission. However, the authors noted that it was not possible to reconstruct a radio halo profile for this object due to its low surface brightness. 
In the LoTSS image, an elongated structure is located in the West direction, apparently not associated with radio galaxies or other compact sources.
%
\subsubsection*{Abell 1758}\label{sec:a1758}
Abell 1758 (A1758) is a cluster pair located at z=0.278 and where both the two components (North and South) are undergoing major merger events \citep[e.g.][]{Schellenberger2019}.
Because of its higher mass ($M_{500} = 7.80 \times 10^{14} {M_{\odot}}$
\footnote{Our result is based on SZ observation from \cite{Planck2016}. However due to the proximity of the two clusters on the plane of the sky, the estimate of both A1758-N and A1758-S mass is a non-trivial exercise. Therefore, the mass estimate of A1758N can be overestimated, even though of a small factor since it remains the most massive object of the A1758 system.}) 
A1758-N is the most studied cluster of the system. Lensing studies found a bimodal mass distribution of two subclusters, A1758-N East and A1758-N West \citep{M-O17}.
In the radio band, the radio halo emission in A1758-N was investigated from 1.4 GHz to 54 MHz
\citep{Kempnere&Sarazin01,Giovannini09,Venturi13,Botteon18,Botteon20}. Instead, A1758-S shows a faint radio halo emission discovered by \cite{Botteon18}. This diffuse emission is also difficult to disentangle from the inter-cluster emission present in this system, which is prominent in the 144 MHz images \citep{Botteon20}.\\
%
A1758-NW shows evident diffuse halo emission. In the East region, instead, there are two emission blobs apparently not associated with any optical galaxy and that might be regions where the plasma might has been somehow locally compressed or re-accelerated. Given their complex nature, it is non-trivial to determine what has originated them and if and how they are associated with larger scale emission \citep[e.g. halo substructures, AGN fossil plasma; see][for more information on the diffuse sources in A1758-NE.]{Botteon18}. Therefore, we exclude these cluster regions from our analysis, focusing our study only on A1758-NW. 
The difference between the A1758-NW and A1758-NE components radio emission emerges also in this work, where we find a better agreement between radio and X-ray surface brightness when excluding the NE part and A1758-S.
\section{Data analysis and Results}\label{sec:data_analysis}
%
%
\subsection{Point-to-point analysis}\label{sec:ptp}
The pioneering work by \cite{Govoni2001} showed the importance of the study of the correlation between the X-ray and radio brightness and posed the basis of its use to derive information on the mechanism responsible for radio emission. This approach has been developed in many subsequent studies up to the recent developments exploiting the strength of low frequency data \citep[e.g.][]{Kamlesh-M0717}.
%
Past works have shown that perturbed galaxy clusters tend to present a positive correlation between radio and X-ray emission which, in general, is found to be sub-linear. \citep[e.g.][]{Giacintucci2005,Cova2019,Xie2020,Hoang2021}.
The majority of the results point toward a weaker radial decline of the non-thermal component with respect to the thermal one. However, differences in the instrument, affecting the observing frequency and resolution, and the adopted analysis technique can impact the final results, making it difficult to draw general conclusions. \citep[e.g.][]{Shimwell2014,Botteon2020-A2255,Kamlesh2021-A2744,Bonafede2022,Bullet-MeerKAT-2022}\\
Thanks to LoTSS and CHEX-MATE high sensitivity images in radio and X-ray bands, we can now perform a homogeneous thermal--non-thermal analysis on a spatially resolved basis for a sample of clusters. This also allows us to compare the results for different clusters and perform a statistical study on cluster radio -- X-ray relations.\\
To perform such a study on our sample, we extract the average surface brightness from radio ($I_R$) and X-ray ($I_X$) images by constructing a grid \citep[following ][]{Govoni2001} that covers the whole radio halo emission, which, in general, appears less extended than the thermal one. Each box of the grid has an area equal to the radio beam resolution, which is typically larger than the XMM-Newton point spread function in the X-rays.
Following \cite{Botteon2020-A2255}, we consider only the cells where the average radio surface brightness is above $2 \times \sigma_{\rm RMS}$. 
Prior to extracting the values from the grid, we took particular care in the excision of sources which are unrelated to the diffuse radio halo emission, as detailed in Sec.~\ref{sec:radio_data}.\\
%
Once the grids are set, we proceed with the surface brightness estimation in each box from both radio and X-ray images. In particular, we derive the cluster X-ray count rate from the background-subtracted and exposure-corrected images.
Given the homogeneous coverage in both the X-ray and radio bands of the sample and the uniformity of the data, we aim to directly compare all the $I_R - I_X$ correlations of all clusters. Since we want to use physical units for the X-ray images, we perform the conversion from detector units $(count/s)$ to physical units ($erg/cm^2/s$). We calculate the conversion from count-rates to fluxes using a thermal spectral model of the source, with the mean temperature as defined in \cite{Bartalucci23}, abundance fixed at $Z=0.3$ (using \citealt{Asplund2009} solar tables) modified by Galactic absorption. The response files used are the ones for the MOS2 camera on-axis as this is the reference camera for the merged image \cite[see][]{Bartalucci23}, where the exposure map is scaled according to the response of that camera. Finally, the brightness values we report are the unabsorbed ones and k-corrected to have a 0.7-1.2 keV rest-frame flux \citep[see][]{Jones1998}.
For the radio band, instead, the k-correction is accounted for by dividing the extracted brightness for the term $(1+z)^{1+\alpha}$ assuming a spectral index $\alpha$ = - 1.3, as made by \cite{Botteon2022}. Finally, we also account in both bands for the cosmological dimming factor $(1+z)^4$.\\
We have investigated the thermal -- non-thermal relation fitting $I_X$ and $I_R$ with a power law :
\begin{equation}
    {\rm log} I_R = {\rm A} ~  {\rm log} I_X + {\rm B},
\end{equation}
where the slope A tells us whether the radio plasma components (CRe and magnetic fields) decline faster or slower with radius than the thermal ICM distribution.\\
Using the python package \text{PyMC} \citep{py_mc16,pymc_v4.2}, we developed a Bayesian regression model that allows us to perform a linear fit of ${\rm log} I_R$ on ${\rm log} I_X$ accounting for x and y errors, selection effects (i.e. Malmquist bias, \citealt{Malmquist1922}) and the intrinsic scatter in the regression relationship (see also Appendix~\ref{appendix: test_ptp}).
%
%
%
%
\begin{table}
    \centering
    \begin{tabular}{c|c|c}
         Object & $\rm r_S$ unmasked & $\rm r_S$ masked\\
         \hline
          A2069 & 0.50 & 0.54\\
          A2244 & 0.63& 0.68\\
          PSZ2G066.41+27.03 & 0.71 & 0.78\\
          A2409 & 0.45 & 0.76\\
          A1758-NW & 0.57 & 0.84\\
    \end{tabular}
    \caption{Spearman correlation coefficients for the $I_R - I_X$ relations of each cluster with and without considering "contaminant" regions.}
    \label{tab:2}
\end{table}
%
%
\begin{figure}
    \centering
    \includegraphics[scale=0.23]{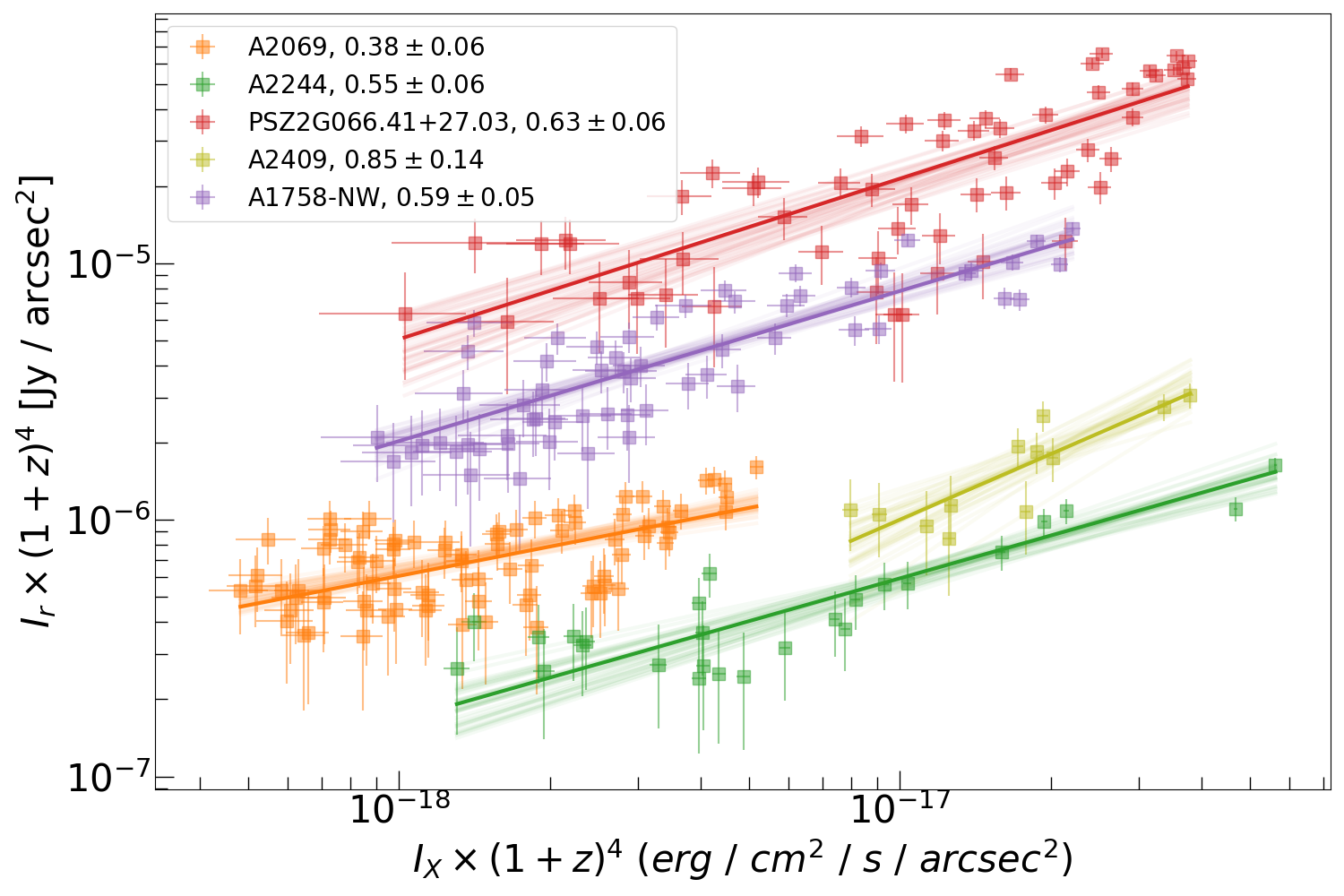}
    \caption{$I_R - I_X$ relations of all the clusters in a single plot. The label reports the cluster names together with the best-fit slopes and their errors. The best-fit lines are the colored solid lines.}
    \label{fig:Ix-Ir_tot}
\end{figure}
\begin{table}[]
    \small
    \centering
    \begin{tabular}{c|c|c|c}
         Object & $ \rm A \pm \sigma_A $ & $ \rm B \pm \sigma_B $ & $\sigma_{int}$ \\
         \hline
          A2069 & $0.38 \pm 0.06$ & $0.61 \pm 0.02$ & $0.20 \pm 0.05$ \\
          A2244 & $0.55 \pm 0.06$ & $3.19 \pm 0.03$ &  $0.14 \pm 0.08$\\
          PSZ2G066.41+27.03 & $0.63 \pm 0.06$ & $5.97 \pm 0.04$ & $0.50 \pm 0.12$\\
          A2409 & $0.77 \pm 0.15$ & $7.22 \pm 0.03$ & $0.16 \pm 0.11$\\
          A1758-NW & $0.59 \pm 0.05$ & $4.84 \pm 0.02$ & $0.30 \pm 0.05$\\
    \end{tabular}
    \caption{Correlation slopes, intercepts and recovered intrinsic scatter of the $I_X-I_R$ relation for the five studied targets.}
    \label{tab:3}
\end{table}
Figure \ref{fig:Ix-Ir_tot} (see also Fig.~\ref{fig:PtP-plot_single} in Appendix~\ref{appendix:ptp-single}) shows the point-to-point comparisons between the radio brightness at 144 MHz and the X-ray brightness in the range 0.7-1.2 keV for the five objects.\\
We observe a strong correlation between X-ray and radio brightness. The strength of those correlations is shown also by high Spearman correlation coefficients ($\rm r_S$), which are reported in Tab.~\ref{tab:2}. 
We also notice that the correlations appear stronger when excluding the contaminant regions from the images.
Figure \ref{fig:Ix-Ir_tot} also shows that, for each of the clusters, a positive $I_R - I_X$ correlation exists. All correlations are sub-linear, implying a shallower radial decline of the non-thermal ICM component with respect to the thermal one. We report in Tab.~\ref{tab:3} the best-fit values for the slope, intercept and intrinsic scatter of linear relation.\\
In Fig.\ref{fig:Ix-Ir_tot}, we see the surface brightness differences of our objects. In particular, the two higher mass clusters (A1758-N and PSZ2G066.41+27.03) present brighter radio emission.
The brightest X-ray, emitting regions are found in A2244 and A2409 which are also the two more relaxed objects of our sample \citep{Campitiello2022} and then a higher central peak might be expected. We also note the absence of any correlations between mass, morphological parameters \citep[M-parameter, $c$ and $w$, from][]{Campitiello2022} and the recovered $I_R - I_X$ correlation slope. Due to the small size of our sample, we cannot make any general claim on this particular lack of correlation and we will further investigate this kind of connection with the full sample.

\subsection{Radial analysis}
Past works have shown how the radial analysis of thermal and non-thermal properties provides additional information about the ICM environment \citep[e.g. interaction between different ICM components and on the properties of CRe and magnetic fields at large radii;][]{Govoni2001-altro, Pearce2017, Kamlesh2021-A2744, Bruno2021, Botteon2020-A2255, Cuciti22, Kamlesh2023-A2256, Bruno2023}. 
Therefore, we proceed to analyze the radial emission properties in both radio and X-ray bands.
\subsubsection{Radial profiles}\label{sec:radial_prof}
We extracted radial profiles in both X-ray and radio bands. Given the fact that the radio emission is generally less extended than the X-ray one, the radial profile extraction is guided by the halo emission and starts from the radio halo centre.\\
We fit the radio profiles using the Halo-Flux Density CAlculator \citep[Halo-FDCA,][]{HFDCA-2021}. This code directly fits the two dimensional surface brightness profile with an exponential model (see also Sec.~\ref{sec:rad_prof_discussion}) and uses a Markov chain Monte Carlo method for uncertainties estimation. Since this code allows also us to choose the exponential model geometry (circle, ellipse, rotated ellipse and skewed), we exploit the information provided by \cite{Botteon2022} on the halo profile to select the best-fitting model. \\
The surface brightness model is given by:
\begin{equation}
    I(r) = I_0 ~ e^{-{G(r)}},
\end{equation}
where $I_0$ is the central surface brightness and $G(r)$ is a radial function that takes different forms depending on the selected model. For our cases of circle and rotated ellipse models we adopt, respectively:
\begin{align}
    & G_{circ}(r) = \frac{r}{R_e},\\
    & G_{ell}(r) = \left [ \left( \frac{X_{\phi}}{r_x} \right)^2 + \left( \frac{Y_{\phi}}{r_y} \right)^2 \right ] ^{0.5},
\end{align}
where:
\begin{equation}
\begin{pmatrix}
    X_{\phi}\\
    Y_{\phi}
\end{pmatrix}
= 
\begin{pmatrix}
    \cos \phi ~ \sin \phi \\
    -\sin \phi ~ \cos \phi
\end{pmatrix}
\begin{pmatrix}
    x\\
    y
\end{pmatrix},
\end{equation}
where $R_e$ is the characteristic e-folding radius and $r^2 = x^2 + y^2$. For the data preparation process we followed the procedure described by \cite{Botteon2022}. We have masked contaminant regions (see Sec:~\ref{sec:ptp} and Fig.~\ref{fig:5-targets}) and reduced the Field of View (FoV) of each radio image to approximately $1.5 R_{500} \times 1.5 R_{500}$ before passing them as input to the code. From the Halo-FDCA output, we determine the centre of each halo. Then we extract the radio profile using linearly spaced annuli up to $2.5~R_e$ (the outer limit where we expect to find halo emission, \citealt{Bonafede2017}). Following \cite{Cuciti21-1}, we choose the width of the annuli to be half of the Full-Width at Half Maximum of the image beam.  We performed the profile extraction based on the shape of the halo, using circular or elliptical annuli depending on what has been used for the halo fitting.\\
\begin{table*}[th!]
    \centering
    \begin{tabular}{c|c|c|c|c|c}
         Object & Halo model &$\chi_r^2$ & $I_0 ~ (\mu Jy ~ arcsec^{-2}) $ & $r_x$ (kpc) & $r_y$ (kpc) \\
         \hline
          A2069 & Circular & $0.96$ & $1.17 \pm 0.09$ & $ 516 \pm 19$ & / \\
          A2244 & Circular & 0.87 & $0.96 \pm 0.10$ & $ 256 \pm 23$ & / \\
          PSZ2G066.41+27.03 & Elliptical & $1.4$ & $15.7 \pm 0.53$ & $ 468 \pm 17$ & $220 \pm 8$ \\
          A2409 & Circular & $0.93$ & $1.45 \pm 0.22$ & $ 315 \pm 42$ & / \\
          A1758-NW & Elliptical & $1.25$ & $ 6.06 \pm 0.24$ & $ 322 \pm 13$ & $261 \pm 13$\\
    \end{tabular}
    \caption{Results of the radio exponential fit with Halo-FDCA.}
    \label{tab:HFDCA_fit}
\end{table*}
\begin{figure*}[h!]
    \centering
    \includegraphics[width=8cm, height=6.2cm]{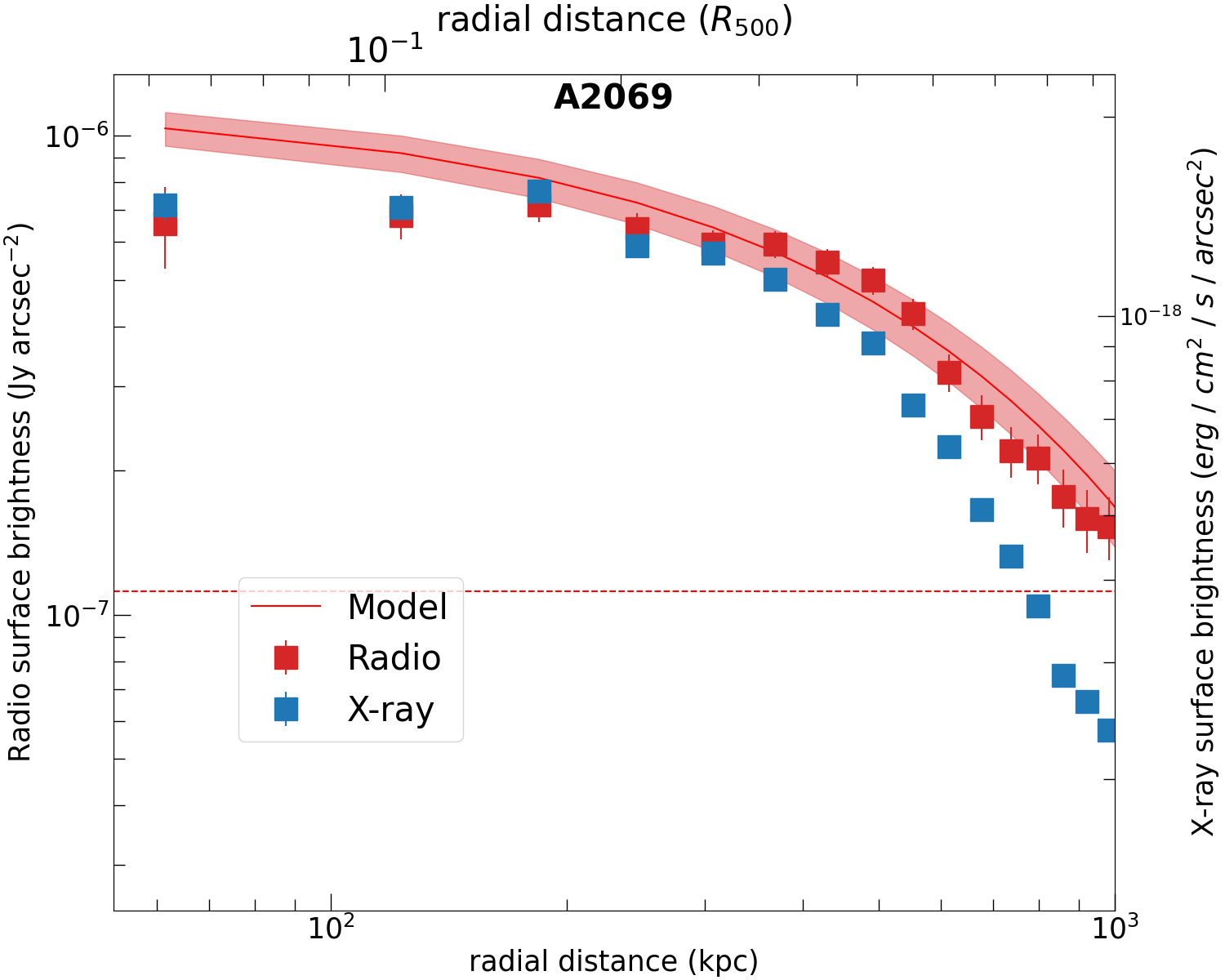}
    \includegraphics[width=8cm, height=6.2cm]{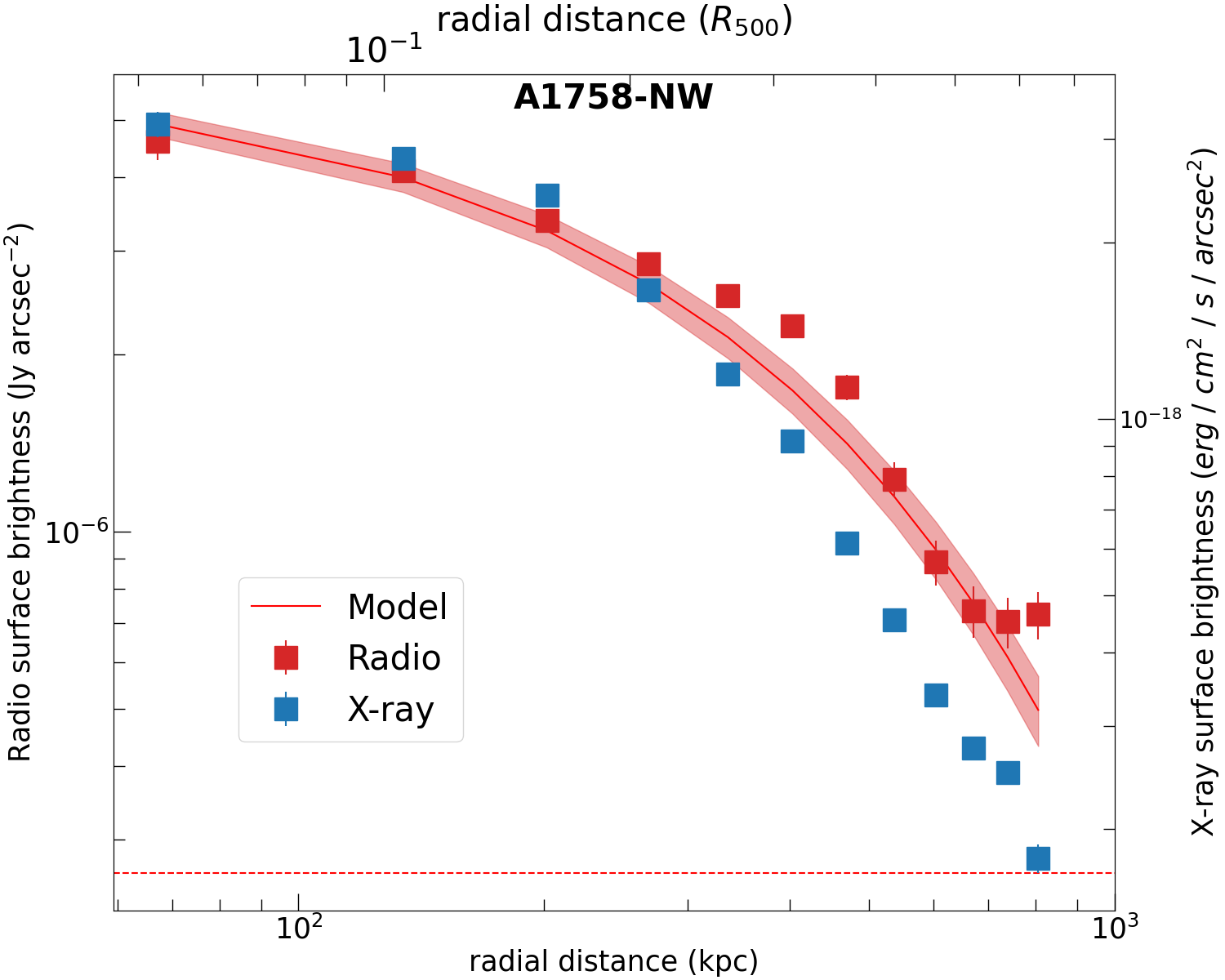}
    \includegraphics[width=8cm, height=6.2cm]{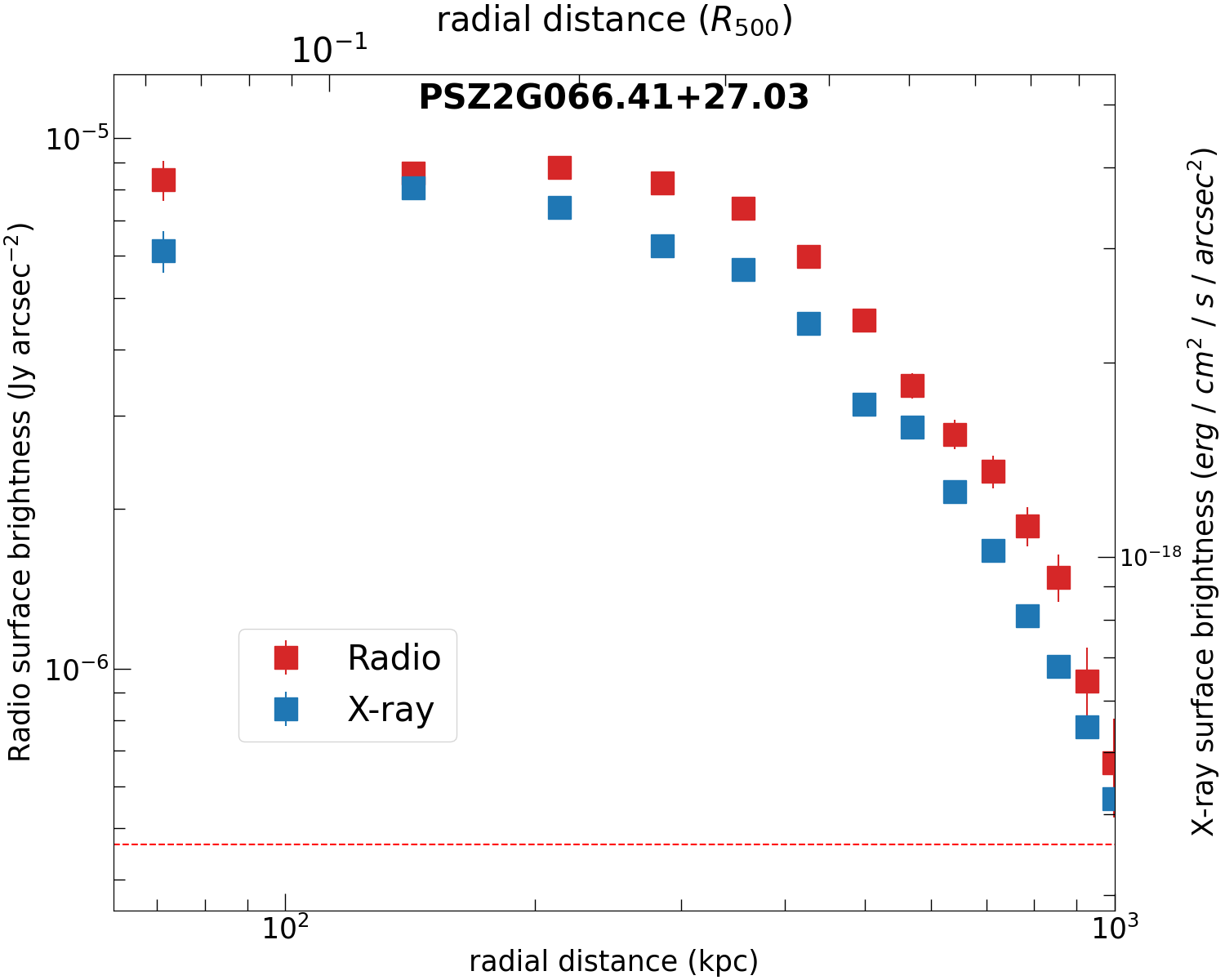}
    \includegraphics[width=8cm, height=6.2cm]{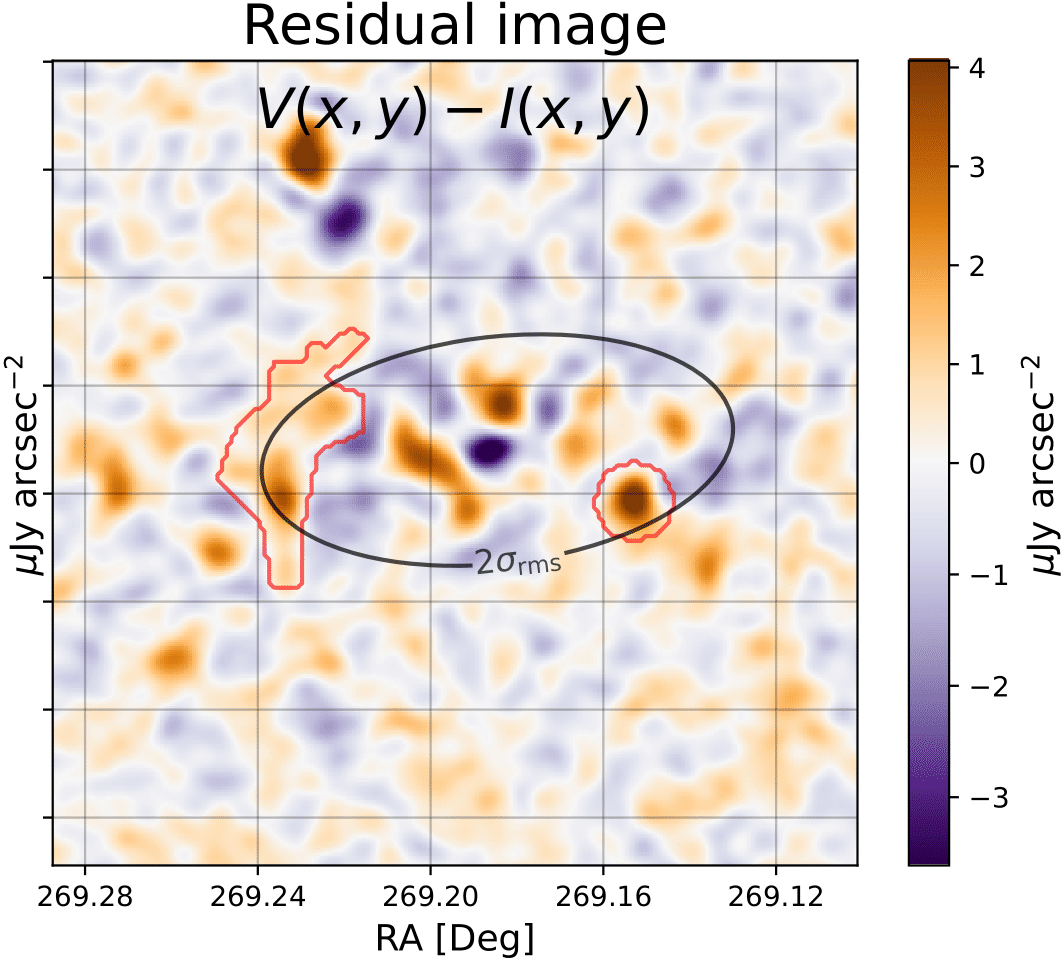}
    \includegraphics[width=8cm, height=6.2cm]{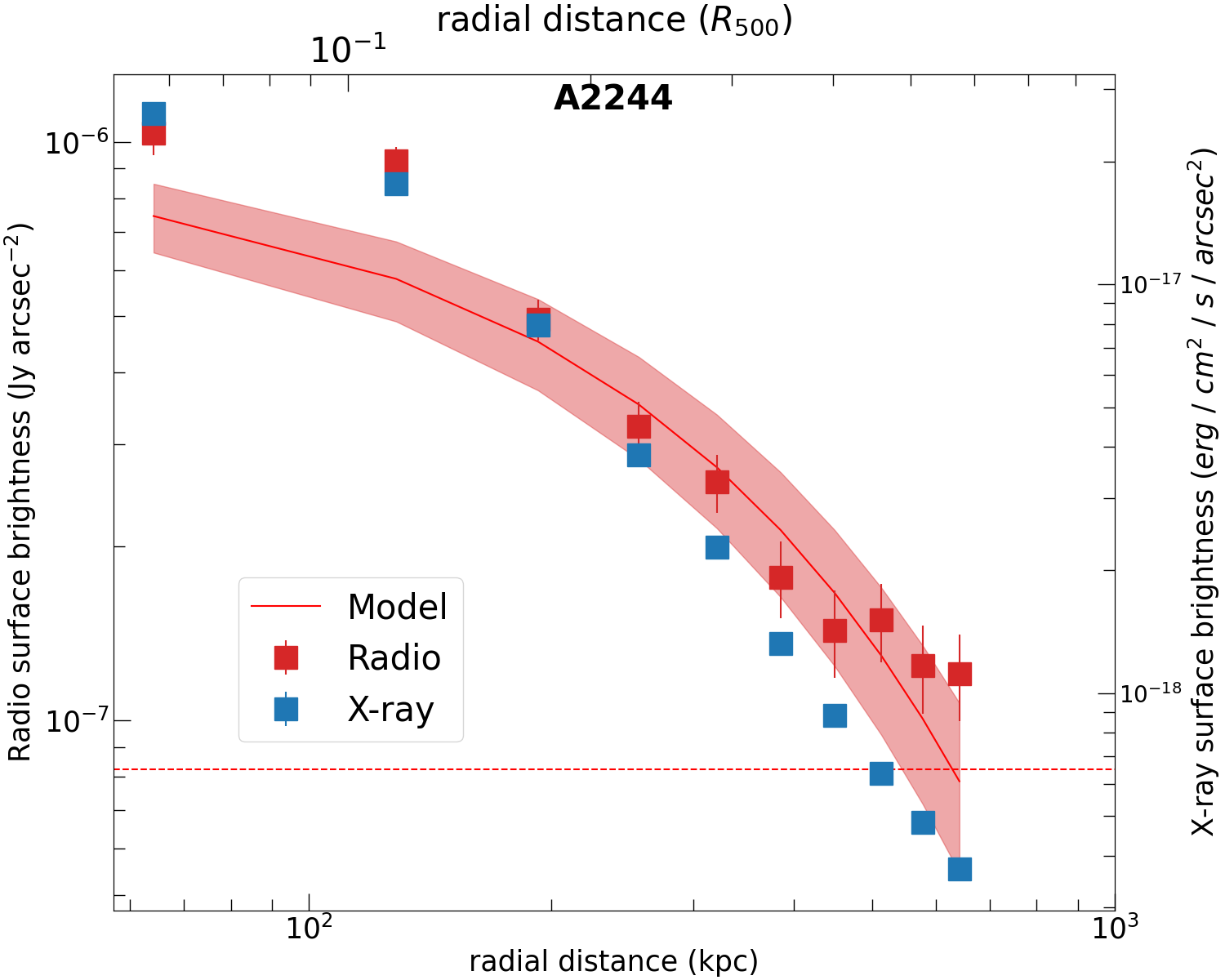}
    \includegraphics[width=8cm, height=6.2cm]{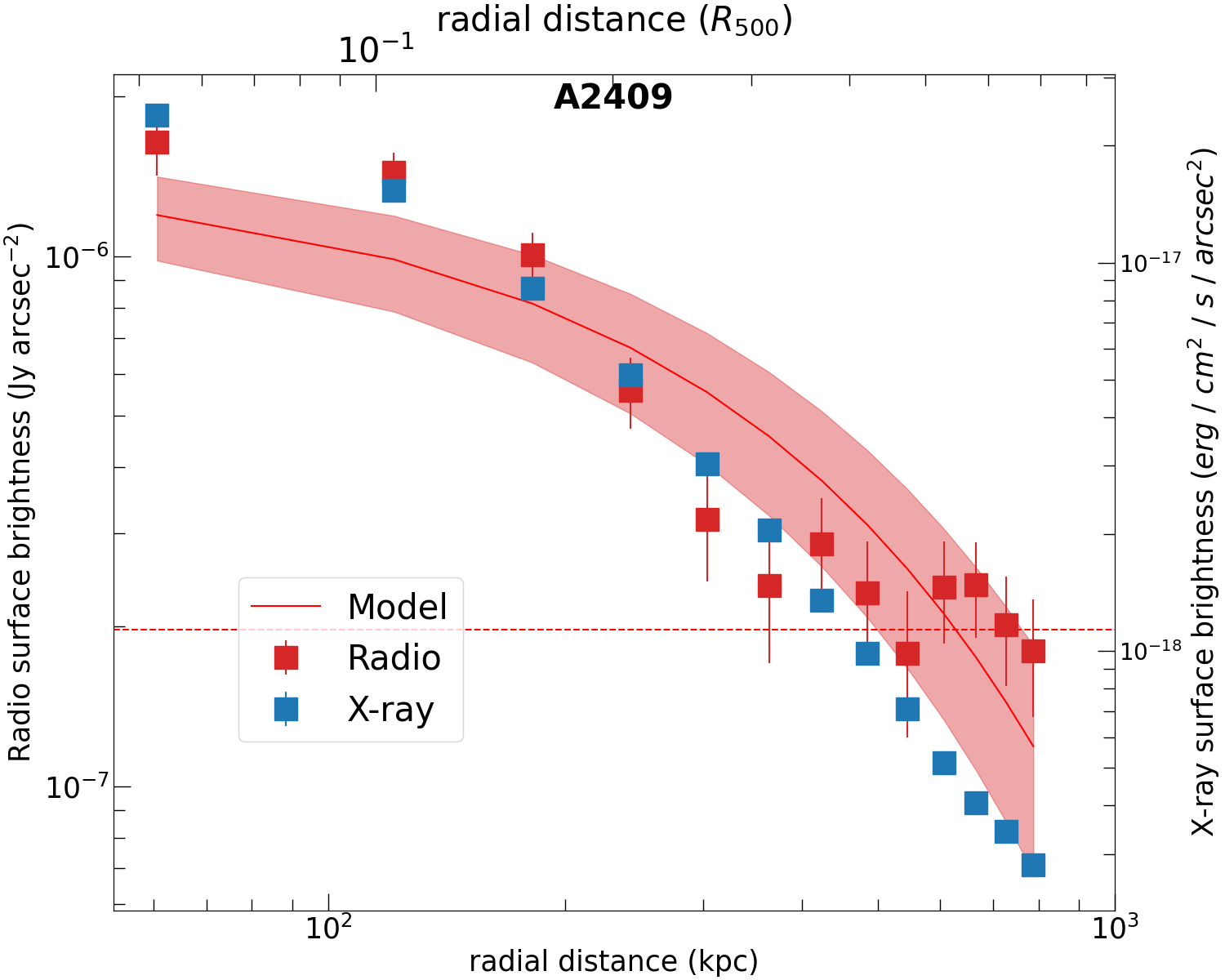}
    \caption{Radial profiles for the five studied objects (in the cases of elliptical annuli, we consider the major axis for the radial distance). The radio emission with the best-fit model found from Halo-FDCA and errors (shaded) is shown in red; the X-ray emission extracted in the same regions is shown in blue.  The red dotted line is the 1$\sigma$ threshold of the radio images. For PSZ2G066.41+27 we report the 1D profile and the residual image after the subtraction of the elliptical exponential model, with the masked regions in red.}
    \label{fig:radial_profiles}
\end{figure*}\\
The results of the fits are presented in Table~\ref{tab:HFDCA_fit}. In Fig.~\ref{fig:radial_profiles} we report the observed 1-D radio and X-ray (extracted in the same annuli) radial profile of the five studied clusters. For each annulus, we compute the mean of the radio brightness and associated an error of $\delta S = { {\rm \sigma_{RMS}} / \sqrt{N_{beam}}}$ (where $\rm N_{beam}$ is the number of beam in the annulus). The solid red line represents the best-fit halo profile obtained with Halo-FDCA. 
We see that all our objects show a monotonically decreasing profile, in both radio and X-ray data and with roughly comparable trends in both bands. 
These 1-D plots provide a good representation of the cluster emission only for those with circular shapes. Therefore, for a highly elliptical halo as PSZ066.41+27.03, we do not report the best-fit line in the 1-D plot, instead, we show the residual image after the subtraction of the elliptical exponential model.
We also note that the bump of radio emission at $r \sim 800$ kpc in A1758-NW is probably due to residual contaminant emission from the masked regions. 
\\
Here we see how, in general, this simple exponential model manages to reproduce the overall radial trend of the halos in our sample (as well as for many of the halos in the past, which is one of the reasons why this simple model was chosen, \citealt{Murgia2009}). However, when observed in detail, it seems that there are some systematic discrepancies in reproducing simultaneously inner and outer halo regions with a single exponential profile \citep[see also ][]{Cuciti21-1,Botteon2022}. Although it does not affect the rest of our analysis, we will better discuss this point in Sec.~\ref{sec:rad_prof_discussion}.
%
%
\subsubsection{Radial slope's trend}\label{sec:slope_trend}

\begin{figure*}[th!]
    \centering
    \includegraphics[width=8cm, height=6.8cm]{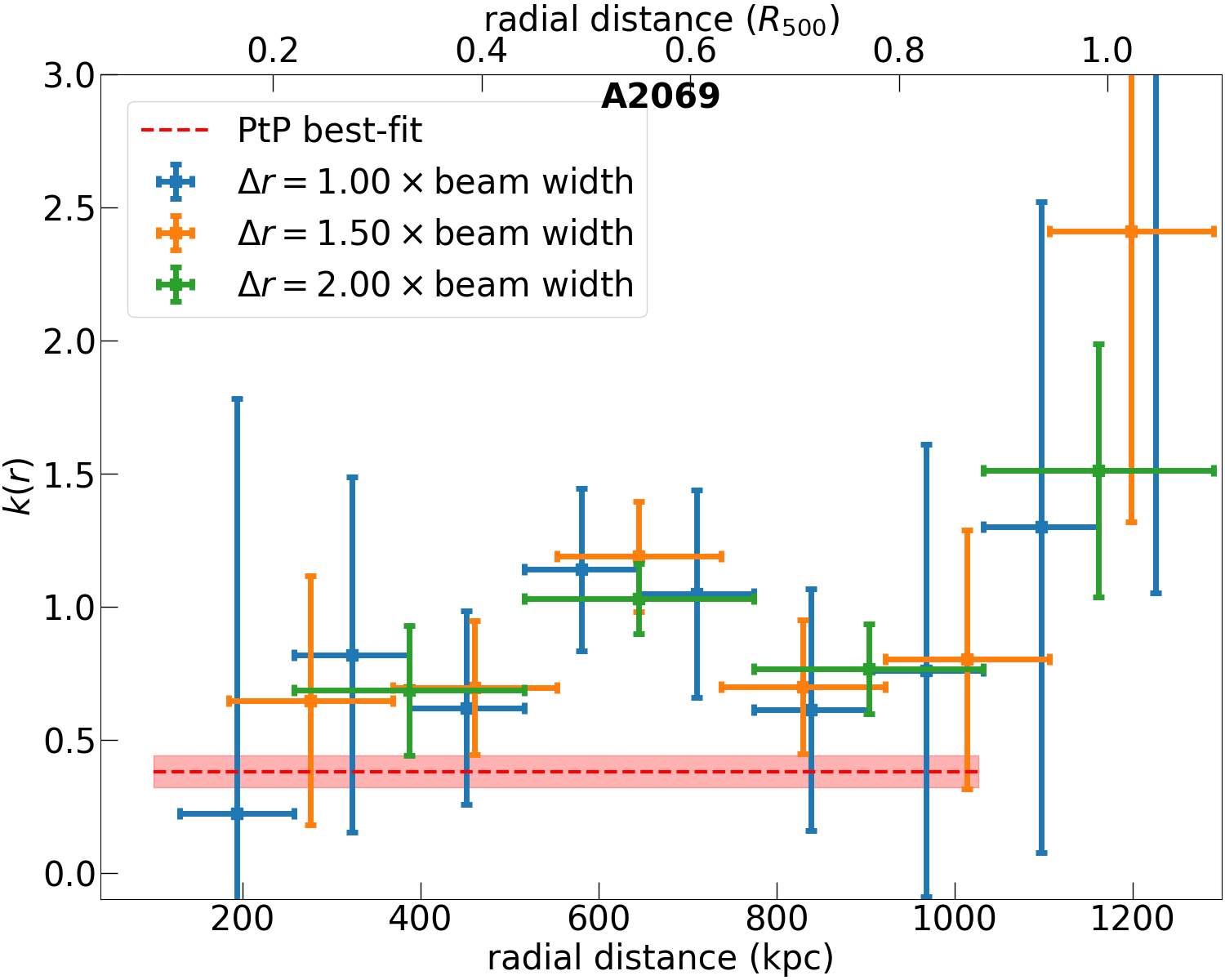}
    \includegraphics[width=8cm, height=6.8cm]{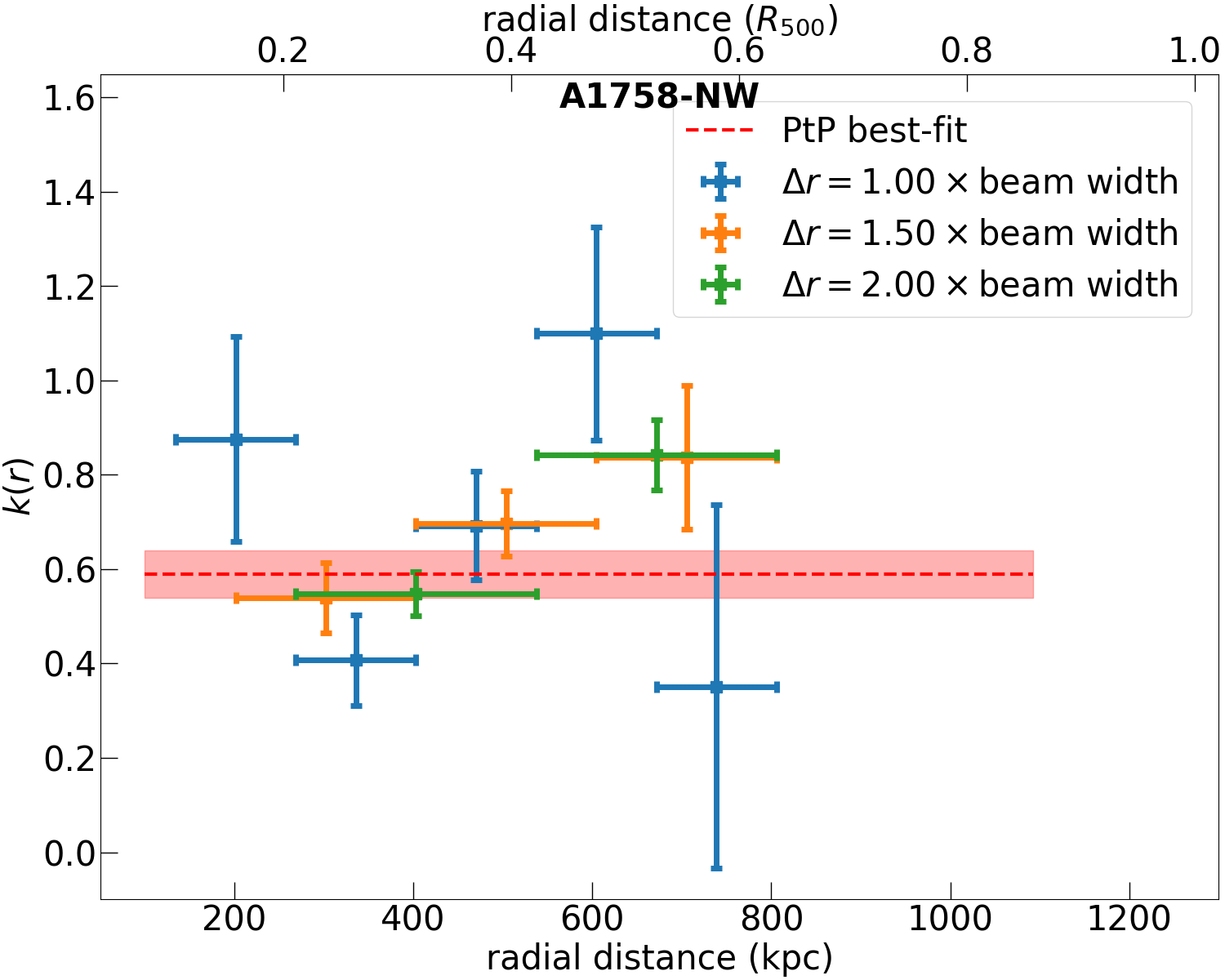}
    \includegraphics[width=8cm, height=6.8cm]{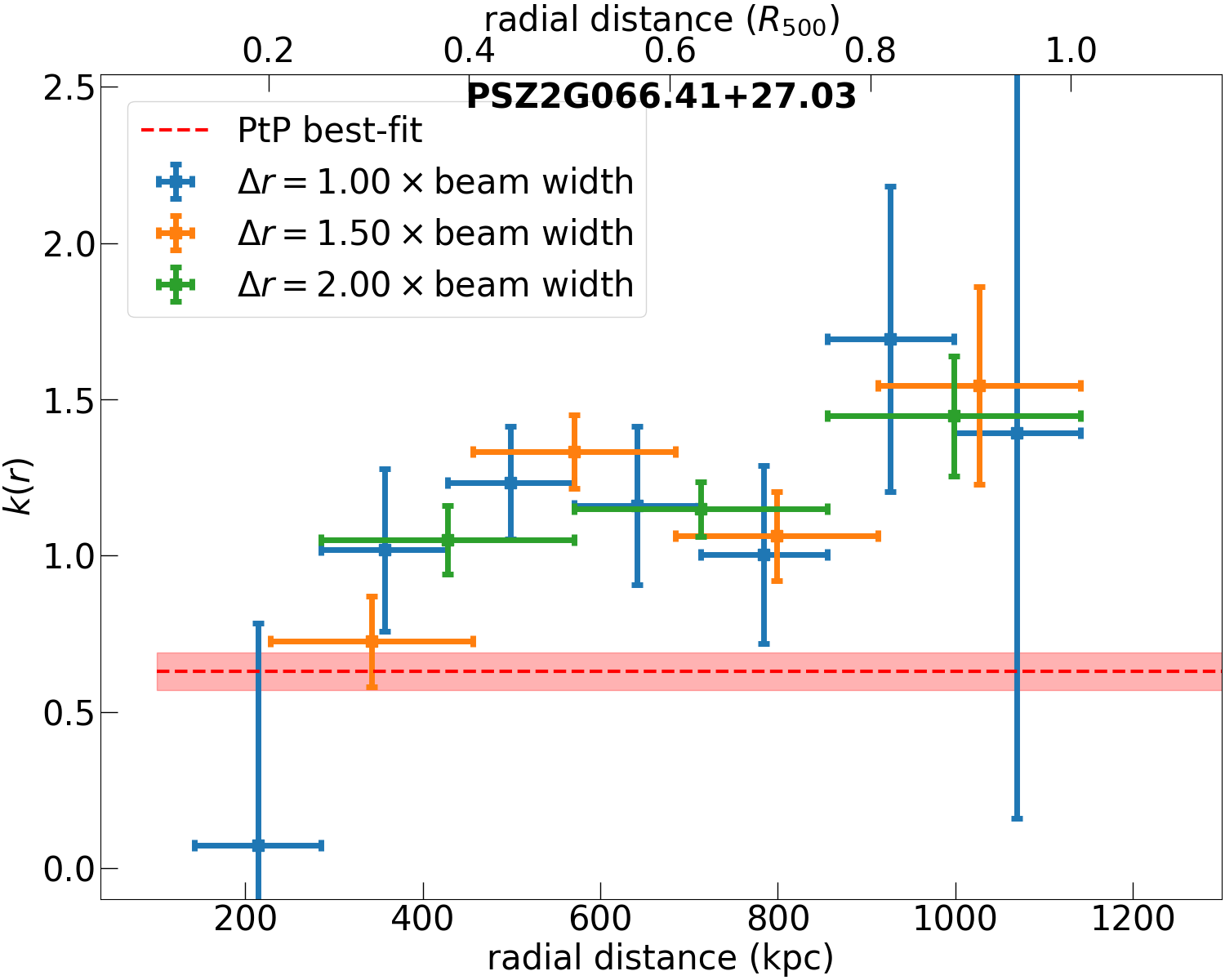}
    \includegraphics[width=8cm, height=6.8cm]{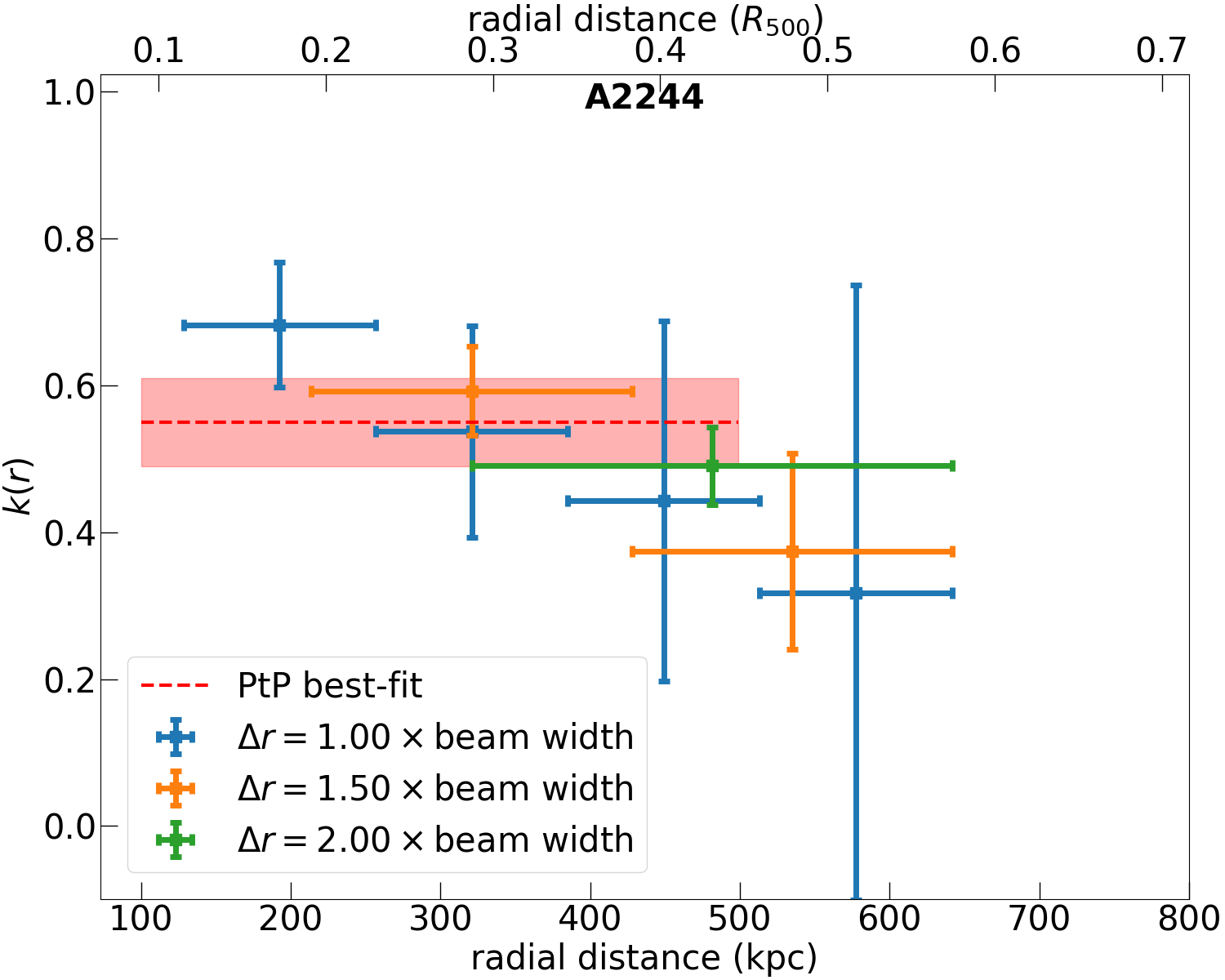}
    \includegraphics[width=8cm, height=6.8cm]{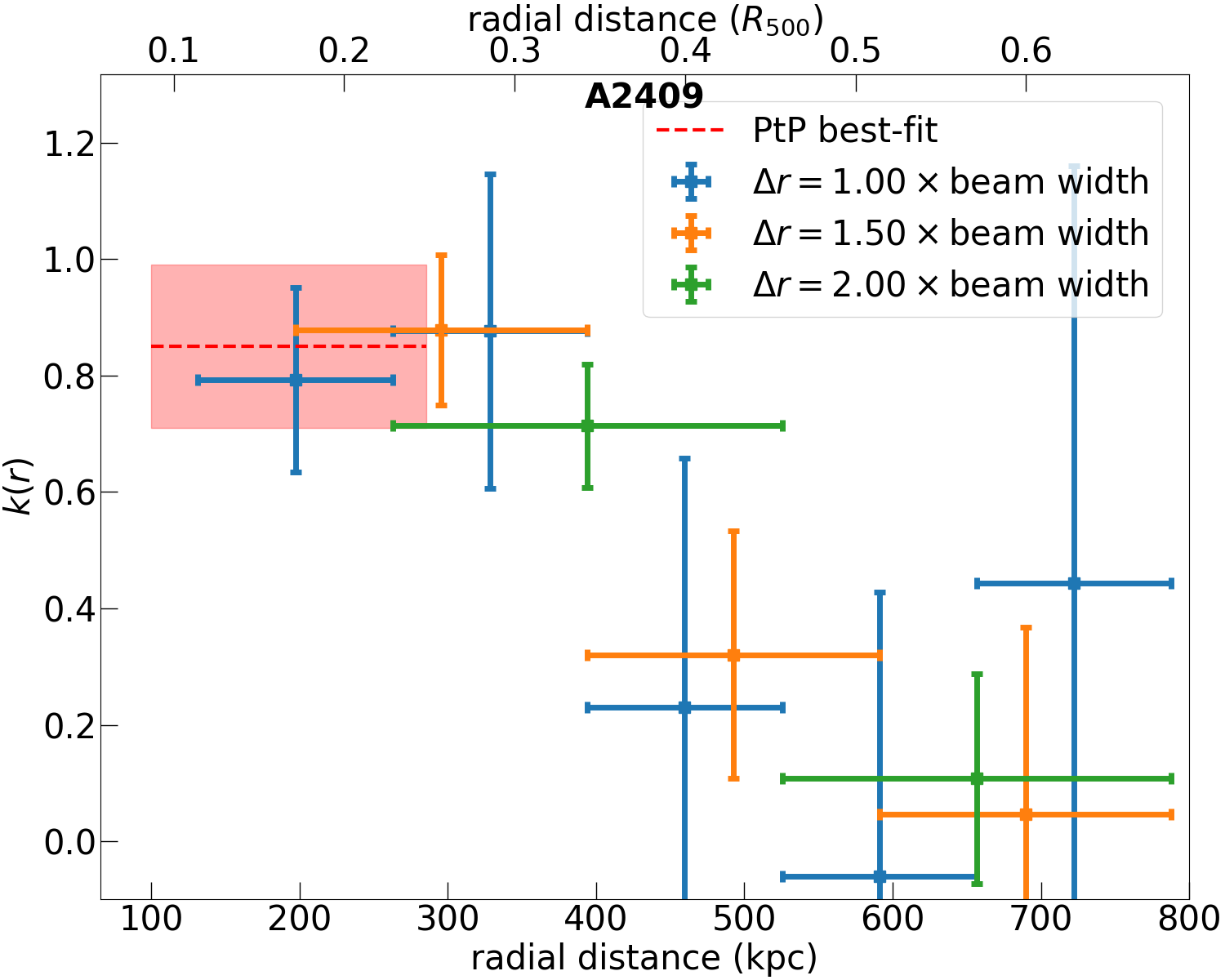}
    \caption{$I_R - I_X$ slope as a function of the radius (i.e. distance from the halo centre) for the five presented targets. The $k$ values are computed using Eq.~\ref{eq:k} among the consequent annuli. To reduce possible dependencies on the annulus width, we report the results for three different sizes of the annuli ($1,1.5,2 \times $ beam size). The red dotted horizontal line indicates the best-fit slope found by the point-to-point analysis (with its error) up to the radius considered for that analysis.}
    \label{fig:slope_trend}
\end{figure*}

Finally, we combine the two analyses presented so far to investigate if the correlation slope of the $I_R - I_X$ relation changes with the radius \citep[e.g.][]{Bruno2023}. Such studies provide information on how the ratio between X-ray and radio components varies across the cluster size. The value of the slope tells us how radio and X-ray components relate. 
Therefore, a constant slope throughout the whole cluster extension would indicate that their relation is constant, implying that the radial variations of the thermal and non-thermal components are the same. Instead, if we find fluctuations in the slope values, it means that their relation is changing and the value of the slope provides information on which component is increasing/decreasing with respect to the other.
Here we investigate the radial variation of the $I_R - I_X$ relation, both by deriving the correlation slope and studying directly the $I_R/I_X$ ratio as already done in other works \citep[][]{Biava_inprep,Bonafede2022,Kamlesh2023-A2256}. Here we present and discuss only slope studies and report the results of the ratios in Appendix~\ref{sec:appendix_ratios}.
To reduce as much as possible the introduction of biases \citep[as also made by][]{Bruno2023}, here we choose to study these possible slope changes computing it as:
\begin{equation} \label{eq:k}
    k(r) = \frac{\Delta (\ln I_R)}{\Delta (\ln I_X)},
\end{equation}
where $k(r)$ is the correlation slope of the $I_R \propto I_X^k$ scaling, $\Delta \ln I_R$ ($I_X$) is the difference between the logarithm of the radio (X-ray) brightness in two consecutive annuli (the annuli are centred in the radio peak as for the brightness profile). Again, to reduce the dependence from the chosen separation of the annuli, we try three different widths for them, respectively $1,1.5$ and $2$ times the beam width. 
As shown in Fig.~\ref{fig:slope_trend}, we obtain good agreement among the three different binning choices, finding that they reproduce similar features in the radial profile. The main difference lies in the fact that smaller bins retain larger errors and more scatter but are capable of resolving finer features than the larger ones (e.g. the case of A1758-NW). 
Figure~\ref{fig:slope_trend} shows the evident
complex trends in the presented halos, displaying how a uniform $k$ is generally not representative of the whole cluster extension.
\section{Discussion}\label{sec:discussion}
\subsection{Comments on point-to-point analysis}\label{sec:disc_ptp}
As pointed out by \cite{Govoni2001}, the point-to-point analysis can provide not only information on the spatial distribution of the thermal and non-thermal ICM but also on the mechanism responsible for halo emission.
They showed how, under some reasonable assumptions on the magnetic field profile, the expected correlation between $I_X$ and $I_R$ in (re-)acceleration and hadronic models is different. In particular, the sublinear scaling often observed in radio halos is hardly reproduced by the hadronic scenarios, as they would require an unphysical amount of CRp in the cluster peripheries to justify such $I_R-I_X$ relation \citep[e.g.][]{Brunetti14}.\\
%
Here we provide a consistent comparison among five radio halos whose results confirm the general view that radio halos show sub-linear $I_R - I_X$ relations. It implies a flatter profile of the radio components with respect to the X-ray one, suggesting the distribution of CRe and/or the magnetic field to be rather flat compared to the thermal ICM. This potentially indicates a larger size of the radio emission than the X-ray one when not limited by sensitivity, as already observed in a few cases \citep{Shweta2020,Botteon2020-A2255, Botteon2022-a2255,Kamlesh2021-A2744, Bruno2023}.
Additionally, we note that if we mask a larger region in A2069, similarly to what is made for the sky regions around A1758-NW (the yellow region in the first panel of Fig.~\ref{fig:5-targets}), the recovered slope becomes steeper ($\sim 0.5$). This may indicate that there is a projected diffuse emission between A2069-A and A2069-B. 
Here we do not speculate further on its origin since is beyond the scope of this work. However, we remark on the potential of these novel analyses in discovering new features thanks to the combination of radio and X-ray data.\\
Regime changes of the thermal--non-thermal relation throughout the cluster can be investigated through variations of the correlation slope. To test a possible change in the slope, indicating a departure from a single power law relation, we fit the $I_R - I_X$ relation with a \emph{broken power law }. As we show in Appendix~\ref{appendix: test_ptp}, it is worth investigating slope changes if the intrinsic scatter is $\lesssim 30 \%$.
All targets except PSZ2G066.41+27.03 have an intrinsic scatter (as retrieved by the power law fit) lower than $30 \%$.
Therefore, for these objects, we can search for slope changes. We use the same approach as described in Sec.~\ref{sec:ptp}, but this time searching for a relationship like:
\begin{align} 
    \large
   {\rm log} I_R = 
    \begin{cases}
            \beta + ~ {\alpha_1}~ {\rm log}I_X   &    {{\rm if ~ log}I_X < x_c} \\
            \beta + ({\alpha_1-\alpha_2}) {\rm log x_c}  + {\alpha_2}~{\rm log}I_X   &  { {\rm if ~ log}I_X > x_c}
    \end{cases}
\end{align}    
where $x_c$ is the X-ray surface brightness value at which the change from the slope $\alpha_1$ to $\alpha_2$ occurs.
With this analysis, we find that for three (A1758-NW, A2069 and A2244), out of five, objects we found that a broken power law is preferred rather than a simple power law, but with a low significance
\footnote{We performed the comparison and model evaluation using the \texttt{aesara.compare} module. See more on this comparison in Appendix~\ref{appendix: test_ptp} }.
We also note that all best-fit results, also with a broken power law, remain sub-linear.
\subsection{Discussion on radio profiles}\label{sec:rad_prof_discussion}
As pointed out in Sec.~\ref{sec:radial_prof}, there are discrepancies when fitting the radio halo profiles with a single exponential profile.\\
The choice of this profile shape is motivated by two reasons: it is a simple function with only two free parameters and it provides a reasonable description of many radio halos (albeit not being physically motivated).
Thanks to the new generation of radio interferometers, it is now possible to observe radio halos at a high signal-to-noise ratio with relatively high angular resolution. This allows to recover features in the emission that were invisible with past instruments, showing that the radio emitting structure is difficultly reproduced by such a simple profile. Indications of a departure from the single exponential profile were already present in \cite{Cuciti21-1} and \cite{Botteon2022}. Recently, \cite{Botteon2023} pointed out the presence of substructure in radio halos using MeerKAT data.\\
Here we find similar results. We see a common departure from the exponential profile in the central regions for the most perturbed systems. Specifically, the model tends to overestimate the central emission while recovers optimally the most extended one (this is also true for PSZ2G066.41+27.03, for which in the model-subtracted image negative residuals are found in the central parts while close to zero in the outer ones).\\
Instead, for the two more relaxed clusters, A2244 and A2409, the model seems to systematically underestimate the emission in the cluster centre. This suggests a centrally peaked distribution (Sec.~\ref{sec:ptp}) of the non-thermal ICM component for both objects, alongside a more concentrated thermal component. 
Instead, the outer regions seem to have the opposite behaviour. A flattening of the radio profile is observed in both objects (even though without high significance for A2409, mainly due to its faint diffuse emission).\\
%
Therefore, our results support the emerging idea that, despite the overall agreement between the exponential model and the radial halo profile (both on our sample and previous works), radio halos are more complex structures than the ones observed with past radio telescopes. Extreme cases of this have been shown in \cite{Cuciti22} and \cite{Biava_inprep} where the authors clearly showed how the single exponential model fails to describe the entire halo profile.
%
%
%

\subsection{Radial slope's trend}\label{sec:slope_trend_disc}
From Fig.~\ref{fig:slope_trend} it is evident that a single value of $k$ can not be representative for the $I_R \propto I_X^k$ scaling throughout the whole halo extension.\\
We also notice a different slope $k$ found in the point-to-point analysis with respect to the radial analysis. This difference can be due to the fact that (i) the radial profile reaches a distance of 2 $R_e$ from the cluster centre, while, for some objects, the point-to-point analysis stops at smaller radii; (ii) the boxes used for the point-to-point analysis have a smaller area than the annuli used in the radial analysis, possibly retaining real physical fluctuations on smaller spatial scales; (iii) high signal to noise regions affect more the fitting process, i.e. the fitted slope in the point-to-point analysis is more representative of central regions than for the outer ones.\\
In the following, we discuss our results separately for the more relaxed (A2244 and A2409) and more disturbed (A2069, PSZ2G066.41+27.03 and A1758-NW) objects, starting with the former.\\
In Fig.~\ref{fig:slope_trend}, we see that for A2244 and A2409 the slopes show a clear decreasing trend with $r$. This means that there is a flattening of the $I_R-I_X$ relation moving in the outer regions. Therefore, the non-thermal component radial decline becomes weaker and weaker with respect to the thermal one as outer radii are considered
Such a result indicates, for these two less disturbed objects, an increasingly important contribution of the non-thermal component with respect to the thermal one with the radius.\\
The three more disturbed clusters, instead, display different behaviours.\\
A first common feature is that they do not show a decreasing slope trend as the less disturbed systems. These three objects display an overall increasing value of $k$, indicating that the radio emission steepens with the radius. In particular, $k$ reaches linear and even superlinear values, indicating that the non-thermal component is decreasing faster than the thermal one (in contrast to what was found for A2244 and A2409).
%
\footnote{We note how the first and the last blue points in A1758-NW departs from a rather monotonic profile. The behaviour of the former, and in particular the rapid decrease after it, can be due to an offset between the radio and X-ray peak, as also found by \cite{Bruno2023}. The latter, instead, is probably contaminated by residual radio emission from the masking process as also pointed out in Sec.~\ref{sec:rad_prof_discussion}.}
Hence, the overall picture of these three more disturbed objects is similar to what has been recently observed in Coma by \cite{Bonafede2022}. Their slope's profiles indicate a steeper trend in the outer regions than in the inner ones.\\
For a direct comparison with previous literature results, we also show in Appendix~\ref{sec:appendix_ratios} the $I_R/I_X$ ratio.
%
%
%
\subsection{Modeling the $I_R - I_X$ radial trend}\label{sec:modeling}
\begin{figure*}
    \centering
    \includegraphics[width=8cm, height=6.8cm]{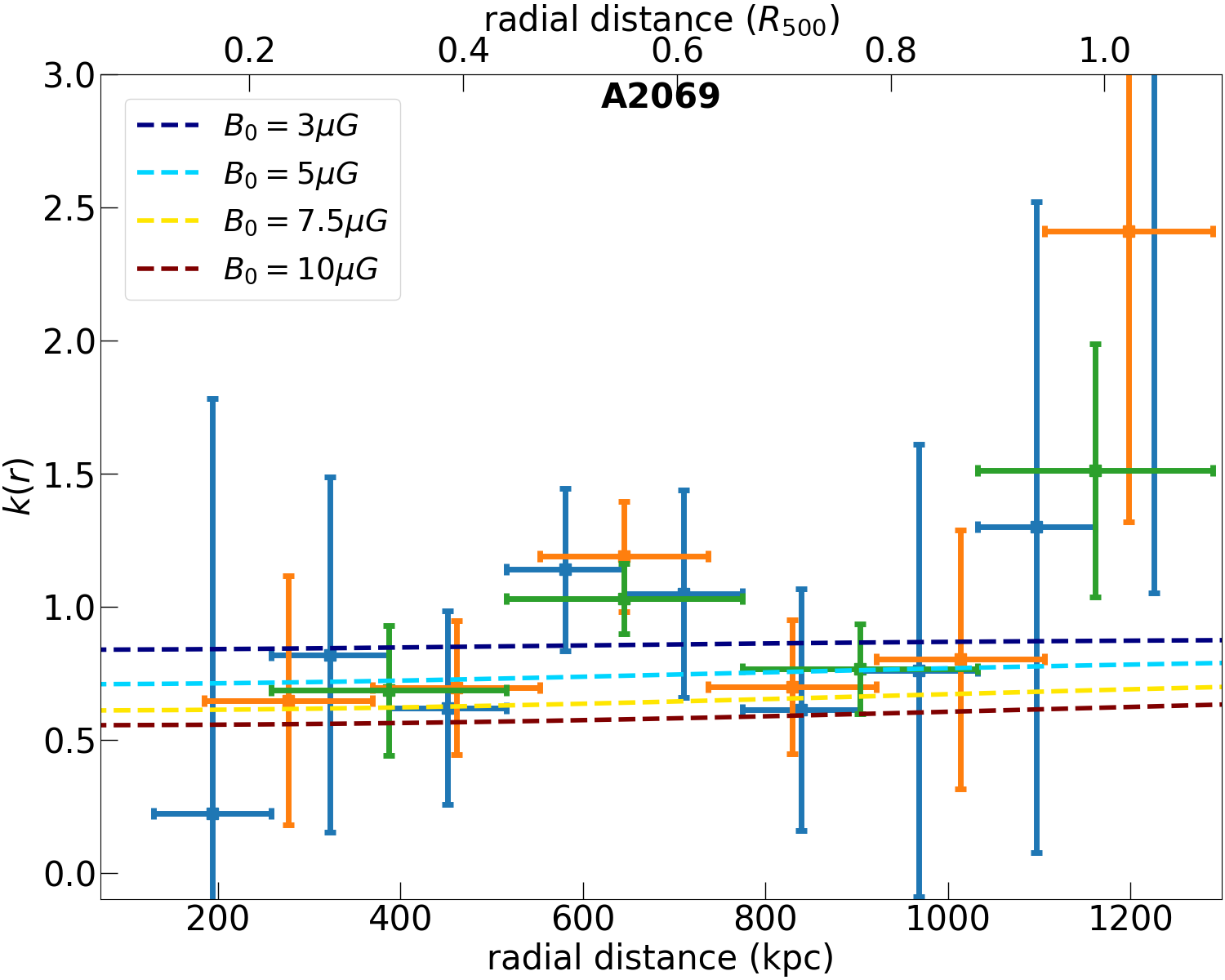}
    \includegraphics[width=8cm, height=6.8cm]{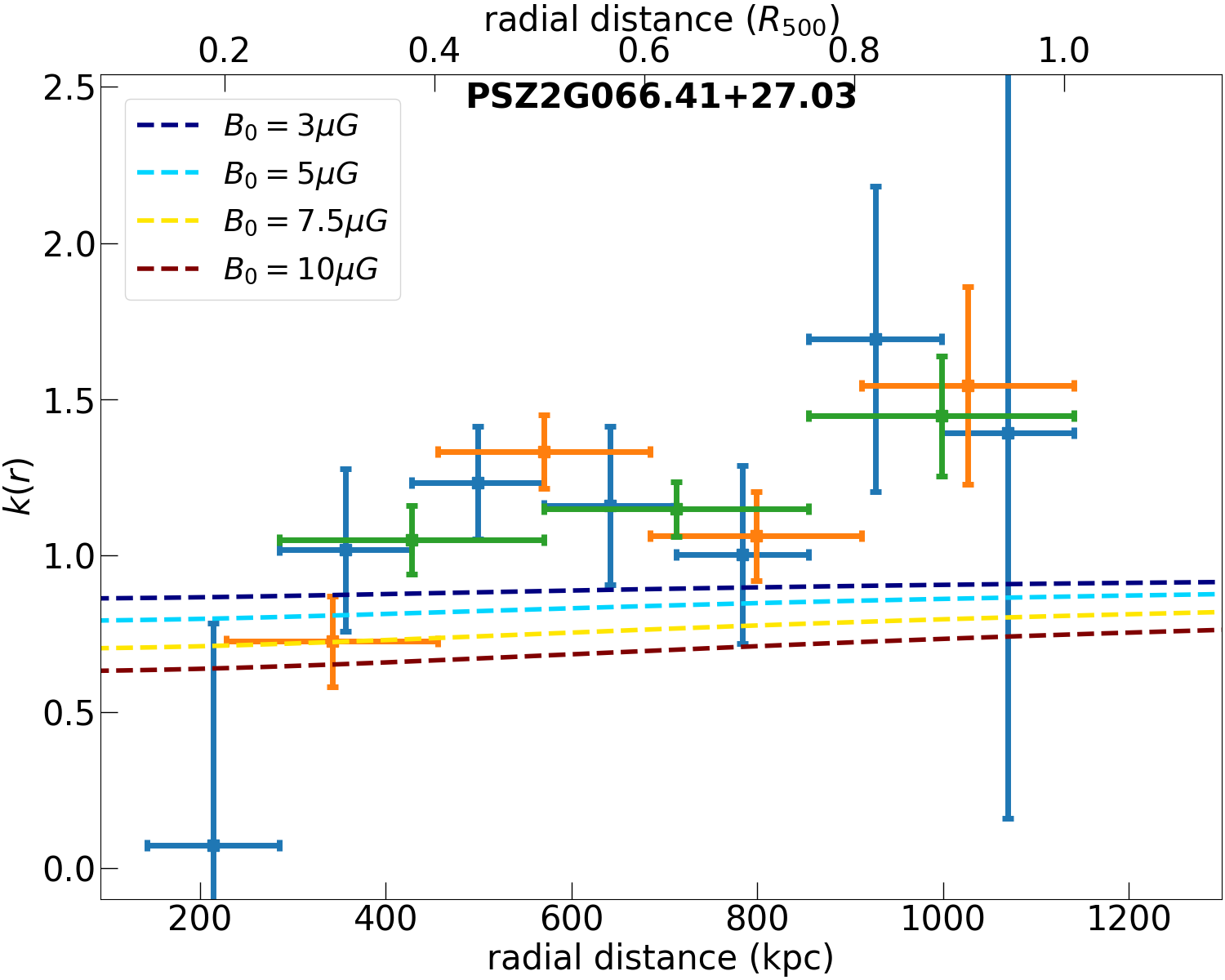}
    \includegraphics[width=8cm, height=6.8cm]{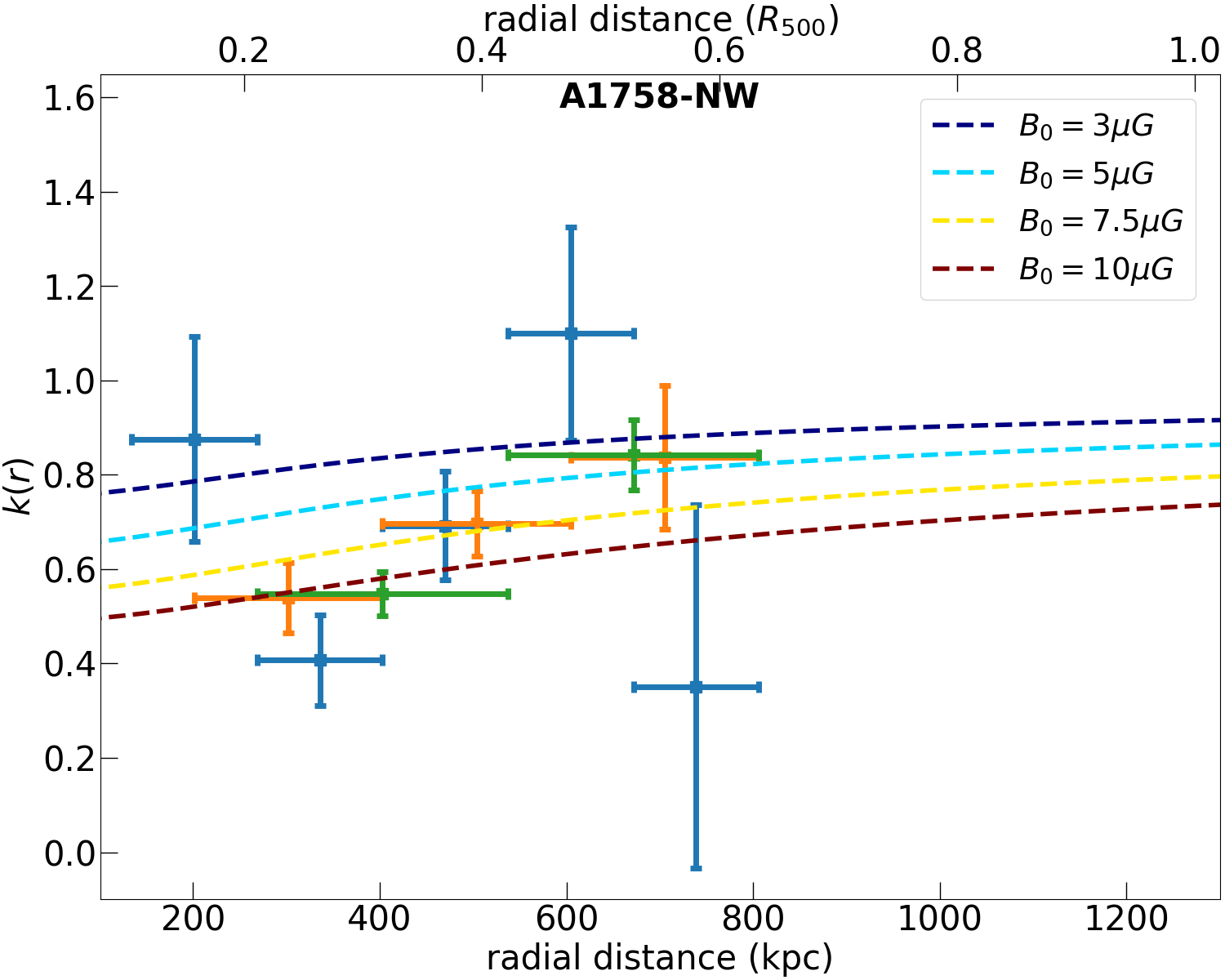}
    \includegraphics[width=8cm, height=6.8cm]{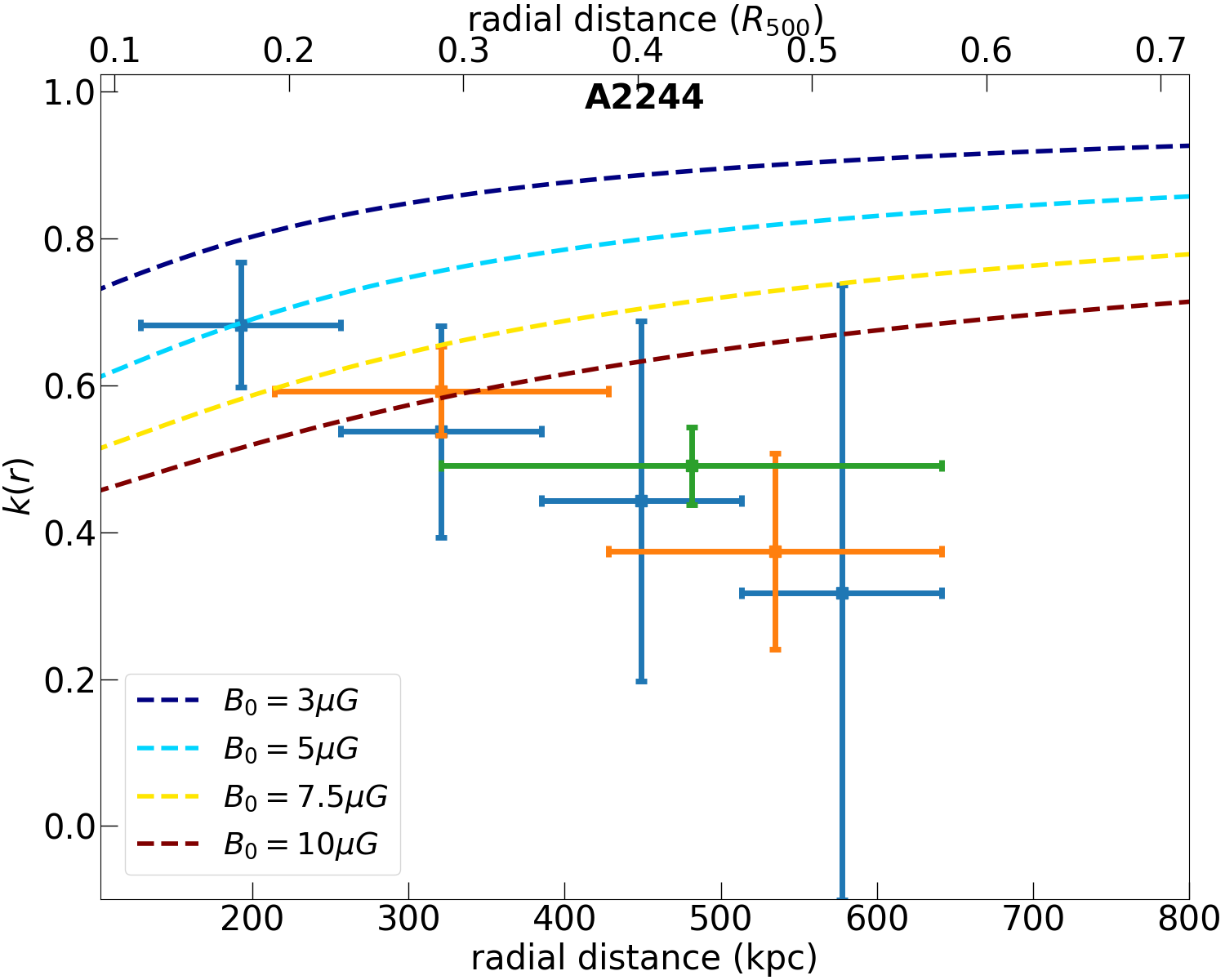}
    \includegraphics[width=8cm, height=6.8cm]{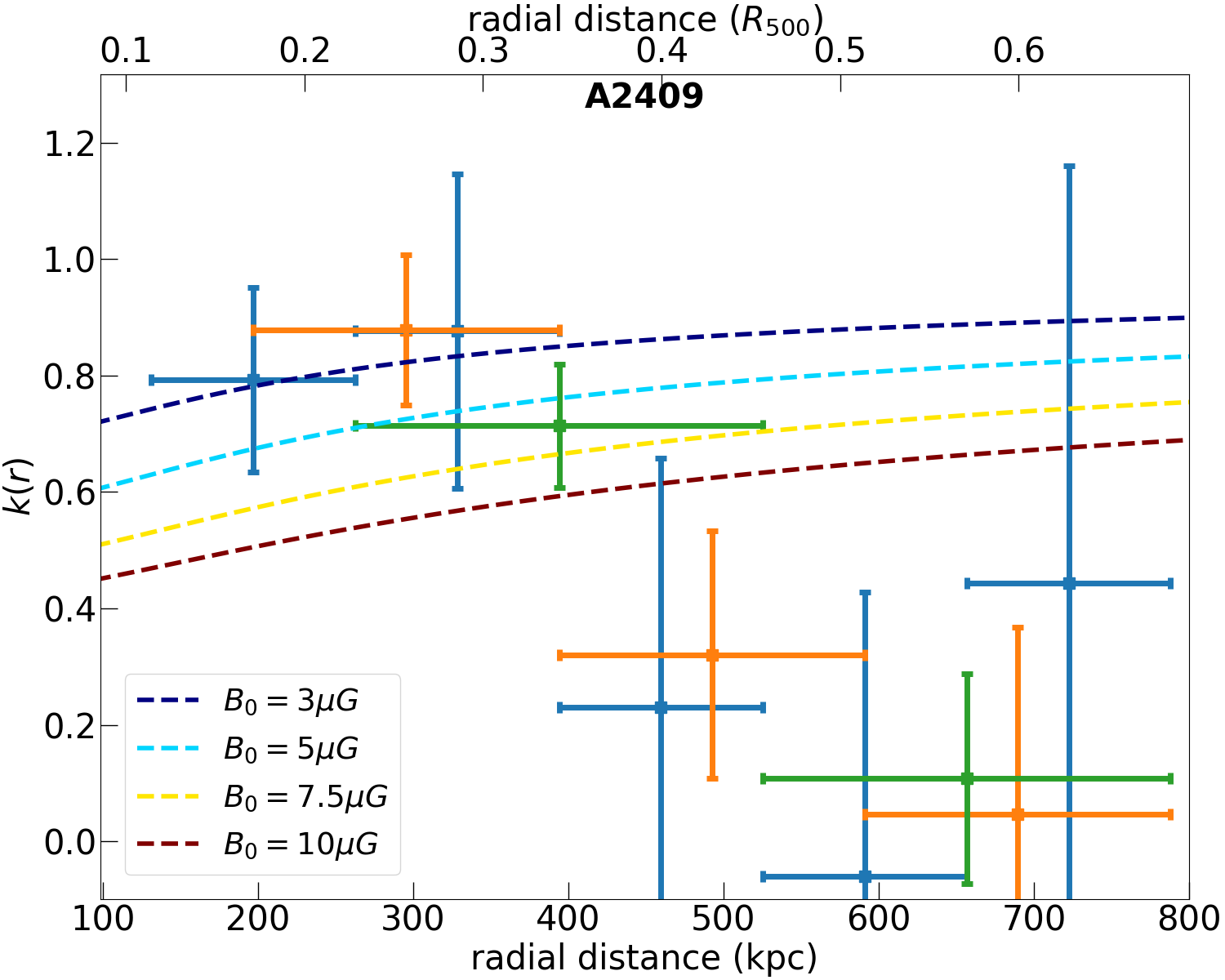}
    \caption{The same $k$ value of in Fig.~\ref{fig:slope_trend} with overlayed model's predictions for different values of $B_0$ ranging between 3 - 10 $\mu$G.}
    \label{fig:slopes_trend+model}
\end{figure*}

\begin{figure}[h!]
    \centering
    \includegraphics[scale=0.4]{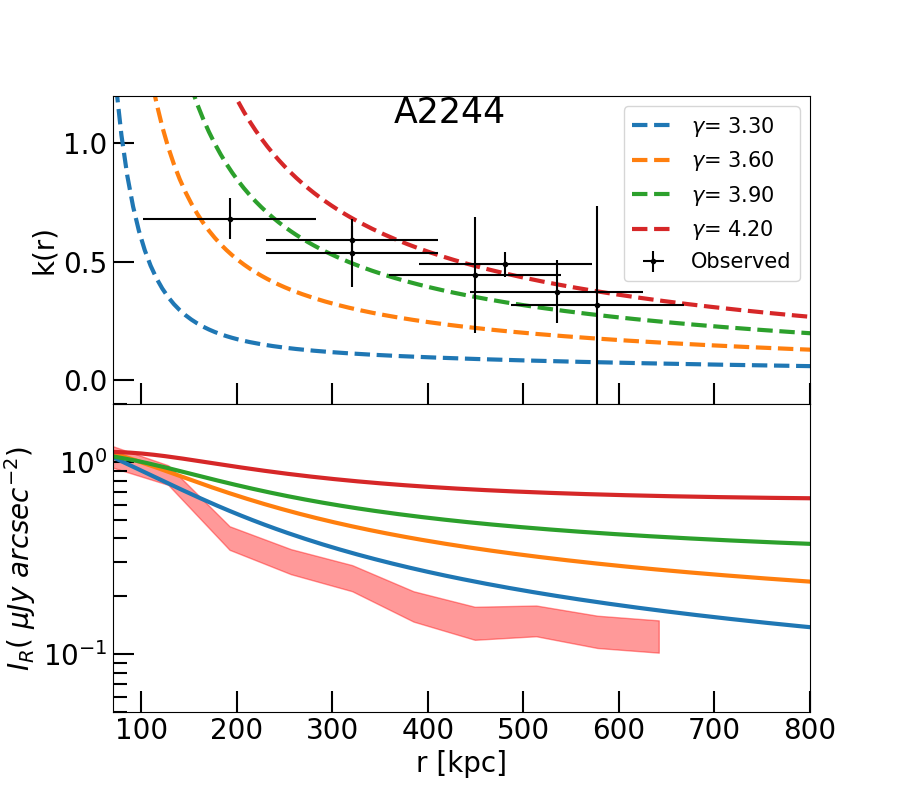}
    \includegraphics[scale=0.4]{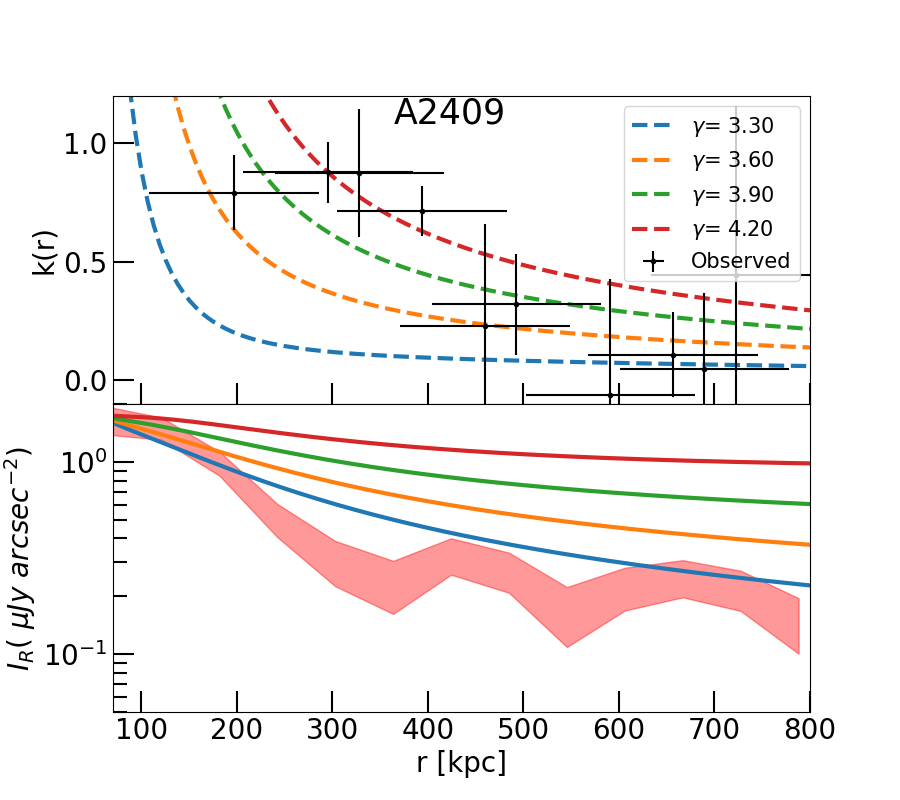}
    \caption{Comparison between predicted and observed profiles for different values of $\gamma$, assuming a central magnetic field $B_0 = 3 \mu G$. Upper panels: observed (black points) and predicted (dashed), radio-X-ray slopes as a function of the radius. Lower panels: predicted and observed (shaded) radio brightness profiles.}
    \label{fig:slope_varX}
\end{figure}

Following the work made by \cite{Bonafede2022}, we now consider a simple model for the radio emissivity to be compared with our observational results. 
Since this is a novel analysis, we stress that here we are not performing a fit of the model on our profiles. We just compare a simple model for radio emission with observations to see if general properties can be reproduced and explained with rather basic assumptions. Here we use the model for radio emission adopted also in \cite{Cassano2023} which can reproduce halo statistical properties, such as the halo fraction as a function of the mass.
%
To derive such a model we assume a magnetic field scaling with the thermal gas as $B(r) \propto B_0 ~ n_e^{0.5}$ \citep[according to the best constrained magnetic field profile made by][]{Bonafede2010}. In the context of turbulent re-acceleration scenarios, the radio emissivity is \citep[e.g.][]{Brunetti-Vazza2020}:
\begin{equation}
    \epsilon_R \propto F \eta_e \frac{B^2}{B^2 + B^2_{IC}},
\end{equation}
where $B_{IC} \approx 3.25 (1+z)^2 ~ {\rm \mu 
G}$ is the CMB equivalent magnetic, $\eta_e$ is the re-acceleration efficiency and $F$ is the turbulent energy flux. The latter is expressed as:
\begin{equation}
    F \sim \frac{1}{2} \rho \frac{\sigma_v^3}{L},    
\end{equation}
with $\sigma_v$ the turbulent velocity dispersion on a scale $L$ and $\rho$ is the gas density.
If we consider an isotropic distribution of the electrons' momentum space ($f(p)$), $\eta_e$ can be written as \citep{Brunetti-Lazarian07}:
\begin{equation}
    \eta_e \sim F^{-1} \int d^3p \frac{E}{p^2} \frac{\partial}{\partial p} \left ( p^2 D_{pp} \frac{\partial f}{\partial p} \right ) \approx \frac{U_{CRe}}{F} (D_{pp}/p^2).
\end{equation}
Here, $U_{CRe}$ is the energy density of re-accelerated electrons and $D_{pp}/p^2$, considering the case of second-order acceleration with super-Alfvenic solenoidal turbulence \citep{Brunetti-Lazarian16}, can be expressed as:
\begin{equation}
    \frac{D_{pp}}{p^2} \propto \frac{c^3_s M^3_t}{L v_A},
\end{equation}
where $M_t$ is the turbulent Mach number and $v_A$ is the Alfvén velocity.
Finally, if we consider a constant temperature profile and a constant turbulent Mach number, the synchrotron emissivity is\footnote{We found a typo in the final equation presented by \cite{Bonafede2022} in the term $\left(\frac{B_{IC}}{B(r)} \right )^2$ in the denominator. Here we report the corrected version.}:
\begin{equation}
    \label{eq:em_radio}
    \epsilon_R(r) = \frac{X(r)}{X(0)}\epsilon_R(0) \left ( \frac{\epsilon_X(r) }{\epsilon_X(0)} \right )^{1/2} \frac{1+ \left ( \frac{B_{IC}}{B_0}\right )^2 } { 1 + \left (\frac{B_{IC}}{B_0} \right )^2 \left ( \frac{\epsilon_X(0) }{\epsilon_X(r)} \right )^{1/2}},
\end{equation}
where $\epsilon_X$ is the X-ray emissivity and $X(r) = \frac{U_{CRe}}{U_{th}}$ is the ratio of CRe energy ($U_{CRe}$) to the thermal gas energy ($U_{th}$).
We compute this emission for different values of the central magnetic field $B_0$, in the range 3-10 $\mu$G and assuming a constant $X$. We derive $\epsilon_X$ by fitting a $\beta$-model to our X-ray images using the \texttt{PYPROFFIT} package presented in \cite{Eckert2020}. 
%
Using this model, we derive the expected correlation slope between the radio and X-ray surface brightness as defined in Eq.~\ref{eq:k}. Finally, we compare model predictions with observational results (i.e. the slopes' trends in Fig.~\ref{fig:slope_trend}). The results of this comparison are presented in Fig.~\ref{fig:slopes_trend+model} (we note how all the theoretical $k$ profiles asymptotically tend to 1 by construction).\\
From Fig.5, we note a generally poor agreement between the data and the model. Only in A1758-NW, we see a marginal agreement of both the trend and the values of $k$.
This indicates that more complicated models should be considered. For instance, a non-constant value of X and/or more complex magnetic field profiles should be taken into account.
The observed differences, however, are not the same in every cluster and a separation can be made, again, based on their dynamical disturbance.\\
In A2244 and A2409, the observed $k$ decreases with the radius, implying a flattening of the radio emission, which is opposite to the slopes' trends predicted by the model. A possible explanation of this behaviour is a non-constant value of $X(r)$, implying changes in the energy content of CRe with respect to the one of the thermal gas. In particular, a weaker decline of the radio emission as observed here, requires an increment of $X(r)$ with the radius, i.e. the ratio between CRe and thermal energy density increases.
%
There is no shortage of physical motivation for an increasing $X(r)$. For instance, the occurrence of more energetic merger shocks in the outskirts originates a higher amount of CRe \citep[e.g.][]{Vazza2017}, as well as the amount of turbulence can be more substantial in the external regions due to an off-axis geometry of the merger (which can be found in clusters where the central core has not been disrupted by merging cluster or group passage).\\
We therefore proceed to model a variable $X(r)$ by assuming the following functional form:
\begin{equation}
    \frac{X(r)}{X(0)} = \left( 1+\frac{r}{r_c} \right)^{\gamma}
\end{equation}
where $r_c$ is the core radius of the best-fit beta model for the X-ray brightness. We tested different values of $\gamma$ and searched for the one which could minimize the difference between the model and the observed trend of Fig.~\ref{fig:slopes_trend+model}. For both A2244 and A2409, we find the best agreement 
for $\gamma \sim 3.9$. In Fig.~\ref{fig:slope_varX}, we compare the observed trends and model expectations considering different values for $\gamma$ and assuming a central magnetic field $B_0 = 3 \mu$G.
%
To check whether the value of $\gamma=3.9$ alone could explain all the observed properties of the halo, we compute the radio emission originated by this new model. We assume the proposed scaling for $X(r)$ and use Eq.~\ref{eq:em_radio} to derive the radio surface brightness profile, as done for observational data.
We find that such a steep $X(r)$ behaviour would originate a surface brightness profile which, at large ($r\gtrsim 200$~kpc) radii, is brighter than the observed one (lower panels in Fig.~\ref{fig:slope_varX}).\\
These results indicate that (i) an increasing $X(r)$ is indeed required to produce a decreasing $I_R-I_X$ slope and (ii) that only a varying $X(r)$ is not enough to explain the radial trend. In fact, other factors (e.g. diverse behaviour of $B(r)$ and, in turn, a different expression for the factor $D_{pp}$) likely play a role in these cases, causing the observed shallower decline of the radio emission with respect to the X-ray thermal emission.\\

%
In A2069 and PSZG066.41+27.03, $k(r)$ does not display a monotonic trend,  though we notice a global increase of it with the radius. Again, this could be related to the dynamical status of the clusters and a non-constant $X(r)$ parameter.
However, for these objects we do not perform the same analysis made for A2244 and A2409. The complex $k(r)$ trend (likely due to their dynamical disturbance) cannot be associated with a monotonic variation of $X(r)$. In addition, in these clusters the thermal gas distribution will be naturally less smooth than in the more relaxed systems, making the magnetic field profile more uncertain. Departures of the magnetic field profile from the assumed $B(r) \propto n_{thermal}^{0.5}$ will cause an increase in the model complexity (e.g. non-constant Alfv\'en velocity) that is beyond the scope of this work.\\

%
%
%
%
%
%
\subsection{Comparison with mini halos results}\label{sec:mini_halo}
We studied the thermal -- non-thermal connection for a homogeneously covered sample of five clusters. In particular, we focused our attention on dynamically disturbed clusters which show radio halo emission detected at low frequencies.\\
A similar analysis has been carried out on mini-halos by \cite{Ignesti-2020-MH}, even though at $\sim 1$ GHz, and by \cite{Biava_inprep} on a sample of 12 cool-core clusters with LOFAR.\\
Mini-halos, are diffuse radio sources on hundreds of kpc, with steep spectra ($\alpha>1$), located at the centre of massive, relaxed cluster \citep[e.g][]{vanWeeren19}. Their emission is often confined in the cool cores of the clusters, which, as for giant halos, suggests a connection between non-thermal and thermal plasma \citep[e.g.][]{Gitti2004, Gitti2007, Mazzottaegiacintucci2008,Giacintucci2024}. The origin of synchrotron emitting electrons in mini-halos is yet unclear. In fact, as for giant radio halos, leptonic and hadronic models have been proposed to explain such emission.
In particular, the hadronic scenario for mini-halo production is not ruled out \citep{Ignesti-2020-MH}. \\%
The new-generation radio telescopes are highlighting the underlying complexity of these systems, with an increasing number of multi-component mini-halo findings \citep[e.g.][]{Venturi2017,Savini2018,Biava21,Riseley2022-MH,Riseley2022-RH}. In the context of the thermal--non-thermal connection, the point-to-point analysis on mini-halos has shown a recurrent linear or super-linear relation between radio and X-ray surface brightness \citep[e.g.][]{Ignesti-2020-MH,Biava21,Riseley2022-MH}. This is in general in antithesis to what is observed for radio halo sources (see Sec~\ref{sec:ptp}). It suggests, as other results have shown, a physical difference in the origin of these diffuse sources.\\
Here, we remark such a difference providing the first consistent comparison of radio halos $I_X-I_R$ trend. All our halos show sub-linear slopes, while the mini-halos presented in \cite{Biava21} and \cite{Riseley2022-MH} show linear or super-linear slopes at LOFAR HBA frequencies. This evidences how the distribution of plasma is different in these two classes of sources. In radio halo clusters, the non-thermal plasma is more extended and less tightly connected to the thermal plasma. Instead, in cool-core clusters, the radio emission seems to be more tightly related to the thermal emission and, in some cases, even more peaked in the cluster centre. The same results are found if we compare our sample with the one of \cite{Ignesti-2020-MH}, with cool-core clusters which show a linear $I_X-I_R$ behaviour. We however note that this latter comparison can be only qualitative since the observing frequencies are different.\\
It is worth mentioning that for the two more relaxed clusters of our sample, a close-to-linear behaviour is observed in the cluster centre (especially for A2409). This might be an indication of a significant contribution of hadronic processes in the core. However, the point-to-point relations of both objects are sub-linear and they are classified as perturbed clusters following their \emph{c} and \emph{w} parameters \citep{Botteon2022, Campitiello2022, Zhang2023}. This indicates that, even though in the central region they may look like mini-halo objects, they are different sources originated by different re-acceleration processes.\\
Finally, we compare our results of Fig.~\ref{fig:radio-X_ratios} with the recent work of \cite{Biava_inprep}. The authors compute the ratio between radio and X-ray surface brightness as a function of the radius for their four cool-core clusters which present diffuse radio emission in the central regions at 144 MHz. The main analogy with our work comes from the fact that also in the central cluster regions of their objects, there are clear changes in the correlation slopes, as found for our five radio halos. This indicates that, even for the mini-halo emission, there is a non-trivial connection between the thermal and non-thermal components. In particular, in their work, the authors find that objects that show slope changes also have strong departures from a single exponential profile in the radio emission. This indicates the presence of more than a single non-thermal component embedded in the studied regions, which, might cause also slope changes.\\
However, in their sample of four objects, \cite{Biava_inprep} find both increasing and decreasing $I_X/I_R$ ratios, while we mainly find increasing ones which are more evident in outer regions. Even though the samples studied so far are small, this might be an indication of a more complex environment in central regions (e.g. due to the stronger impact of radio galaxies on thermal and non-thermal components) than in outer ones, where turbulence is the main responsible for radio emission. In our future work (Balboni et al. in preparation), we plan to shed light on these connections by using the complete sample of CHEX-MATE clusters covered by LoTSS-DR2.
\section{Summary}\label{sec:summary}
In this work, we performed a pilot study investigating the radio and X-ray emission of five galaxy clusters from the CHEX-MATE sample. These objects have been selected to be representative of the 18 CHEX-MATE, radio halo clusters observed with LOFAR in LoTSS DR2.\\
%
The considered targets, according to the classification made by \cite{Campitiello2022}, \cite{Botteon2022} and \cite{Zhang2023}, show signs of disturbance in the ICM. Thanks to the combination of high-quality data provided by the CHEX-MATE and LoTSS DR2 datasets we have been able to perform the first consistent comparison of spatially resolved properties in X-ray and radio bands. Our results can be summarized as follows:
\begin{itemize}
    \item We found a strong connection between $I_X$ and $I_R$ in all clusters (confirmed by a high Spearman ranking). 
    All the analyzed objects presented sub-linear slopes in the $I_R - I_X$ plot, suggesting a slower decline of the non-thermal (CRe + B) component with respect to the thermal one.
    \item The five objects we have analysed do not show any particular trend of the best-fit slope with other parameters like mass, dynamical disturbance, temperature, total power, etc. Instead, we report the possible presence of a projected inter-cluster emission among A2069-A and A2069-B, remarking how this novel analysis can be used to discern among clusters' sub-components. 
    \item We found that for three clusters of the sample, a broken power law might be preferred, rather than a simple power law, to describe the $I_X-I_R$ relation.
    We aim to further investigate these features for a wider sample, searching for general behaviours.
    \item Through the radial analysis, we found a rather smooth trend of the radio halo profile and the X-ray emission in all objects.
    In agreement with \cite{Botteon2022, Botteon2023}, we find that, when studied in detail, the single exponential model struggles to reproduce the whole radio halo profile.
    Interestingly, the two more relaxed objects, A2244 and A2409, also display a peaked radio profile, which is similar to what is observed in the X-ray, suggesting a tighter connection between the thermal and non-thermal components in central regions.
    \item Investigating the radial behaviour of the $I_R - I_X$ correlation slope we found departures from a constant value indicating radial shifts in the thermal and non-thermal contribution in all targets. However, we found differences between more and less perturbed objects: less disturbed clusters show a decreasing slope profile, indicating, indicating a flatter $I_R \propto I_X^k$ scaling in external cluster regions, whereas clusters with a higher degree of disturbance show a more complex $k$ radial behaviour and a general increasing slope which was also found in the Coma cluster \citep{Bonafede2022} 
    \item We compared our findings with the expectations from turbulent re-acceleration models, under simplified assumptions (e.g. homogeneous conditions, a constant temperature, a constant turbulent Mach number, and a constant ratio of the energy density of CRe to thermal gas). Although the model is simplified, it has been proven successful in reproducing statistical properties of radio halos \citep[e.g.][]{Cassano2023}. We find that this model poorly reproduces almost every observed $k$ profile, suggesting that more complicated models should be considered.
    \item We showed how the model cannot reproduce the increasing amount of the non-thermal component with respect to the thermal one observed in more relaxed objects. This supports an increment in the amount of CRe energy density over the thermal one, for which different physical motivations can be presented. 
    However, we showed that an increasing ratio of the energy densities is not sufficient to fully reproduce the observed properties and other model parameters (e.g. different $B(r)$ or a non-constant acceleration efficiency) need to be taken into account.
    %
    \item Among the disturbed systems, A2069 and PSZ066.41+27.03 display complex $k(r)$ profiles. In these cases, we did not search for a best-fit $X(r)$ since their high dynamical disturbance will further increase the uncertainties of the analysis results (e.g. uncertain $B(r)$ profile). A1758-NW, instead is the only one marginally consistent with the model, possibly suggesting smaller variations of $X(r)$.
    We also note that, to the best of our knowledge, PSZ066.41+27.03 is the highest redshift cluster for which a radio--X-ray spatially resolved study has been performed.
    %
    \item By comparing our homogeneous study with the ones made on samples of mini-halos, we remarked the difference between the sub-linear $I_R-I_X$ scalings of radio halos and the linear or super-linear ones found in mini-halos. This indicates a tighter relation between thermal and non-thermal components in the latter. 
    However, the comparison of the radial trend of the $I_R/I_X$ ratio showed that for both mini-halos and radio halos a uniform trend is not evident.
    %
\end{itemize}
Our work supports the idea of a non-thermal component that has a more uniform distribution with respect to the thermal one, despite not being completely decoupled from it and with a connection ($I_X-I_R$) that changes throughout the cluster extension. In a forthcoming publication, we will investigate these results on a larger sample as may have important implications on radio halo origins.
%
%
%
%

\section*{Acknowledgements}
We acknowledge the developers of the following Python packages which were used in this work: \textsc{ASTROPY} \citep{astropy:2013,astropy:2018,astropy:2022}, \textsc{MATPLOTLIB} \citep{matplotlib}, \textsc{SCIPY} \citep{SciPy-NMeth}, \textsc{NUMPY} \citep{numpy} and \textsc{}{PYABEL} \citep{PyAbel2019,PyAbel_project}. 
LOFAR \citep{LOFAR2013} is the Low Frequency Array designed and constructed by ASTRON. It has observing, data processing, and data storage facilities in several countries, which are owned by various parties (each with their own funding sources), and thatare collectively operated by the ILT foundation under a joint scientific policy. The ILT resources have benefited from the following recent major funding sources: CNRS-INSU, Observatoire de Paris and Université d'Orléans, France; BMBF, MIWF-NRW, MPG, Germany; Science Foundation Ireland (SFI), Department of Business, Enterprise and Innovation (DBEI), Ireland; NWO, The Netherlands; The Science and Technology Facilities Council, UK; Ministry of Science and Higher Education, Poland; The Istituto Nazionale di Astrofisica (INAF), Italy. This research made use of the Dutch national e-infrastructure with support of the SURF Cooperative (e-infra 180169) and the LOFAR e-infra group. The Jülich LOFAR Long Term Archive and the GermanLOFAR network are both coordinated and operated by the Jülich Supercomputing Centre (JSC), and computing resources on the supercomputer JUWELS at JSC were provided by the Gauss Centre for Supercomputinge.V. (grant CHTB00) through the John von Neumann Institute for Computing (NIC). This research made use of the University of Hertfordshirehigh-performance computing facility and the LOFAR-UK computing facility located at the University of Hertfordshire and supported by STFC [ST/P000096/1], and of the Italian LOFAR IT computing infrastructure supported and operated by INAF, and by the Physics Department of Turin university (under an agreement with Consorzio Interuniversitario per la Fisica Spaziale) at the C3S Supercomputing Centre, Italy.\\
ABotteon acknowledges financial support from the European Union - Next Generation EU.
We acknowledge the financial contribution from the contracts ASI-INAF Athena 2019-27-HH.0,
``Attivit\`a di Studio per la comunit\`a scientifica di Astrofisica delle Alte Energie e Fisica Astroparticellare''
(Accordo Attuativo ASI-INAF n. 2017-14-H.0), and from the European Union’s Horizon 2020 Programme under the AHEAD2020 project (grant agreement n. 871158).
This research was supported by the International Space Science Institute (ISSI) in Bern, through ISSI International Team project \#565 ({\it Multi-Wavelength Studies of the Culmination of Structure Formation in the Universe}).\\
EP acknowledges support from CNRS/INSU and from CNES.
RJvW acknowledges support from the ERC Starting Grant ClusterWeb 804208.	
HB, PM, and FDL acknowledge financial contribution from the contracts ASI-INAF Athena 2019-27-HH.0, "Attività di Studio per la comunità scientifica di Astrofisica delle Alte Energie e Fisica Astroparticellare" (Accordo Attuativo ASI-INAF n. 2017-14- H.0), from the European Union’s Horizon 2020 Programme under the AHEAD2020 project (grant agreement n. 871158), support from INFN through the InDark initiative, from “Tor Vergata” Grant “SUPERMASSIVE-Progetti Ricerca Scientifica di Ateneo 2021”, and from Fondazione ICSC, Spoke 3 Astrophysics and Cosmos Observations. National Recovery and Resilience Plan (Piano Nazionale di Ripresa e Resilienza, PNRR) Project ID CN\_00000013 ‘Italian Research Center on High-Performance Computing, Big Data and Quantum Computing’ funded by MUR Missione 4 Componente 2 Investimento 1.4: Potenziamento strutture di ricerca e creazione di "campioni nazionali di R\&S (M4C2-19 )" - Next Generation EU (NGEU).
MB and GR wish to thank the "Summer School for Astrostatistics in Crete" for providing training on the statistical methods adopted in this work.
\bibliographystyle{aa}

\bibliography{biblio.bib} 

\begin{appendix}
\section{Test on point-to-point fit}\label{appendix: test_ptp}
The aim of this section is to address the issues and systematics when performing a point-to-point study between the radio and X-ray surface brightness.\\
What is usually done when facing radio brightness emission, is to introduce a threshold below which all the emission is considered not reliable. This is done because, at the end of the calibration and imaging processes, artifacts can still be present in the images making the recovery of low surface brightness emission non-trivial. In particular, the typical criteria adopted to keep (discard) the emission is to check if the considered emission lies above (below) 2-3 times the noise of the whole radio image. However, this choice, almost independently on how conservative it is, will introduce a bias on the final result when in presence of intrinsic scatter \citep[see also ][]{Botteon2020-A2255}. In this section, we try to quantify how much this effect depends on the intrinsic scatter, threshold set and distribution of the data.\\
\\
The idea is to generate a known dataset according to a predefined model, introduce an intrinsic scatter, cut the data at a given threshold (to mock the selection effect) and fit the resulting data with a model. Specifically, the fitting models of interest are power law and broken power law functions.
In addition, since our aim is to see how selection effects and intrinsic scatter affect the final results, we consider, for the same dataset, different cases of intrinsic scatter and threshold/cut on the dependent variable values, $Y_i$.\\
Here we produce both a power law and a broken power law dataset. The power law one allows us to verify how much selection the effects (i.e. a cut on $Y_i$ data) bias the fit. In particular, what is the trend of the power law slope retrieved by the model as a function of the threshold set on data. Using the broken power law dataset, instead, we are going to determine which are the values of intrinsic scatter and threshold at which a broken power law fitting model is no longer preferred to a power law one.\\
\\
Moving to the more practical details, once we produce the desired dataset, we add the intrinsic scatter to $Y_i$, randomly sampling the new $Y_i$ from a Gaussian distribution with mean $Y_i$ and scatter $\sigma_j$.
We also insert errors in the mock dataset considering the error-measurement trend found in the data. Finally, we consider different cases of selection effects by cutting the dependent variable at increasing values (i.e. using different thresholds below which discard the data), excluding up to 80\% of the data.\\
As anticipated in Sec.~\ref{sec:ptp}, exploiting the python \texttt{PYMC} package \citep{py_mc16, pymc_4.2}, we developed a regression algorithm which creates Bayesian models and fit them through Markov chain Monte Carlo methods. It accounts for errors, selection effects and intrinsic scatter. To formally describe our analysis we follow the notation of \cite{Sereno2016}.\\
In the $X-Y$ dataset, the quantities $X$ and $Y$ fall exactly on the regression model. However, real data will always carry intrinsic scatter and measurement uncertainties. Observable properties are usually log-normally distributed around the mean scaling relation \citep[see][and reference therein.]{Sereno2016}, hence we assume the measured ($x$ and $y$) and true ($X$ and $Y$) values being related through:
\begin{equation}
    P(x_i, y_i | X_i, Y_i) = N^{2D}( \{X_i, Y_i\}, C_{\sigma, i} )
\end{equation}
where $C_{\sigma, i}$ is the covariance whose diagonal elements are: $\sigma_x$, given only by the measurement uncertainties on $X_i$, and $\sigma_Y$, due to both measurement uncertainties and intrinsic scatter. The off-diagonal elements are equal to zero since $X$ and $Y$ are assumed to be uncorrelated.\\
Finally, we account for selection bias by truncating the probability distribution below a certain threshold:
\begin{equation}
    P(x_i, y_i | X_i, Y_i) \propto N^{2D}( \{X_i, Y_i\}, C_{\sigma, i} ) \mathcal{U}(y_{th})
\end{equation}
where $\mathcal{U}$ is the uniform distribution null for $y_i \le y_{th}$.
Although we modelled for such effect in the scientific analysis (Sec.~\ref{sec:ptp}), here we do not account for it in the fit algorithm to show its impact on the results.\\
To assess when a broken power law model fits better the data with respect to the simple power law one, we look at the accuracy of model's predictions of future data (i.e. out-of-sample data). To do so, we exploit the \emph{expected log pointwise predictive density} (elpd) for a new dataset ($\tilde{y}_i$) introduced by 
\cite{Gelman2014} and \cite{Vehtari2017}:
\begin{equation}\label{eq:elpd}
    elpd = \sum^n_{i=1} \int f(\tilde{y}_i) ~ {\rm log} p_{post}(\tilde{y}_i | y) ~ d\tilde{y}_i
\end{equation}
where
\begin{equation}
    p_{post}(\tilde{y}_i | y) = \int p(\tilde{y}_i | \theta) p_{post}(\theta) d\theta 
\end{equation}
In the above expressions, $n$ is the total data number, $p_{post}(\tilde{y}_i | y)$ is the posterior predictive distribution for a new data $\tilde{y}_i$ induced by the posterior distribution $p_{post} (\theta)$ given the parameters $\theta$, $p(\tilde{y}_i | \theta)$ is the likelihood/sampling distribution for $\tilde{y}_i$ given $\theta$, $f(\tilde{y}_i)$ is the (generally unknown) true data distribution which will be approximated using cross-validation or information criteria \citep{Gelman2014}.\\
In the case of Leave-One-Out (LOO) cross-validation, the estimate of out-of-sample predictive fit becomes \citep{Vehtari2017}:
\begin{equation}
    elpd_{loo} = \sum^{n}_{i=1} {\rm log} p_{post}(y_i | y_{-i})
\end{equation}
where
\begin{equation}
    p_{post}(y_i | y_{-i}) = \int p(y_i | \theta) p_{post}(\theta | y_{-i}) d\theta
\end{equation}
Expressed in this way, the $elpd_{loo}$ is the probability to predict the $i$-th data point with the current data set without the $i$-th data point. Analogous results can be obtained if considering the Widely Applicable Information Criterion (WAIC) \citep{Watanabe10}) to approximate \ref{eq:elpd}, see \cite{Gelman2014} and \cite{Vehtari2017}.\\
In particular, we compare the $\rm elpd_{loo}$ of the two models on the deviance scale \citep{Vehtari2017}. We apply the same rule provided by \cite{Jorgensen-book} for Akaike’s information criterion (AIC) and deviance information criterion (DIC), considering the elpd difference significant if it is greater than 4 ($ {\rm \Delta elpd_{loo} = elpd^{broken}_{loo} - elpd^{powlaw}_{loo} > 4 }$). 
%
%
\subsection{Test with power law dataset}
\begin{figure*}
    \centering
    \includegraphics[width=9cm, height=8cm]{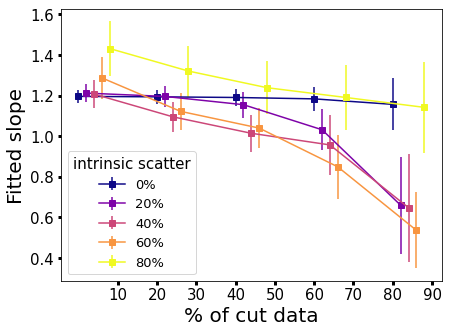}
    \includegraphics[width=9cm, height=8cm]{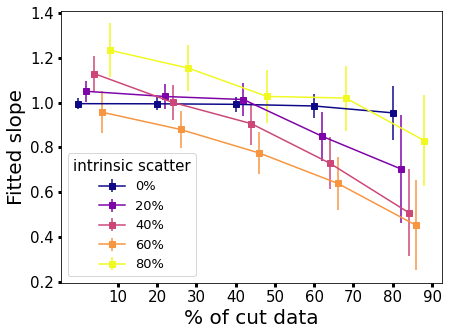}
    \includegraphics[width=9cm, height=8cm]{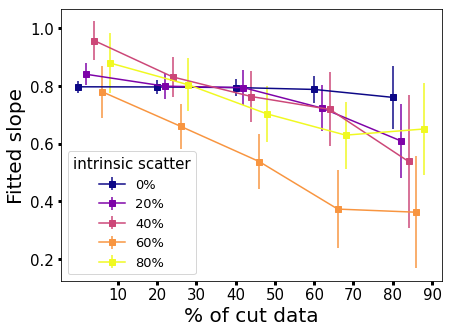}
    \includegraphics[width=9cm, height=8cm]{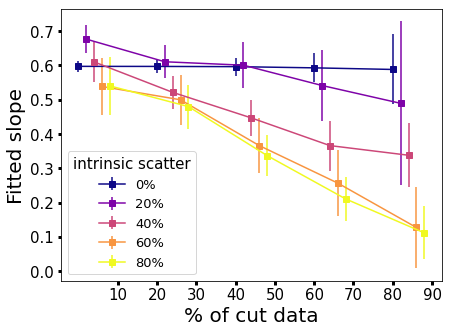}
    \caption{Fitted slope change as a function of the threshold set on data. From upper left to lower right, the real data slope is 1.2, 1., 0.8, 0.6. Points are slightly offset on the x-axis for visualization purposes.}
    \label{fig:slope_changes}
\end{figure*}

Here we consider a power law data distribution, hence we assume a relation like:
\begin{equation}
    y_i = \beta ~ x_i^{\alpha}
\end{equation}
with $\beta = 1$ and $\alpha = 0.6 \div 1.2$, in the range $1 \leq x_i \leq 10$.\\
By studying the elpd difference between the broken power law and power law models, we did not find any particular evidence for preferring a simple power law model to a broken power law  (or vice-versa). This is because the broken power law, in the case of a power law distribution of the data, simply becomes a power law (by finding two similar slopes before and after the break), and so it fits the data as well as the simple power law.\\
Instead, we observe a clear dependence of the power law fitted slope on the dataset used. Fig.~\ref{fig:slope_changes} shows, for four different mock datasets with different $\alpha$, the recovered power law slope as a function of the threshold set on the data (expressed in terms of the percentage of data cut below $y_{th}$) and for different degrees of intrinsic scatter. 
There is a flattening of the fitted slope when increasing the threshold. This is because, when imposing a cut, in the lower end (left) of the relation, we are only considering upscattered values. This effect eventually ends in introducing a bias in the fit for which all the slopes become systematically flatter than the real one. It is also clear how this effect occurs more evidently when a high degree of scatter is present in the data.\\
We also found similar results when considering the slope of the broken power law model before the breaking point. Because this part of the function models the data at lower $y_i$ values, the bias on it will be particularly evident.\\
Therefore, accounting for selection effects is required in these studies to avoid such bias.
\subsection{Test with broken power law dataset}
\begin{figure*}
    \centering
    \includegraphics[width=9cm, height=8cm]{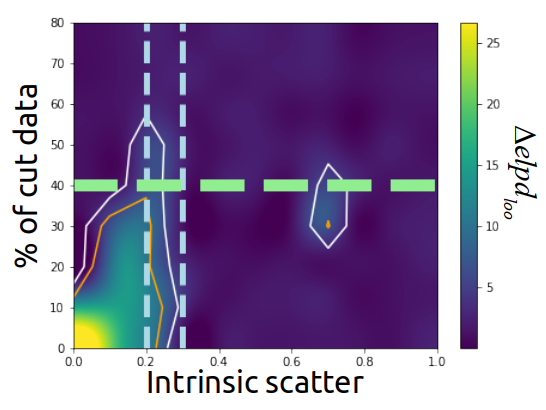}
    \includegraphics[width=9cm, height=8cm]{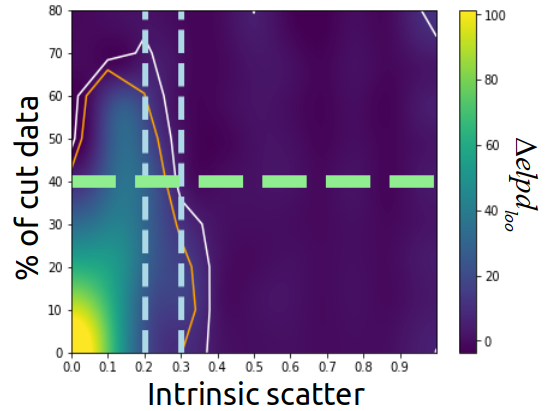}
    \includegraphics[width=9cm, height=8cm]{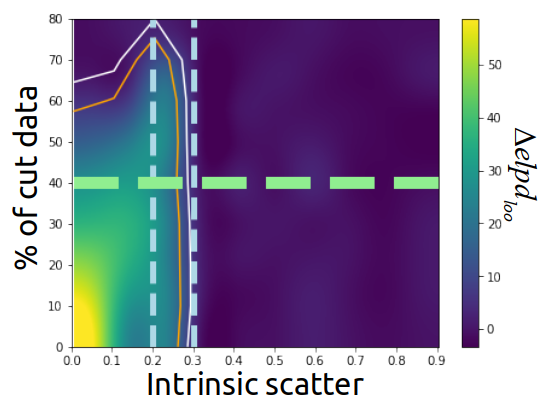}    \includegraphics[width=9cm, height=8cm]{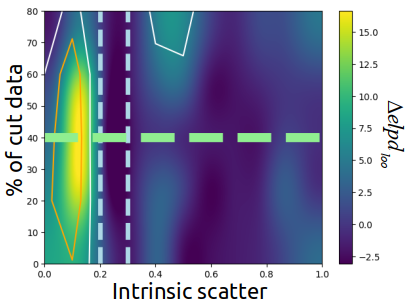}
    \caption{Plots of the ${\rm \Delta elpd_{loo}}$ of the two models (broken and power law ) as a function of both scatter (x-axis) and threshold set (y-axis) in terms of percentage of excluded data. Different figures represent different cases of break point assumption; from upper left to bottom right panel $x_c = 2,4,6,8$. The orange and white contours trace a ${\rm \Delta elpd_{loo}}$ of 10 and 4 respectively. The horizontal dashed line is placed at 40\%, while the two vertical lines are at one scatter of 0.20 and 0.30.}
    \label{fig:test_bkpl}
\end{figure*}

Here we create a power law distribution dataset assuming a slope change from 0.1 ($\alpha_1$) to 0.6 ($\alpha_2$) (similar to the observed point-to-point relations in radio halos) at $x = x_c = 2 \div 8$, with $1 \leq x_i \leq 10$ and with a normalization (beta) of 1:
\begin{align} 
    \large
    y_i = 
    \begin{cases}
            \beta ~ x_i^{{\alpha_1}}   &    {{\rm if ~}x_i < x_c} \\
            \beta x_c^{\alpha_1-\alpha_2} ~ x_i^{{\alpha_2}}   &  { {\rm if ~}x_i \ge x_c}
    \end{cases}
\end{align}   
The outcome is presented in Fig.~\ref{fig:test_bkpl} which shows the $\rm \Delta elpd_{loo}$ between the two models, where different plots are obtained using different $x_c$ values.\\
As might be expected, it shows that for low values of intrinsic scatter of the data and low thresholds (i.e. keeping all the data) there is strong evidence that the broken power law model is the best one (the higher the value in the plot, the stronger the evidence). Increasing the scatter and the cutting threshold, this difference is reduced until it falls below 4 (white lines).\\
%
We see also a dependence on the position of the break $x_c$.
This could also be expected since earlier break points will be sooner excluded when increasing the thresholds and then making the broken power law dataset become a simple power law one. This behaviour, and so the dependence on the chosen thresholds, almost disappears when $x_c$ is "large enough" (roughly, if it lies after $\sim 40\%$ of the $\{x_i\}$ dataset, i.e. $x_c=4$) and the only strong dependence is given by the scatter. In such a regime, when the scatter is above 25-30 \% is difficult to discern between a power law and a broken power law trend.\\
If we account also for threshold dependence, we see that when excluding less than $\sim 40 \%$ of the data it is possible to recognize a broken power law trend, even for "low" $x_c$ cases. Beyond this value is not so straightforward.\\
Given the dependences that we observe by both the threshold set and $x_c$, also for the case of a broken power law dataset, selection effects must be considered when performing a data regression.\\
Our analysis suggests that, once accounted for selection biases, for intrinsic scatters $\lesssim 25-30 \%$ a broken power law distribution can be identified by the fitting algorithm. \\
In our scientific analysis, four out of five objects show an intrinsic scatter less or equal to these limits. For this reason, we investigated possible slope changes in their $I_R - I_X$ relation.

\section{Point-to-point plot for single objects}\label{appendix:ptp-single}

Here we report the single $I_X - I_R$ relations for each studied target, accounting also for the cosmological dimming and k-correction (see Sec.~\ref{sec:ptp}). The points are color coded based on their distance from the radio halo emission peak taken from the Halo-FDCA results.

\begin{figure*}
    \centering
    \includegraphics[width=8cm, height=7cm]{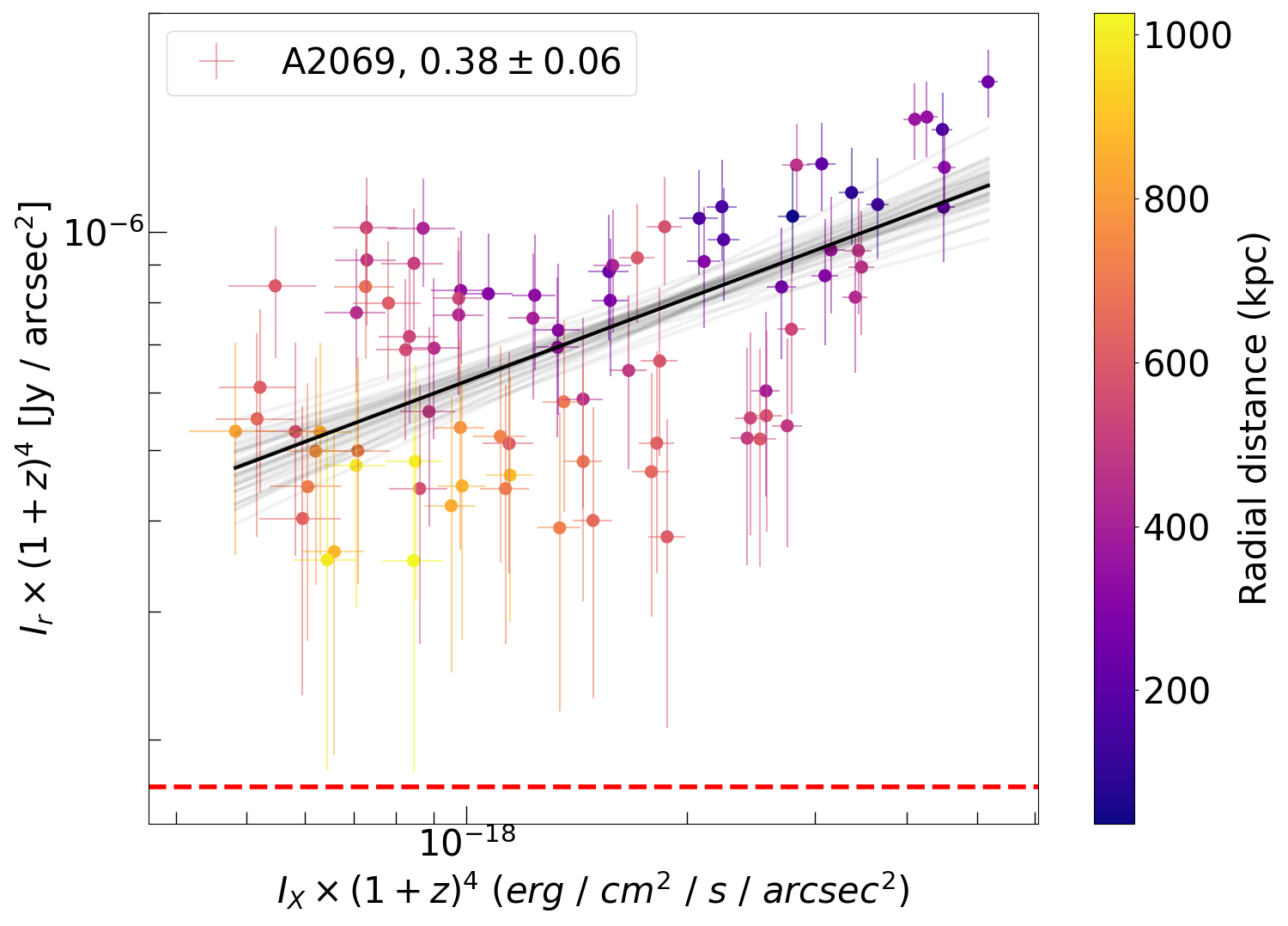}
    \includegraphics[width=8cm, height=7cm]{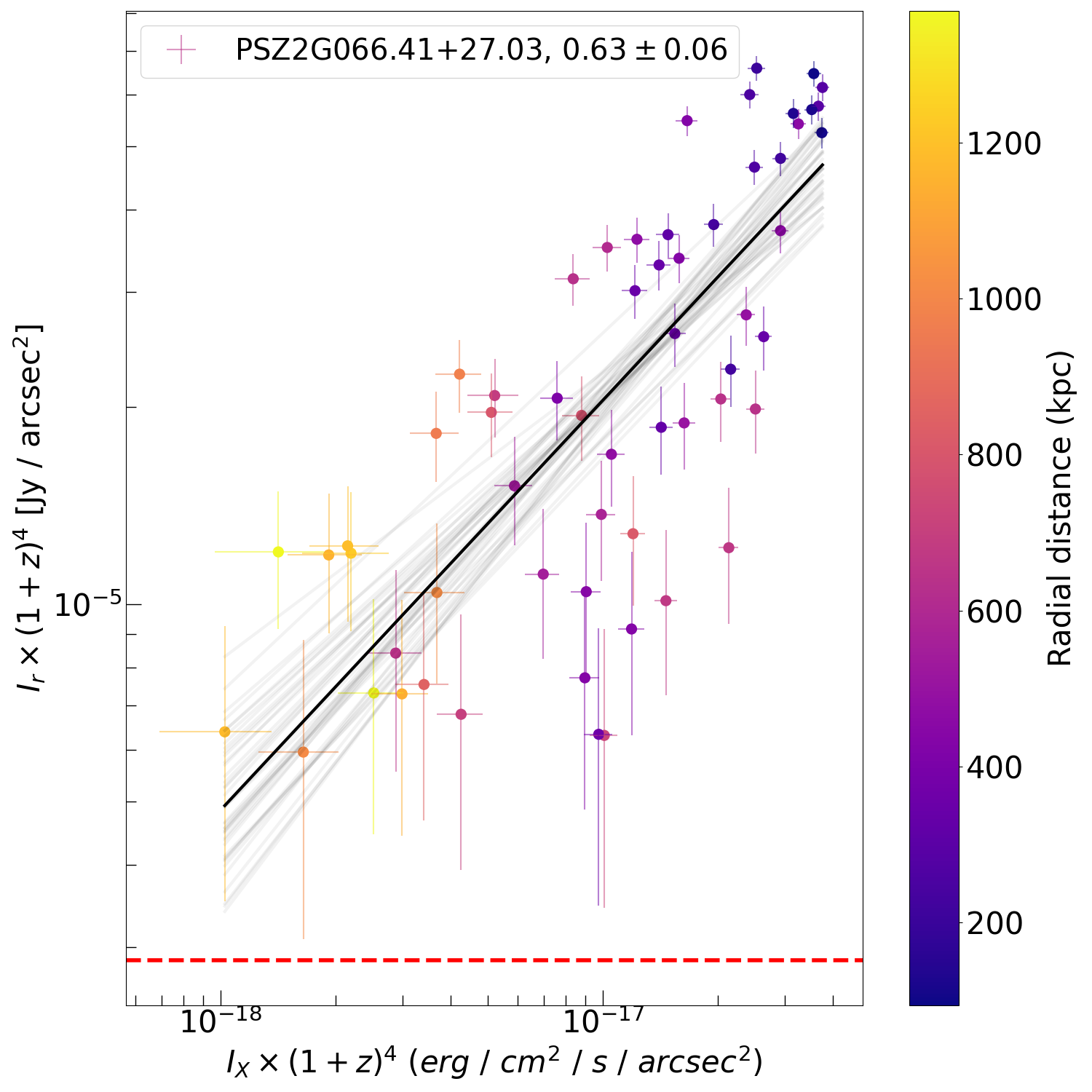}
    \includegraphics[width=8cm, height=7cm]{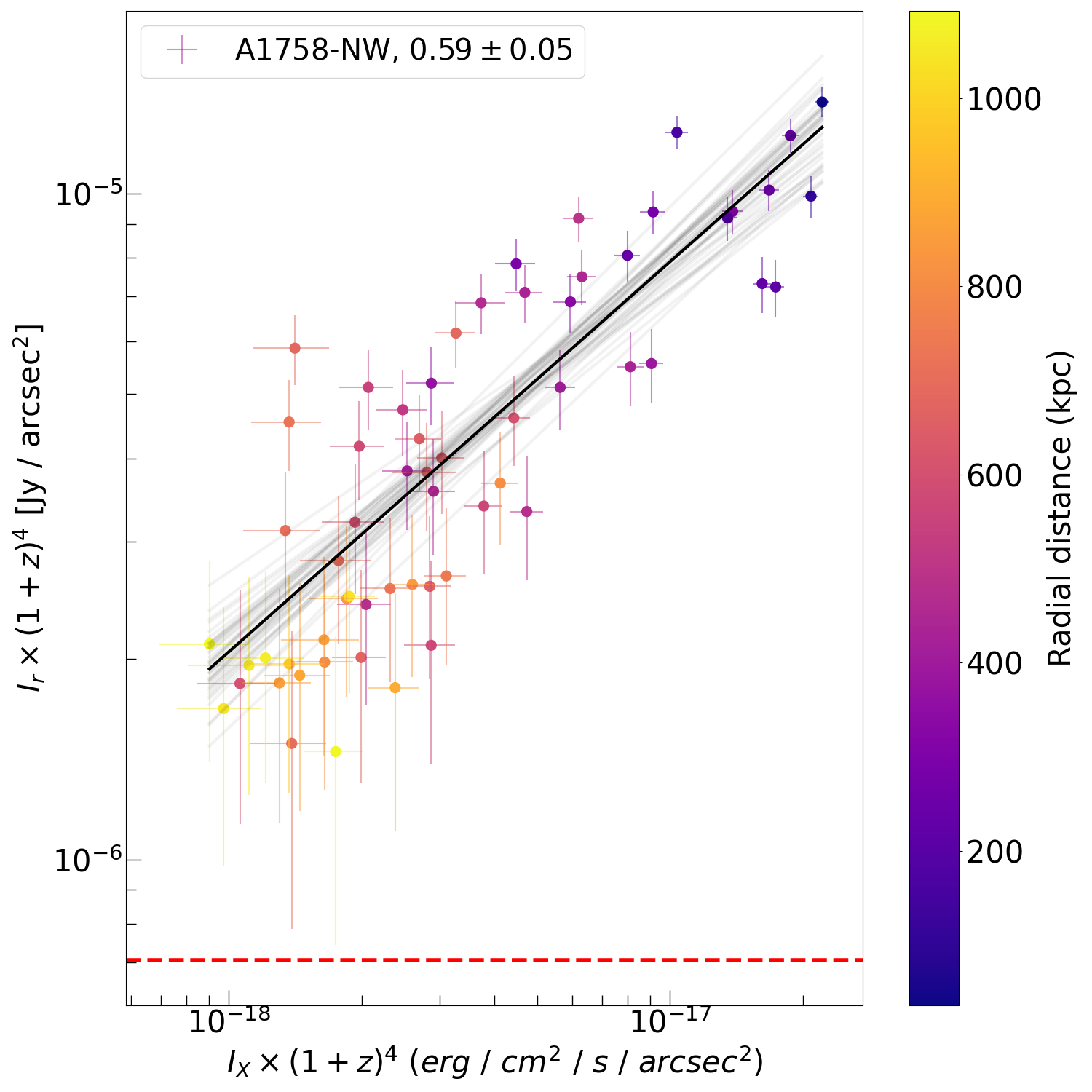}
    \includegraphics[width=8cm, height=7cm]{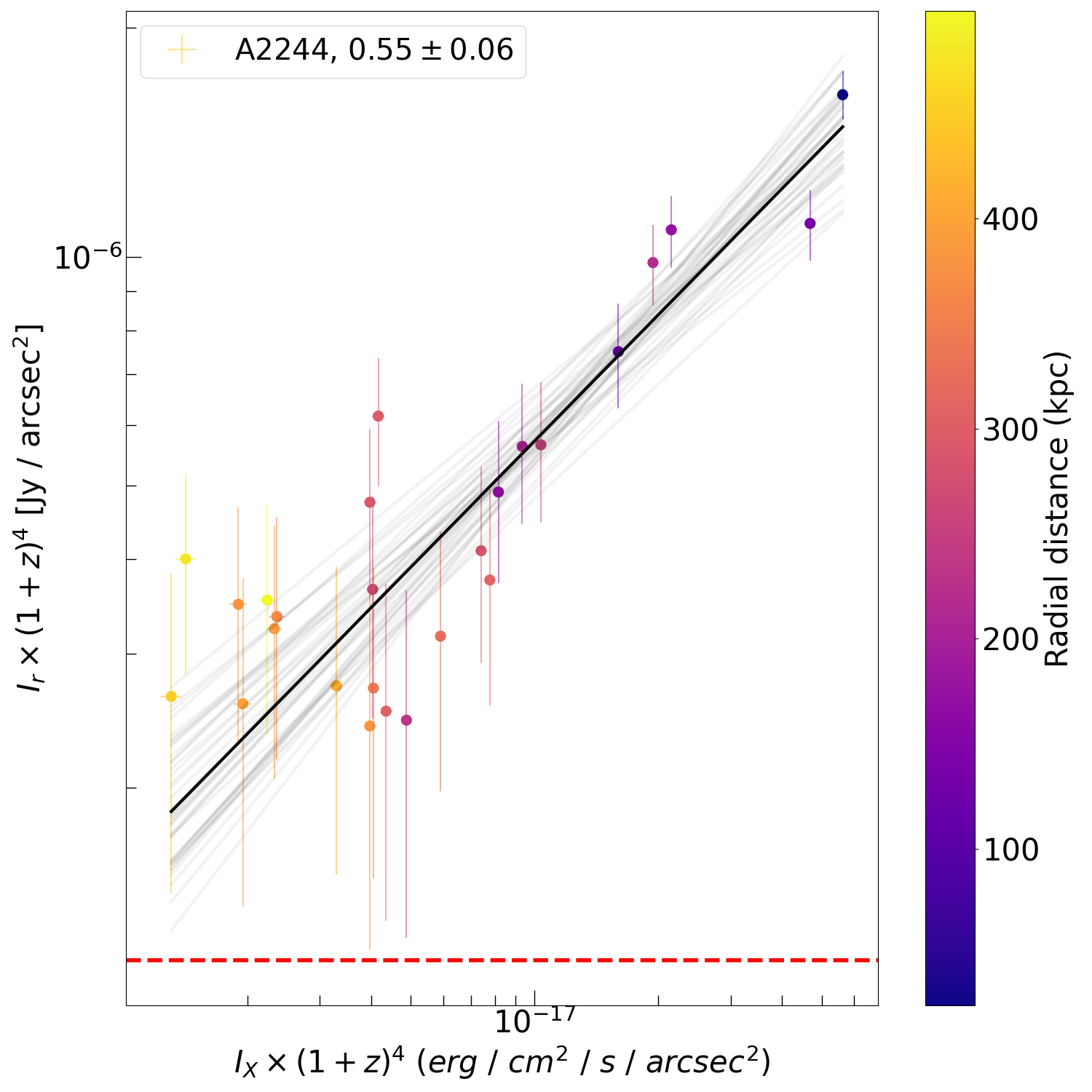}
    \includegraphics[width=8cm, height=7cm]{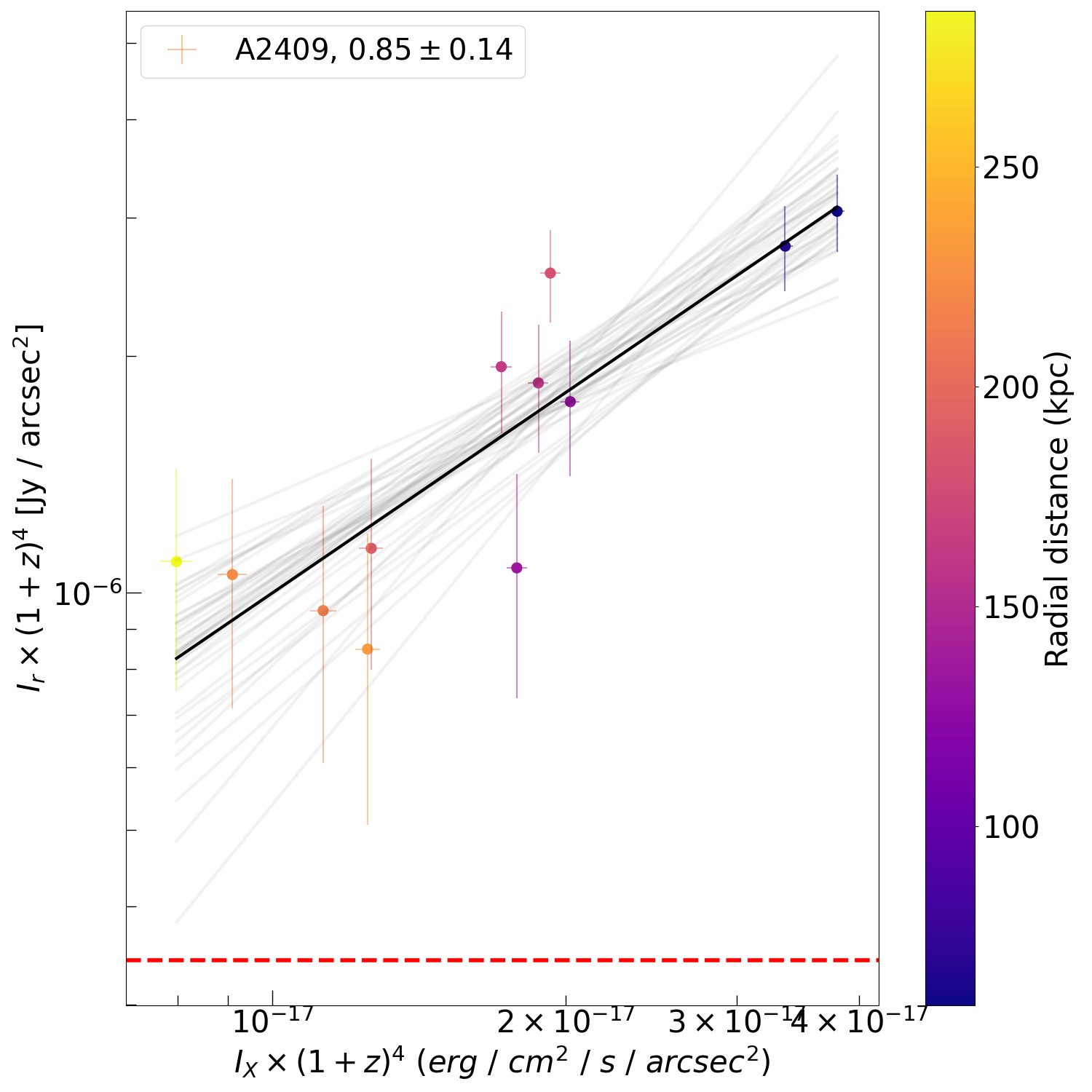}
    \caption{k-corrected $I_R - I_X$ relations for the five studied clusters. The points are color coded based on their distance from the radio halo emission peak. The best-fit line is shown in black and the red dashed line is the ${\rm \sigma_{RMS}}$ of the image.}
    \label{fig:PtP-plot_single}
\end{figure*}
\section{$I_R - I_X$ ratios}\label{sec:appendix_ratios}

In Sec.~\ref{sec:slope_trend} we explain the importance of radial spatially resolved studies. They can be performed both by studying the $I_R - I_X$ correlation slope and the $I_R/I_X$ ratio. We report the ratios between the radio and X-ray surface brightness, normalized to an arbitrary constant, computed using the radial binning described in Sec.~\ref{sec:radial_prof}.
The radial behaviour of the ratio $I_R/I_X$ bears similar information to the study of the slope presented in the main text. In particular, if the ratio $I_R/I_X$ remains constant throughout the cluster extension it means that the thermal and non-thermal components have a similar decline with the radius \citep[e.g.][]{Bonafede2022}. Instead, if this ratio changes it will be an indication of increasing importance of one component with respect to the other. An increasing ratio implies that the non-thermal component has a weaker decline with respect to the thermal one (i.e. more sublinear $I_R-I_X$ relation) and vice-versa for a decreasing profile.\\
We also overlayed the predictions made by the model presented in Sec.~\ref{sec:modeling}. This additionally allows to observe how the energetic contribution of CRe and ICM particles ($X(r)$) changes throughout the cluster extension.\\
\begin{figure*}
    \centering
    \includegraphics[width=8cm, height=6cm]{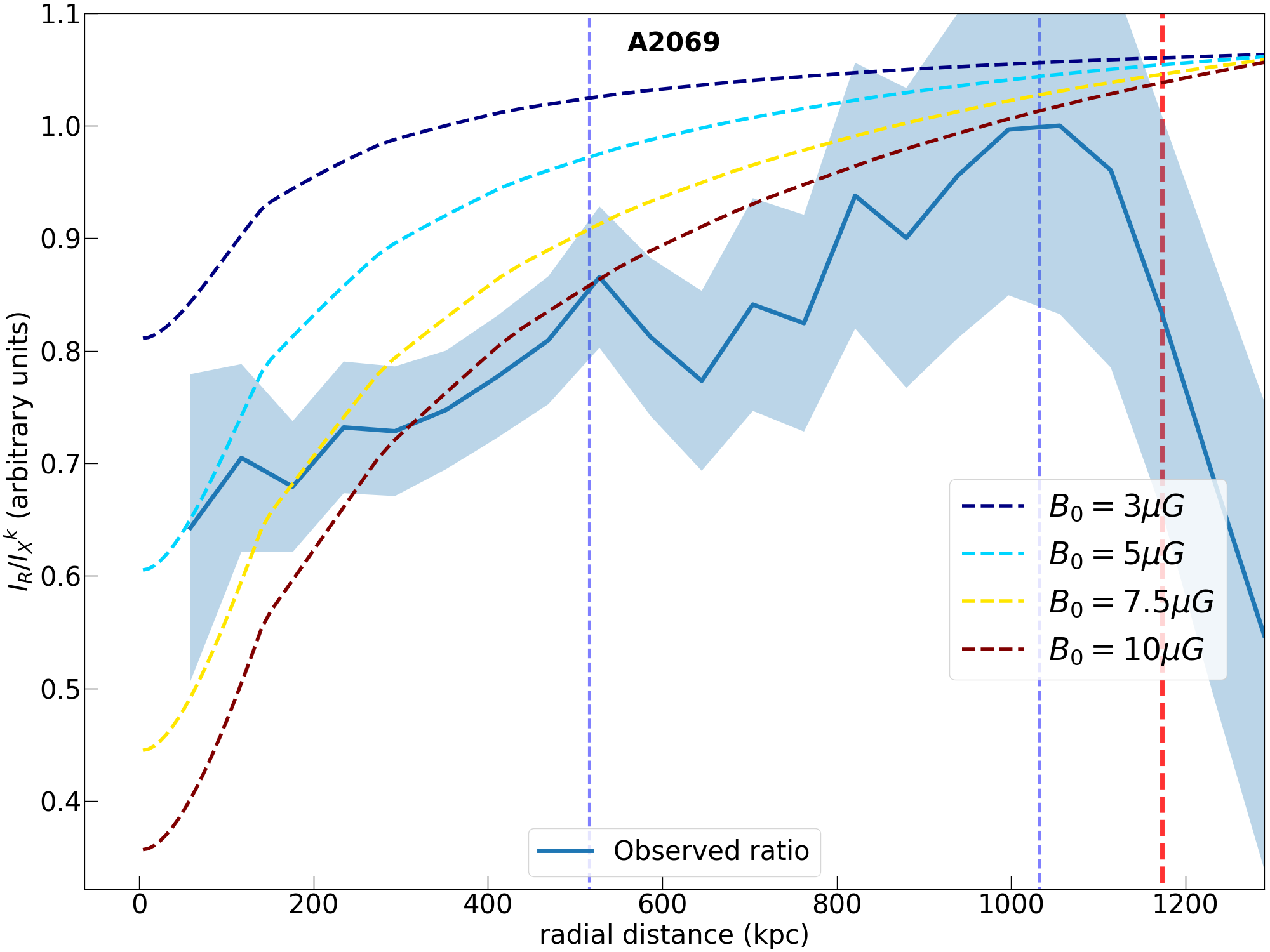}
    \includegraphics[width=8cm, height=6cm]{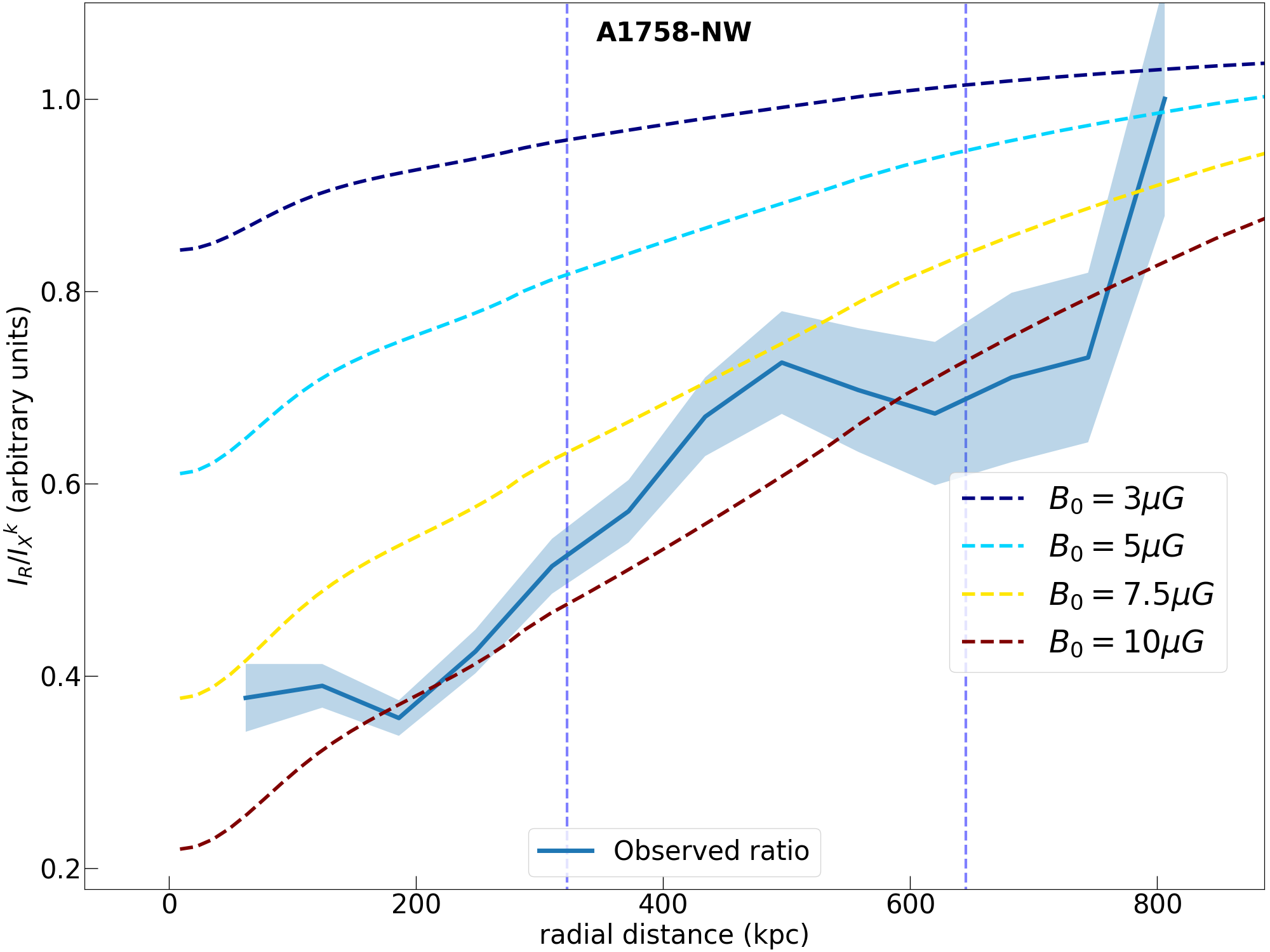}
    \includegraphics[width=8cm, height=6cm]{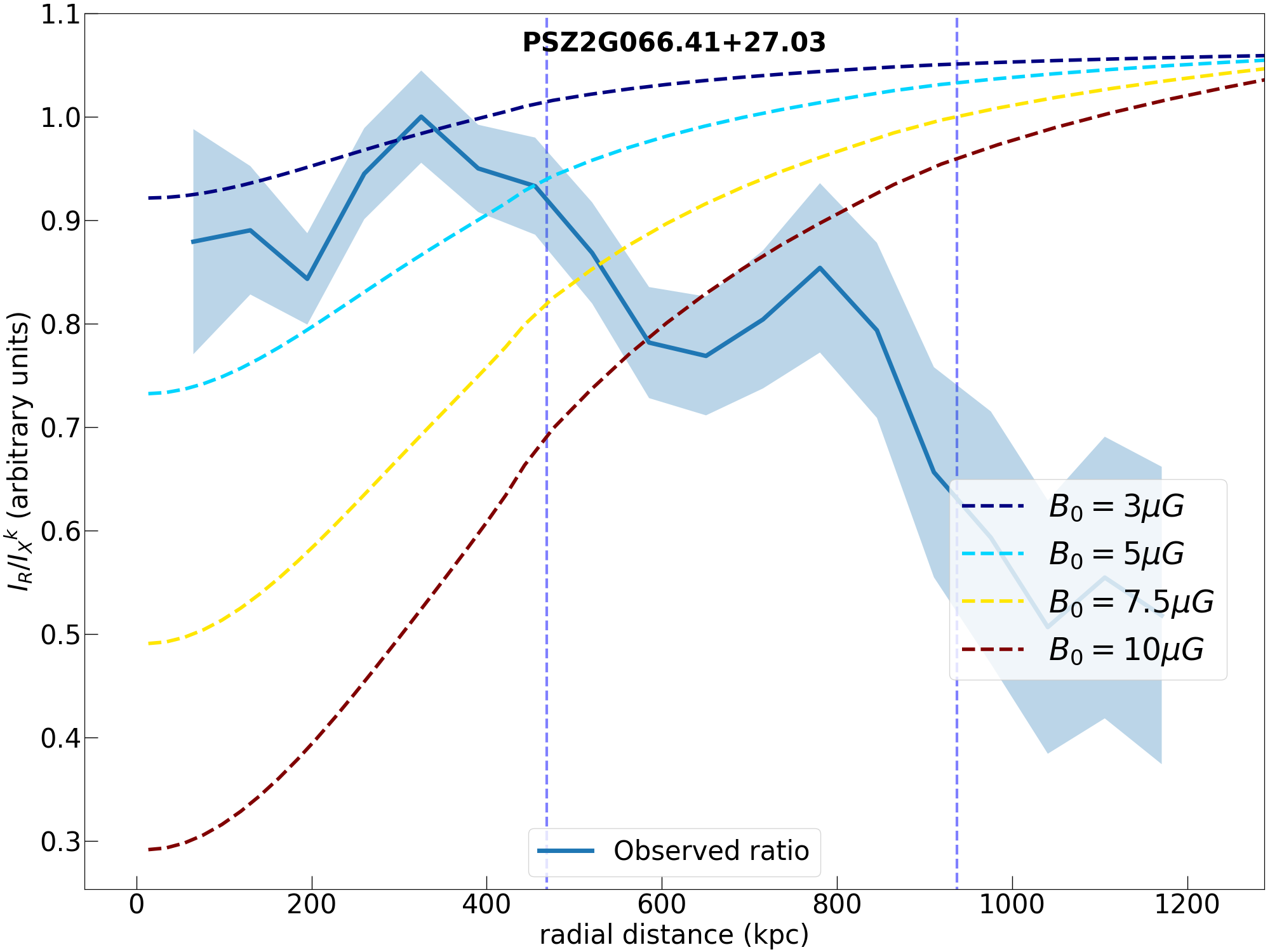}
    \includegraphics[width=8cm, height=6cm]{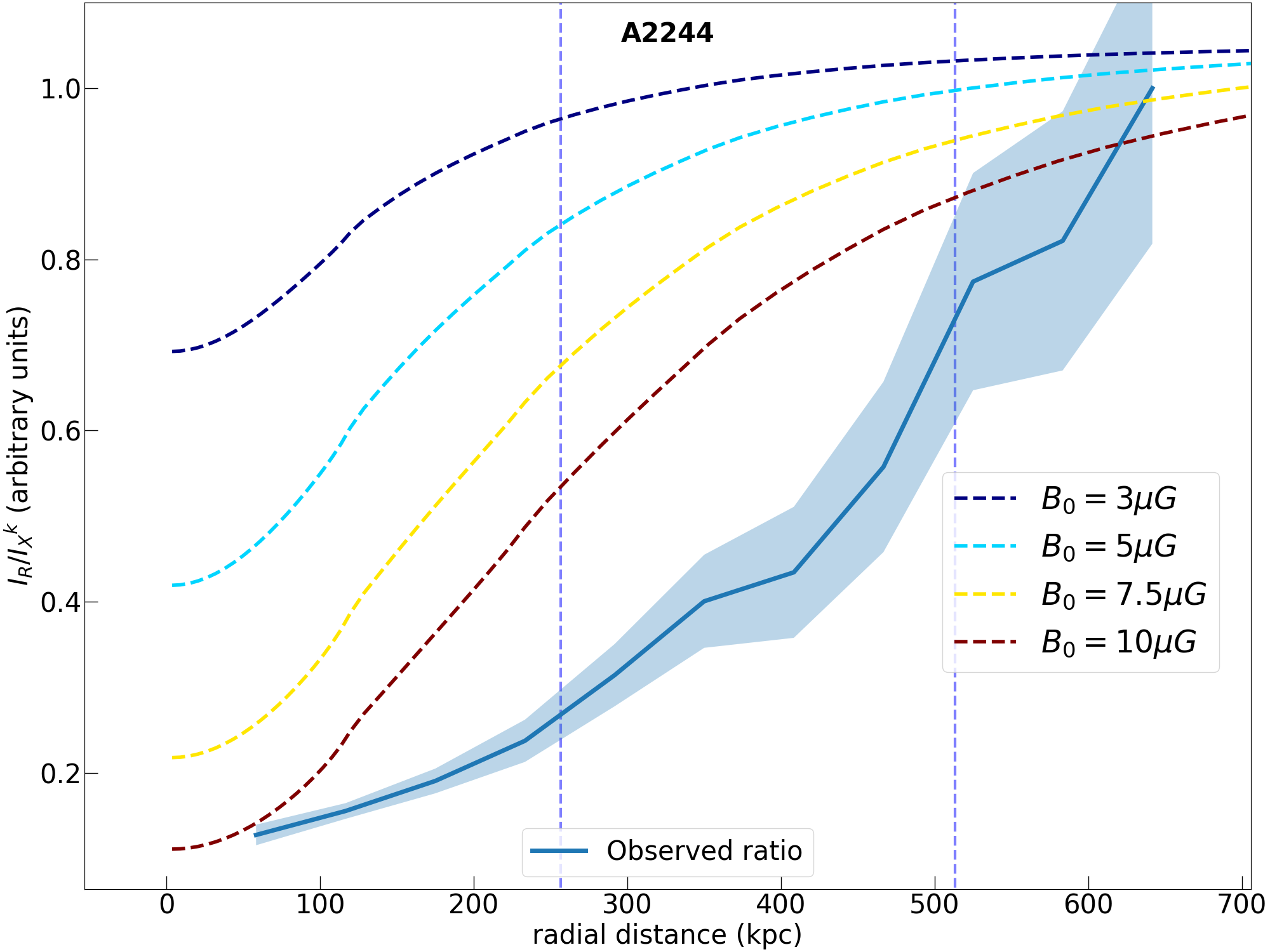}
    \includegraphics[width=8cm, height=6cm]{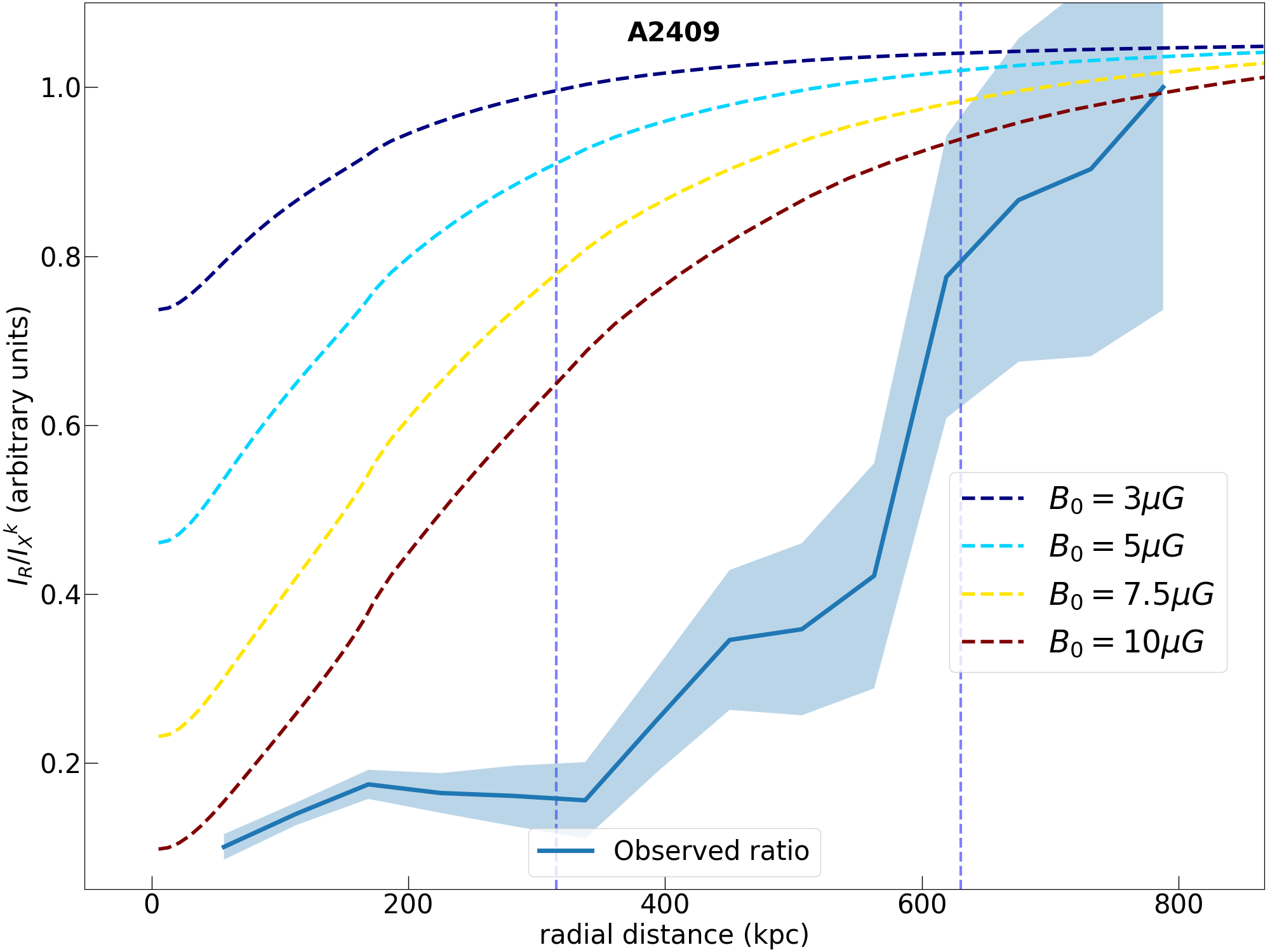}
    \caption{$I_R/I_X$ ratio as a function of the radius (i.e. distance from the halo centre) for the five presented targets. For A2069, the red dotted vertical line defines the radius beyond which the average signal in the annulus falls below 1 $\sigma_{RMS}$. The blue thinner lines are at $\rm R_e$ and $\rm 2 R_e$.}
    \label{fig:radio-X_ratios}
\end{figure*}
As found in Sec.~\ref{sec:slope_trend_disc} and \ref{sec:modeling}, we observe that the two less perturbed 
objects A2244 and A2409 have: (i) an increasing profile $I_R/I_X$ profile indicating more sub-linear trend in outer regions and (ii) an opposite trend with respect to the one predicted 
by the model, suggesting an increasing value of $X(r)$. Instead, the three more disturbed objects
have more complex profiles. In particular, A1758-NW is in line with model predictions while A2069 
has some discrepancies and a rapid decrease at large radii. Finally, PSZ2G066.41+27.03 has a 
profile that departs from the model, suggesting a decreasing value of $X(r)$.
\end{appendix}

\end{document}